\documentclass{jfm}
\usepackage{graphicx}
\usepackage{epstopdf, epsfig}
\usepackage[usenames, dvipsnames]{color}

\usepackage{url}
\usepackage{hyperref}
\usepackage[noabbrev]{cleveref}
\hypersetup{
	colorlinks   = true, %Colours links instead of ugly boxes
	urlcolor     = black, %Colour for external hyperlinks
	linkcolor    = blue, %Colour of internal links
	citecolor    = ForestGreen %Colour of citations
}

\usepackage{siunitx}
\usepackage{xfrac}
\usepackage{nicefrac}

\usepackage{amsmath}
\usepackage{graphicx}
\usepackage[labelsep=period]{caption}
\usepackage{subcaption}

\shorttitle{Dynamics of a liquid plug in a capillary tube under periodic forcings}
\shortauthor{S. Signe Mamba, J. C. Magniez, F. Zoueshtiagh and M. Baudoin}

\title{Dynamics of a liquid plug in a capillary tube under cyclic forcing: memory effect\textcolor{black}{s} and airway reopening}

\author{
	S. Signe Mamba, %\aff{1}
	J. C. Magniez,  %\aff{1,2},
	F. Zoueshtiagh  %\aff{1}
	\and 
	M. Baudoin \corresp{\email{michael.baudoin@univ-lille1.fr}}	
}

\affiliation{
	\aff{}
	 Univ. Lille, CNRS, Centrale Lille, ISEN, Univ. Valenciennes, UMR 8520 - IEMN, International laboratory LIA/LICS, F-59000 Lille, France
	 }

\begin{document}
	
	\maketitle
	
	\begin{abstract}
In this paper, we investigate both experimentally and theoretically the dynamics of a liquid plug driven by a cyclic \textcolor{black}{periodic} forcing inside a cylindrical rigid capillary tube. First, it is shown that depending on the type of forcing (flow rate or pressure cycle), the dynamics of the liquid plug can either be stable and periodic, or conversely accelerative and eventually leading to the plug rupture. In the latter case, we identify \textcolor{black}{the sources of the instability as: (i) the cyclic diminution of the plug viscous resistance to motion due to the decrease in the plug length and (ii) a cyclic reduction of the plug interfacial resistance due to a lubrication effect. Since the flow is quasi-static and the forcing periodic, this cyclic evolution of the resistances relies on the existence of flow memories stored in the length of the plug and the thickness of the trailing film.} Second, we show that contrary to unidirectional pressure forcing, cyclic forcing enables breaking of large plugs in confined space though it requires longer times. All the experimentally observed tendencies are quantitatively recovered from an analytical model. This study not only reveals the underlying physics but also opens up the prospect for the simulation of "breathing" of liquid plugs in complex geometries and the determination of optimal cycles for obstructed airways reopening.
	\end{abstract}
	
	\begin{keywords}
		Liquid plug, Slug, Bolus, Capillary tube, Cyclic forcing, Airway reopening
	\end{keywords}
	
	\section{Introduction}
Since the seminal works by \cite{fairbrother1935}, \cite{taylor1961deposition} and \cite{bretherton1961}, gas-liquid flows in capillary tubes have attracted much interest among different scientific communities due to their widespread occurence in many natural and engineered fluidic systems such as pulmonary flows \citep{grotberg2011respiratory}, oil extraction \citep{havre2000taming,mp_dimeglio_2011}, flow in porous media \citep{lenormand1983mechanisms,hirasaki1985mechanisms,dias1986network,stark2000motion}, or microfluidic systems \citep{gunther2004transport,assmann2011extraction,ladosz2016pressure}. In particular, liquid plugs (also called bridges, slugs or boluses) play a fundamental role in pulmonary flows where they can form in healthy subjects \citep{burger1968airway,hughes1970site} or in pathological conditions \citep{weiss1969acute,griese1997pulmonary,wright2000altered,hohlfeld2001role} due to a capillary or elasto-capillary instability \citep{kamm1989airway,white2005three,duclaux2006effects,heil2008mechanics,grotberg2011respiratory,dietze2015films}. \textcolor{black}{For patients suffering from pulmonary obstructive diseases, these occluding plugs may dramatically alter the distribution of air in the lungs, hence leading to severe breathing difficulties.}

Conversely, liquid plugs can be used for therapeutic purpose \citep{van1998lung,nimmo2002intratracheal}: boluses of surfactant are injected inside the lungs of prematurely born infants to compensate for their lack and improve ventilation \citep{engle2008surfactant,barber2010respiratory}. A thorough understanding of liquid plugs dynamics is therefore mandatory to improve both treatments of patients suffering from obstructive pulmonary diseases and of prematurely born infants.
	
When a liquid plug moves inside a cylindrical airway at low capillary number, deformation of the front and rear menisci occurs near the walls and leads to interfacial pressure jumps at the front and rear interfaces. This deformation also leads to the deposition of a liquid film on the walls. From a theoretical point of view, \cite{bretherton1961} was the first to provide an estimation of the pressure jump and the thickness of the liquid layer at asymptotically low capillary numbers. Bretherton's analysis was later formalized in the framework of matched asymptotic expansions by \cite{park1984two} who extended this work to higher order developments. Later on, the dynamics of a meniscus moving on a dry capillary tube was studied both experimentally and theoretically by \cite{hoffman1975study} and \cite{tanner1979spreading}. 

These pioneering results were later extended to unfold the effects of wall elasticity \citep{howell2000propagation}, the behavior at larger capillary numbers \citep{aussillous2000quick,klaseboer2014extended}, the effects of surfactants \citep{waters2002propagation}, the role of a microscopic or macroscopic precursor film \citep{jcis_jensen_2000,chebbi2003deformation}, the influence of more complex tube geometries \citep{wong1995motion1,wong1995motion2,hazel2002steady}, the influence of gravity \citep{suresh2005effect,zheng2007} or the influence of non-Newtownian properties of the liquid \citep{a_guttfinger_1965,jfm_hewson_2009,jnnfm_jalaal_2016,jfm_laborie_2017}. These key ingredients have then been combined with conservation laws determining the evolution of plug size and estimation of pressure jump in the bulk of the plug to provide analytical models of the unsteady dynamics of liquid plugs in capillary tubes \citep{baudoin2013airway,magniez2016dynamics,fujioka2016reduced}. In particular, \cite{baudoin2013airway} introduced the long range and short range interactions between plugs to simulate the collective behaviour of a train of liquid plugs. These models were in turn used to determine the critical pressure head required to rupture a liquid plug in a compliant \citep{howell2000propagation} or rigid  prewetted capillary tube (\cite{magniez2016dynamics}), or determine the maximum stresses exerted on the walls \citep{fujioka2016reduced}, a fundamental problem for lung injury produced by the presence of liquid plugs in the lung. 
 
 It is interesting to note that the dynamics of bubbles \citep{bretherton1961,ratulowski1989,fries2008segmented,warnier2010pressure} and liquid plugs in capillary tubes look similar from a theoretical point of view, since the interfacial pressure jumps and the deposition of a liquid film on the walls induced by the dynamical deformation of the interfaces can be calculated with the same formula. Nevertheless, there are also fundamental differences, which lead to very different dynamics: Trains of bubble are pushed by a liquid finger whose viscous resistance to the flow is generally higher than the \textcolor{black}{resistance induced by the presence of the bubble}. In this case, a pressure driven flow is stable and the flow rate remains essentially constant over time. In the case of liquid plugs, the resistance of the plugs to motion is higher than the one of the air in front and behind the plug. This leads to an unstable behavior with an acceleration and rupture of the plugs \citep{baudoin2013airway} or a deceleration and the obstruction of the airways \citep{magniez2016dynamics}.
 
From an experimental point of view, Bretherton's interfacial laws have been extensively verified for different systems (bubbles, liquid fingers, foams, ...). Nevertheless, there have been few attempts to compare the unstable dynamics of single or multiple plugs in capillary tubes to models accounting for the interface and bulk pressure jumps along with mass balance.  \cite{baudoin2013airway} showed that their model was able to qualitatively predict the collective accelerative dynamics of multiple plugs in rectangular microfluidic channels. More recently, \cite{magniez2016dynamics} were able to quantitatively reproduce the acceleration and deceleration of a single liquid plug in a prewetted capillary tube. Their model further provided the critical pressure below which the plug slows down and thickens whereas above it accelerates and ruptures. These experiments were particularly challenging owing to the complexity of controlling the prewetting film thickness and performing the experiments before the occurence of Rayleigh-Plateau instability. \cite{huh2007acoustically} measured in realistic experiments the injury caused by the motion of liquid plugs on humain airway epithelia deposited at the surface of an engineered  microfluidic airway. Later on, \cite{zheng2009} quantified the deformation of the walls induced by the propagation of a liquid plug in a flexible microchannel. \cite{song2011air} employed microfluidic technics to investigate single liquid plug flow in a tree geometry and evidenced the role of the forcing condition on the flow pattern. Finally, \cite{hu2015microfluidic} studied the rupture of a mucus-like liquid plug in a small microfluidic channel.

From a numerical point of view, simulations of liquid plugs in capillary tubes are highly challenging. Indeed, the thin layer of liquid left on the walls requires either adaptive mesh or the use of  Boundary Integral Methods to reduce the computational costs. Moreover, the unstable dynamics of the plugs pushed at constant pressure head leads to high variability in the associated characteristic times. \cite{fujioka2004steady} were the first to provide numerical simulations of the steady dynamics of a liquid plug in a two-dimensional channel. Later on, they studied the effects of surfactants \citep{fujioka2005steady}, the unsteady propagation \citep{fujioka2008unsteady} in an axisymmetric tube, the effects of gravity \citep{zheng2007} the role played by the tube's flexibility \citep{zheng2009} and the motion of Bingham liquid plugs \citep{zamankhan2012}. More recently, \cite{vaughan2016splitting} studied numerically the splitting of a two dimensional liquid plug at an airway bifurcation.

In all the aforementioned theoretical, experimental and numerical studies, the liquid plugs are pushed either at constant flow rate or at constant pressure head in a single direction. These driving conditions substantially differ from the one in the lung where liquid plug will experience periodic forcing. In this paper, we investigate both experimentally and theoretically the response of liquid plugs to cyclic and periodic pressure or flow rate forcing. The experiments are conducted in straight cylindrical glass capillary tubes and compared to an extended theoretical model based on previous developments by \cite{baudoin2013airway} and  \cite{magniez2016dynamics}. It is shown that, depending on the type of forcing (flow rate or pressure cycle), the dynamics of the liquid plug can either be periodic with the reproduction of the same cyclic motion over time, or accelerative eventually leading to the plug rupture. In particular, this study discloses the central hysteretic role played by the liquid film deposition on the plug dynamics.

The paper is organized as follows: Section \ref{method} describes the experimental set-up and the mathematical model. Section \ref{Periodic_forcings} is dedicated to the comparison of different type of forcings: pressure head and flow rate. In section \ref{cyclic_motion}, we compare the efficiency of cyclic and unidirectional forcings for obstructed airways reopening. Finally, concluding remarks and future prospects are provided in section \ref{conclusion}.

\section{Methods} \label{method}

\subsection{Experimental set-up}

\begin{figure}
	\centerline{\includegraphics[width=12cm]{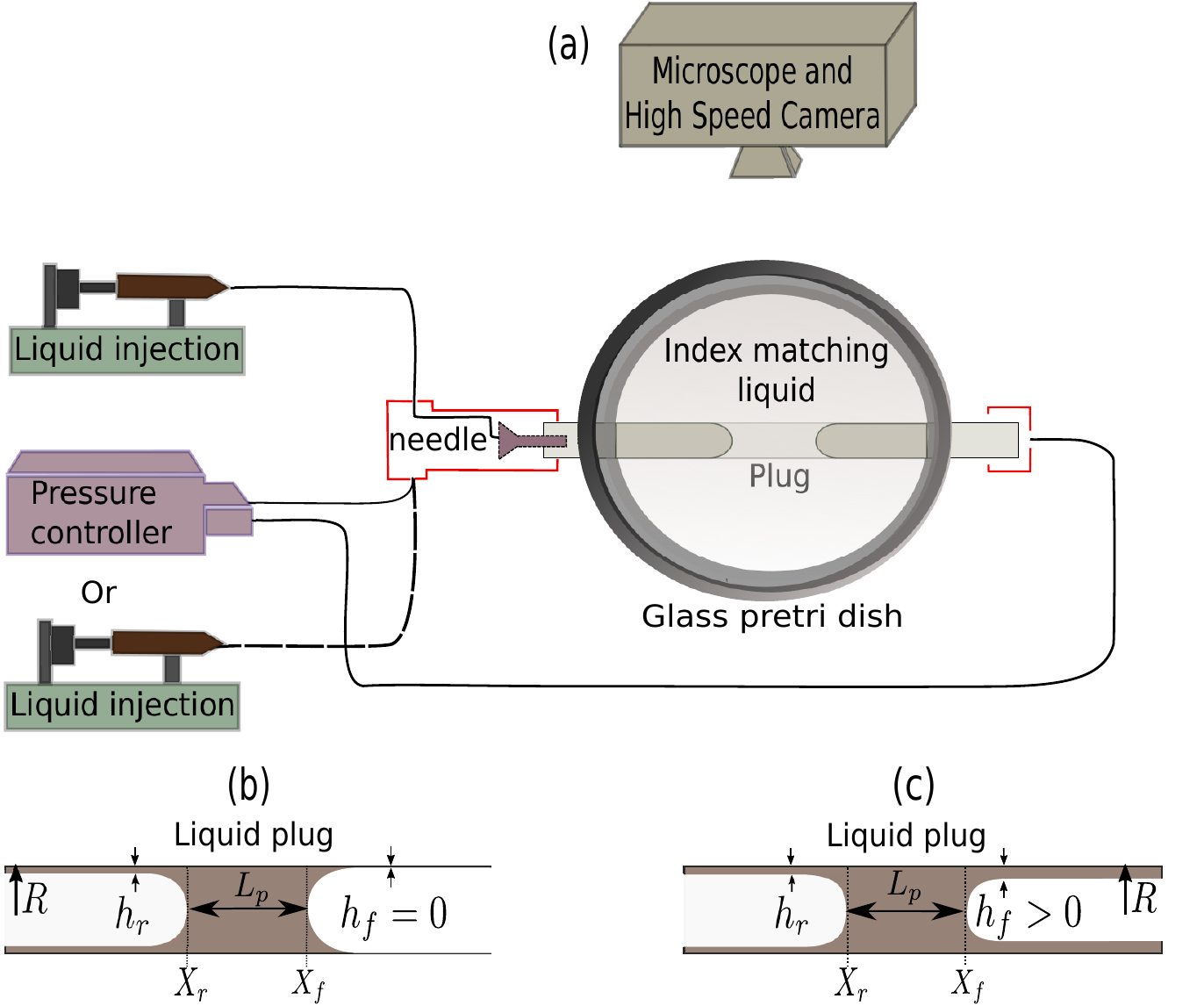}}
	\caption{(a) Sketch of the experimental set-up.  (b) First half cycle : the liquid plug moves on a dry capillary tube.  (c) Following back and forth motions: the liquid plug moves on a prewetted capillary tube.}
	\label{setup}
\end{figure}
The schematic of the experimental set-up is provided in \cref{setup}(a). A perfluorodecalin liquid plug of controlled volume is injected through a needle inside a rigid horizontal cylindrical glass capillary tube (inner radius  $R=\SI{470}{\micro\metre}$). Then, air is blown at low  flow rate ($Q =10 \mu$l/min) to bring the liquid plug to the center of the channel \textcolor{black}{and stopped manually when the plug reaches the target position. Thus, depending on the size of the liquid plug, the creation step can take up to 10s.} Finally, liquid plugs are moved back and forth with either pressure or flow rate cyclic forcings enforced respectively with a MFCS Fluigent  pressure controller or a KdScientific 210 programmable syringe pump. For both pressure driven and flow rate driven experiments, the period of oscillation is fixed at $2T = 4s$, with $T$ the duration of the motion in one direction.  It is important to note that during the first half-cycle, $t \in [ 0, T]$, the liquid plug moves along a dry capillary tube (\cref{setup}(b)). The motion of the plug leads to the deposition of a trailing film on the walls of thickness $h_r$ behind the rear meniscus at position $X_r(t)$. Thus, in the subsequent back and forth motion, the front interface of the liquid plug moves on walls prewetted by a layer of thickness $h_f(X_f)$ (with $X_f(t) = X_r(t) + L_p(t)$ the position of the front meniscus and $L_p(t)$ the plug length) as long as it remains on a portion of the channel already visited by the liquid plug (\cref{setup}(c)) in a previous half-cycle. 

The glass tubes were cleaned prior to the experiments with acetone, isopropanol, dichloromethane and piranha solutions (a mixture of sulfuric acid (H2SO4) and hydrogen peroxide (H2O2)) successively to obtain perfectly wetting surface and prevent dewetting induced by the presence of dust or organic contaminants on the surface. Perfluorodecalin (dynamic viscosity $ \mu = 5.1\times 10^{-3} Pa.s$, surface tension $ \sigma = 19.3 \times 10^{-3} N/m$ and density $ \rho = 1.9\times10^{-3} kg/m^{3}$) was chosen for its good wetting properties and inertness. Experiments are recorded using a Photron SA3 high speed camera mounted on a Z16 Leica Microscope at a frame rate of $125$ images per second, a trigger time of $1/3000 s$ and a resolution of $1024 \times 64$ pixels. To prevent image deformation due to the cylindrical shape of the capillary tube, it is immersed in an index-matching liquid. The image analysis is then performed using ImageJ software and Matlab.

\subsection{Dimensional analysis of the problem}

\label{ssection:da}

The characteristic parameters in this problem are the radius of the tube $R$, the surface tension $\sigma$, the liquid density $\rho$ and viscosity $\mu$ , and the characteristic speed $U$ of the liquid plug. From these parameters, one can derive the characteristic convection time $ \tau _c = R/U$, the characteristic viscous diffusion time $ \tau _v = {{\rho}{R^2}}/{\mu} $, the Reynolds number $\Rey = \rho U R / \mu$ (comparing inertia to viscous diffusion), the capillary number $Ca = {\mu U}/{\sigma}$ (comparing viscous diffusion to surface tension), the Bond number ${\it Bo} = {\Delta {\rho} g {R^2}}/{\sigma}$ (comparing gravity effects to surface tension) and finally the Weber number ${\it We}={\Delta {\rho} {U^2} R}/{\sigma}$ (comparing inertia to surface tension). Table \ref{tab1} summaries the maximum values of these key dimensionless parameters based on the maximal velocity of the liquid plug $U_m = 28mm/s$ observed in the present experiments.

Based on the order of magnitude of these dimensionless parameters, a few primary insights can be drawn. The low Bond number and the horizontal position of the tube suggest weak effect of gravity in this problem. The flow in the bulk of the plug remains laminar owing to the moderate values of the Reynolds number. In addition, \cite{aussillous2000quick} studied the impact of inertia on the deposition of a trailing liquid film behind a moving liquid plug. From dimensional analysis and experiments, they introduced a critical capillary number $Ca_c$ (equal to ${3.6 \times 10^{-1}}$ in the present case) above which the effect of inertia becomes significant. In the present experiments, the capillary number is two order of magnitude smaller than this critical value and thus inertia can be neglected in the film deposition process. Finally, \cite{kreutzer2005inertial} studied numerically the influence of inertia on pressure drops at liquid/air interfaces. They showed that inertia plays no role for $\Rey<10$ at capillary numbers comparable to the present study. Thus, inertial effects can safely be neglected here. Furthermore, the weak capillary and Weber numbers indicate that surface tension is globally dominant over viscous stresses and inertia. Nevertheless, it is to be emphasized that viscous effects must still be accounted for close to the walls, in the so-called "dynamic meniscus" that is the part of the meniscus deformed by viscous stresses. Finally, since the convection and viscous diffusion times $\tau_c$ and $\tau_v$ are two orders of magnitude smaller than the duration of the pressure or flow rate cycles, unsteady term in Navier-Stokes equation can be neglected and the flow can be considered as quasi-static. 

\textcolor{black}{Another phenomenon that may occur during the plug motion is the destabilisation of the trailing liquid film due to a Rayleigh-Plateau instability. The characteristic time associated with the most unstable mode is given by the following formula \citep{d_chandrasekhar_1961,rmp_eggers_1997}:
$$
\tau_{RP} = \frac{12 \mu R^4}{\sigma h^3}
$$
The smallest destabilisation times is thus obtained for the thickest fluid layer. In the experiments conducted in this paper, the thickness of the liquid film remains typically smaller than $5 \%$ of the tube radius leading to $\tau_{RP} = 13$ s, whose value remains significantly larger than the period of the plug motion ($2T = 4$ s). In addition, this time grows rapidly ($\propto 1/h^3$) when the thickness of the layer is decreased ( $\tau_{RP} = 58$ s for $h/R = 3 \%$) and thick films are only deposited close to the plug rupture in the pressure driven experiments so that the destabilisation of the trailing film is expected to play a minor role in the following experiments.}

\begin{table}
	\begin{center}
		\def~{\hphantom{0}}
		\begin{tabular}{lccc}
			parameters  & Formula   &   Maximum value   \\[3pt]
			$ \tau _c$       & $R/U$   & ~~ $1.7 \times {10^{-2}} \; s$ ~ \\
			$ \tau _v$       & $ {{\rho}{R^2}}/{\mu}$   & ~~ $8.2 \times {10^{-2}} \; s$ ~ \\
			\Rey       & $ {\tau _d / \tau _c}$   & ~~ $4.9$ ~ \\
			{\it Ca}   & $ {\mu U}/{\sigma}$ & ~~ $ 7.4 \times {10^{-3}}$ \\
			{\it Bo}   & ${\Delta {\rho} g {R^2}}/{\sigma}$ & ~~ $ 2.1 \times {10^{-1}}$ \\
			{\it We}   & $ {\Delta {\rho} {U^2} R}/{\sigma}$ & ~ $ 3.6 \times {10^{-2}}$ \\
		\end{tabular}
		\caption{Values of the key parameters associated with the maximal velocity $U_m$}
		{\label{tab1}}
	\end{center}
\end{table}

\subsection{Model: pressure driven forcing}
\label{ssection:model}

In the above context, the liquid plug dynamics can be predicted from a quasi-static pressure balance and a mass balance.  We thus adapted a visco-capillary quasi-static model previously introduced by \cite{magniez2016dynamics} to include the motion on both dry and prewetted portions of the tube and also the memory effects resulting from a trailing liquid film deposition. Assuming that the pressure losses in the gas phase are negligible compared to that induced by the liquid plug, the total pressure jump $\Delta P_t$  across a liquid plug can be decomposed into the sum of the pressure jump induced by the presence of the rear interface $\Delta P ^{int}_{rear}$, the front interface  $\Delta P ^{int}_{front}$ and the flow in the bulk of the plug $\Delta P ^{bulk}_{visc}$:
\begin{equation}
	\Delta P_t = \Delta P ^{bulk}_{visc} + \Delta P ^{int}_{rear} + \Delta P ^{int}_{front}
\end{equation}
%where $\Delta P_t = {P_0} exp(-6 exp (-3t))$
In the experiments, $\Delta P_t $ corresponds to the driving pressure head.

Since the flow is laminar, the viscous pressure drop in the bulk of the plug can be estimated from Poiseuille's law:
\begin{equation}
	\Delta P ^{bulk}_{visc} = \dfrac{8 \mu L_p U}{R^2}
	\label{poiseuille}
\end{equation}
with $L_p$, the plug length, that is to say the distance between the front and rear meniscus $L_p(t) = X_f(t) - X_r(t)$ as described in \cref{setup} and $U = d {X_r}/dt$ the liquid plug velocity. \textcolor{black}{This expression relies on two approximations: (i) it assumes a Poiseuille flow structure and thus neglects the fluid recirculation occurring close to the menisci and (ii) it assumes the same speed for the rear and front interfaces. The validity of the first approximation has been tested with direct numerical simulations performed with a Volume of Fluid (VOF) method (see appendix \ref{viscousdrop}). As expected, the accuracy of this approximated formula decreases with the size of the plug. Nevertheless, the difference between approximated and computed values remains below $4.5 \%$ for plugs larger than the tube diameter and below $25 \%$ for plugs larger than $1/4$ of the tube diameter. The approximation is therefore consistent for large plug. The larger discrepancy observed for smaller plugs is not critical since in this case the interfacial pressure drops at the rear and front interfaces are largely dominant over the viscous one. We verified the consistency of the second approximation experimentally by measuring the difference between the front and rear interface speeds. These measurements show that the speed of the two menisci only differ by a few percent. The reason is that the evolution of the plug size $dL_p / dt$ remains much smaller than the plug velocity $d {X_r}/dt$ for the essential part of the plug dynamics. In the following, we keep this assumption (front and rear interfaces velocities approximatively equal) to evaluate the interfacial pressure jump). }

\textcolor{black}{Since the Capillary, Bond and Reynolds numbers are small and the flow is quasi-static (see section \ref{ssection:da})}, the pressure drop across the rear interface of the moving liquid plug can be estimated from Bretherton's formula \citep{bretherton1961} :
\begin{equation}
	\Delta P ^{int}_{rear} = \dfrac{2 \sigma }{R}(1+1.79{(3Ca)^{2/3}})
\end{equation}
Finally, the Laplace pressure drop across the front meniscus depends on the apparent dynamic contact angle $\theta ^{a}_{d}$ according to the formula (in the limit of low capillary number and thus $\theta ^{a}_{d}$):
\begin{equation}
	\Delta P ^{int}_{front} = - \dfrac{2 \sigma  cos{\theta ^{a}_{d}}}{R} \approx - \dfrac{2 \sigma  (1-{\theta ^{a}_{d}}^2/2)}{R} 
\end{equation}
Choosing the Laplace pressure jump ${2 \sigma}/R$ as the characteristic pressure scale, and the tube radius $R$ as the characteristic length scale, the dimensionless pressure jump across the liquid plug becomes: 

\begin{equation}
	\Delta \tilde{P_t} = 4 \tilde{L_p} Ca + 1.79{(3Ca)^{2/3}} +  \frac{{\theta ^{a}_{d}}^2}{2}
\end{equation}
\textcolor{black}{A tilde indicates dimensionless functions.}

In order to achieve a closed set of equation, two additional equations must be derived. They are (i) the relation between the apparent dynamic contact angle of the front meniscus $\theta ^{a}_{d}$ and the capillary number $Ca$ and (ii) an equation determining the evolution of the plug length $\tilde{L_p}$. The first relation depends on the wetting state of the tube walls ahead of the liquid plug: 

When the liquid plug moves on a \textit{dry substrate}, this relation is given by Hoffman-Tanner's law valid at low capillary numbers:
\begin{equation}
	\theta ^{a}_{d} = E {{Ca}^{1/3}} 
\end{equation}
with $E$ a numerical constant of the order of $4€"-5$ for a dry cylindrical capillary tube as reported by \cite{hoffman1975study} and \cite{tanner1979spreading}. For a liquid plug moving on a \textit{prewetted substrate}, $\theta ^{a}_{d}$ can be calculated from Chebbi's law \citep{chebbi2003deformation}, which can be simplified at low capillary number through a second order Taylor expansion \citep{magniez2016dynamics}:
\begin{equation}
	\theta ^{a}_{d} = \dfrac{-1 + \sqrt{1 + 4 CD}}{2C}
\end{equation}
with:
\begin{eqnarray}
A &=& {(3 Ca)}^{-2/3} \tilde{h_f} \\
B &=& {(3 Ca)}^{1/3} \\
C &=& \frac{1}{log(10)} \bigg(\frac{b_1}{2}+ {b_2} log_{10}(A)  + \frac{3 b_3}{2} \left[ log_{10} A \right]^2 \bigg) B\\
D &=& \bigg( b_0 + {b_1} log_{10}(A)  + {b_2} \left[ log_{10}(A) \right]^2 + {b_3} \left[ log_{10} (A) \right]^3 \bigg) B\\
& & b_0 \approx1.4, \; b_1 \approx ˆ'-0.59, \; b_2 \approxˆ' -3.2 \times 10^{-2}, \mbox{ and } b_3 \approx 3.1 \times 10^{-3}
\end{eqnarray}
Since $CD \ll 1$ at low capillary number (in the present experiments $CD < 5 \times 10^{-2}$), this equation can further be  simplified into:
\begin{equation}
\theta_a^d = D = F Ca^{1/3}
\end{equation}
with $F = 3^{1/3}  \bigg( b_0 + {b_1} log_{10}(A)  + {b_2} \left[ log_{10}(A) \right]^2 + {b_3} \left[ log_{10} (A) \right]^3 \bigg)$.

The next step is to determine the dimensionless plug length $\tilde{L_p}$.  A simple mass balance between the fluid collected from the fluid layer lying ahead the plug (of thickness $h_f$) and the trailing liquid film (of thickness $h_r$) deposited behind the plug gives:
\begin{equation}
	dV = {({\pi} R^2 - {\pi}(R-h_f)^2)} {dX_f} - {({\pi} R^2 - {\pi}(R-h_r)^2)} {dX_r}
\end{equation}
with $V = \pi R^2 L_p$ the volume of the plug. Finally, with $d{X_r} = Udt$ and $d{X_f} = {\dfrac{(R-h_r)^2}{(R-h_f)^2}}d{X_r} $, we obtain:
\begin{equation}
	\dfrac{d L_p}{dt} = \bigg[{\dfrac{(R-h_r)^2}{(R-h_f)^2}} - 1 \bigg] U
\end{equation}

Using the capillary time scale, $\mu R/\sigma$, as the characteristic time scale, this equation can be rewritten in the dimensionless form:
\begin{equation} \label{eq:length_plug}
	\dfrac{d \tilde{L_p}}{d\tilde{t}} =  \bigg[ {\dfrac{(1-\tilde{h_r})^2}{(1-\tilde{h_f})^2}} - 1 \bigg] {Ca}
\end{equation}
The last essential point is to determine the thicknesses of the liquid film lying in front and left behind the liquid plug $h_r$ and $h_f$ respectively. The thickness of the trailing film can be calculated from an extension of Bretherthon's law introduced by  \cite{aussillous2000quick}. This thickness only depends on the velocity of the plug $U$, that is to say in dimensionless form only on the capillary number $Ca$:
\begin{equation} \label{eq:film_deposition}
	\tilde{h_r}= \dfrac{1.34 {Ca}^{2/3}}{1 + 2.5 \times 1.34 {Ca}^{2/3}}
\end{equation}
Finally, the thickness $\tilde{h_f}$ depends on the history of the plug motion. Indeed, the capillary tube is initially dry. Thus,  for a cyclic motion, the liquid film lying ahead of the plug at position $X_f^N$ during the half-cycle N comes from the deposition of a trailing film behind the plug at the same position $X_r^{N-1} = X_f^N$ during the half cycle $N-1$. In order to determine $\tilde{h_f}$,  the thickness of the liquid film deposited on the walls must therefore be kept in memory and then taken as an entry when the plug moves back to the same location. If the plug moves to a location never visited before, then the tube is dry,  $\tilde{h_f} = 0$ and the pressure jump for the front interface corresponds to the dry version.This analysis shows that the liquid film acts as a memory of the liquid plug motion. \textcolor{black}{Nevertheless, each back and forth motion of the liquid plug prescribes new values of the liquid film thickness. This means that this memory is a short term memory whose lifetime is a half cycle.}

To summarize, the complete nonlinear system of equations that need to be solved to determine the evolution of the plug is:
\begin{eqnarray}
& & \Delta \tilde{P_t}  =  \begin{cases} { 4 \tilde{L_p} Ca} +\bigg(3.72 + \dfrac{E^2}{2} \bigg)  {Ca^{2/3}}  , & \mbox{if } { dry } \label{eq:ptdry} \\ 
		{4 \tilde{L_p} Ca + \bigg(3.72 +  \dfrac{F^2}{2}  \bigg) {Ca^{2/3}} } , & \mbox{if } { prewetted} \end{cases} \\
& & F = 3^{1/3}  \bigg( b_0 + {b_1} log_{10}(A)  + {b_2} \left[ log_{10}(A) \right]^2 + {b_3} \left[ log_{10} (A) \right]^3 \bigg) \label{eq:lubri2}\\
& & A = {(3 Ca)}^{-2/3} \tilde{h_f}  \label{eq:lubri} \\
&  & \frac{d \tilde{X_r}}{d \tilde{t}} = Ca, \quad \tilde{X_f} = \tilde{X_r} + \tilde{L_p}  \\
& & \dfrac{d \tilde{L_p}}{d\tilde{t}}  =    \bigg[ {\dfrac{(1-\tilde{h_r})^2}{(1-\tilde{h_f})^2}} - 1 \bigg] {Ca} \\
& & \tilde{h_r} =  \dfrac{1.34 {Ca}^{2/3}}{1 + 2.5 \times 1.34 {Ca}^{2/3}} \label{eq:hr} \\
& & \tilde{h_f}(\tilde{X_f}) \mbox{ is obtained from the memory of the liquid film deposition} \label{eq:mem}  
\end{eqnarray}
At each change of flow direction, the front meniscus becomes the rear meniscus and vice versa. The pressure balance in the dry and prewetted tubes share a relatively similar expression, but the coefficient $E$ remains constant while $F$ depends both on $Ca$ and  $\tilde{h_r}$. This system of equations is solved numerically with a \textcolor{black}{first order Euler explicit method}. Since the dynamics is accelerative, an adaptive time step refinement is used. It consists in keeping the spatial displacement over a time step constant: $\Delta{\tilde{t}} = \Delta{\tilde{x}} / Ca$ with $\Delta{\tilde{x}}$ constant. Convergence analysis on $\Delta{\tilde{x}}$ was performed for the calculations presented in this paper.

\subsection{Validation of the model for unidirectional pressure forcing in a dry capillary tube}
\label{ss:uni}

\begin{figure}
\begin{center}
\includegraphics[width=0.5\textwidth]{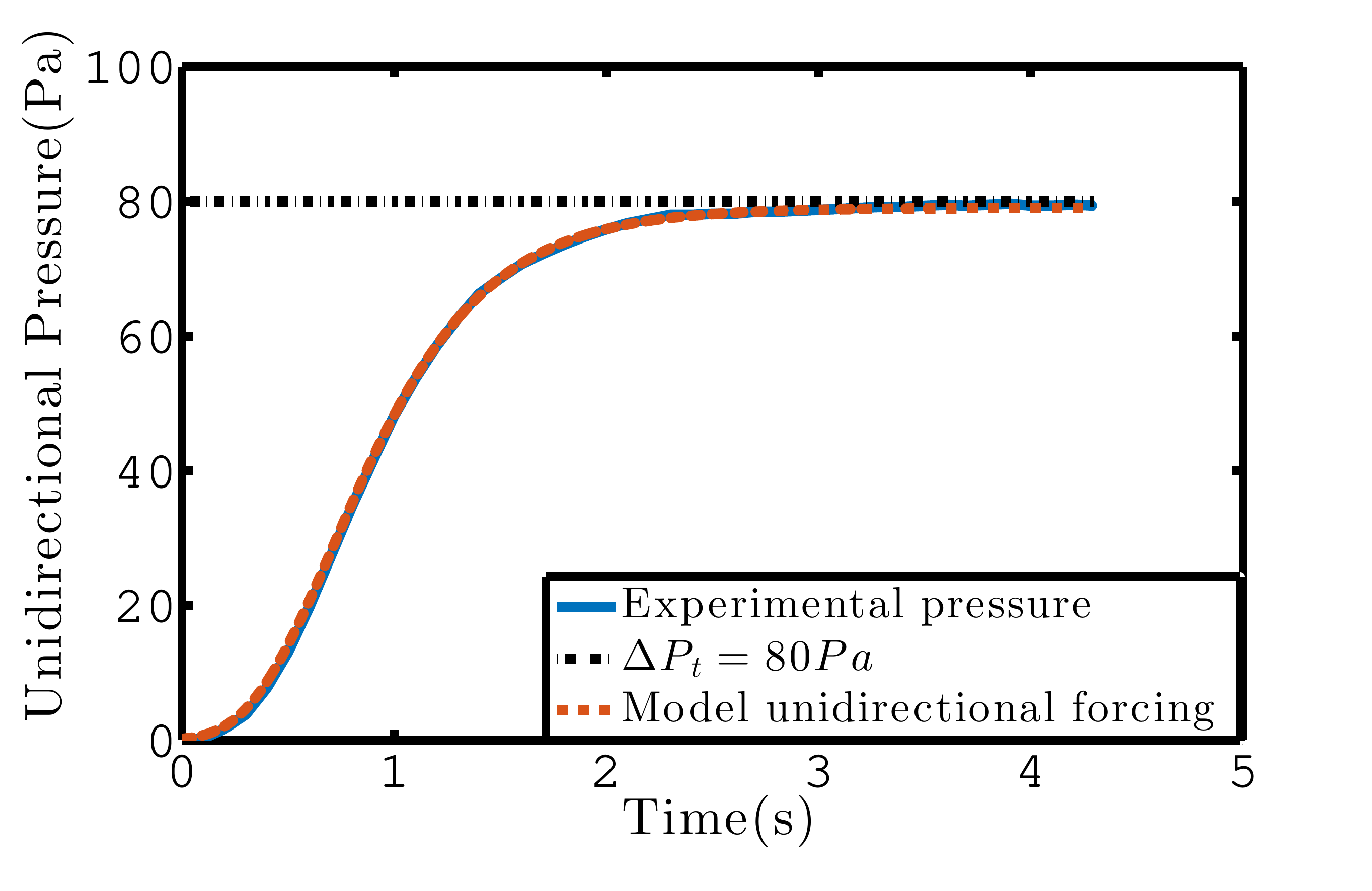} 
\caption{Unidirectional pressure forcing imposed by the pressure controller. Blue solid line: pressure magnitude measured with an integrated pressure sensor. Red dashed line: approximation of the pressure driving by the so-called Gompertz function \textcolor{black}{$\Delta P_t = 78 \exp (-6 \exp(-3t))$} used in the simulations as the driving pressure head since it fits well with the experimental signal. Black dashed-dotted line: asymptote $\Delta P_t = 80$ Pa when $t \rightarrow \infty$.}
\label{pressure_driving}
\end{center}
\end{figure}
	
\cite{magniez2016dynamics} validated the constitutive laws summarized in the previous section  through careful comparison with experiments of the motion of liquid plugs in \textit{prewetted} capillary tubes driven by a constant pressure head. This section is dedicated to the validation of the constitutive laws for the motion of liquid plugs in \textit{dry} capillary tubes driven by an unidirectional pressure driving (represented on \cref{pressure_driving}) and in particular the determination of the Hoffman-Tanner constant $E$ (an essential parameter in the analytical model).  

\begin{figure}
	\centering
	\begin{subfigure}[htbp]{0.45\textwidth}
		\includegraphics[height=5cm]{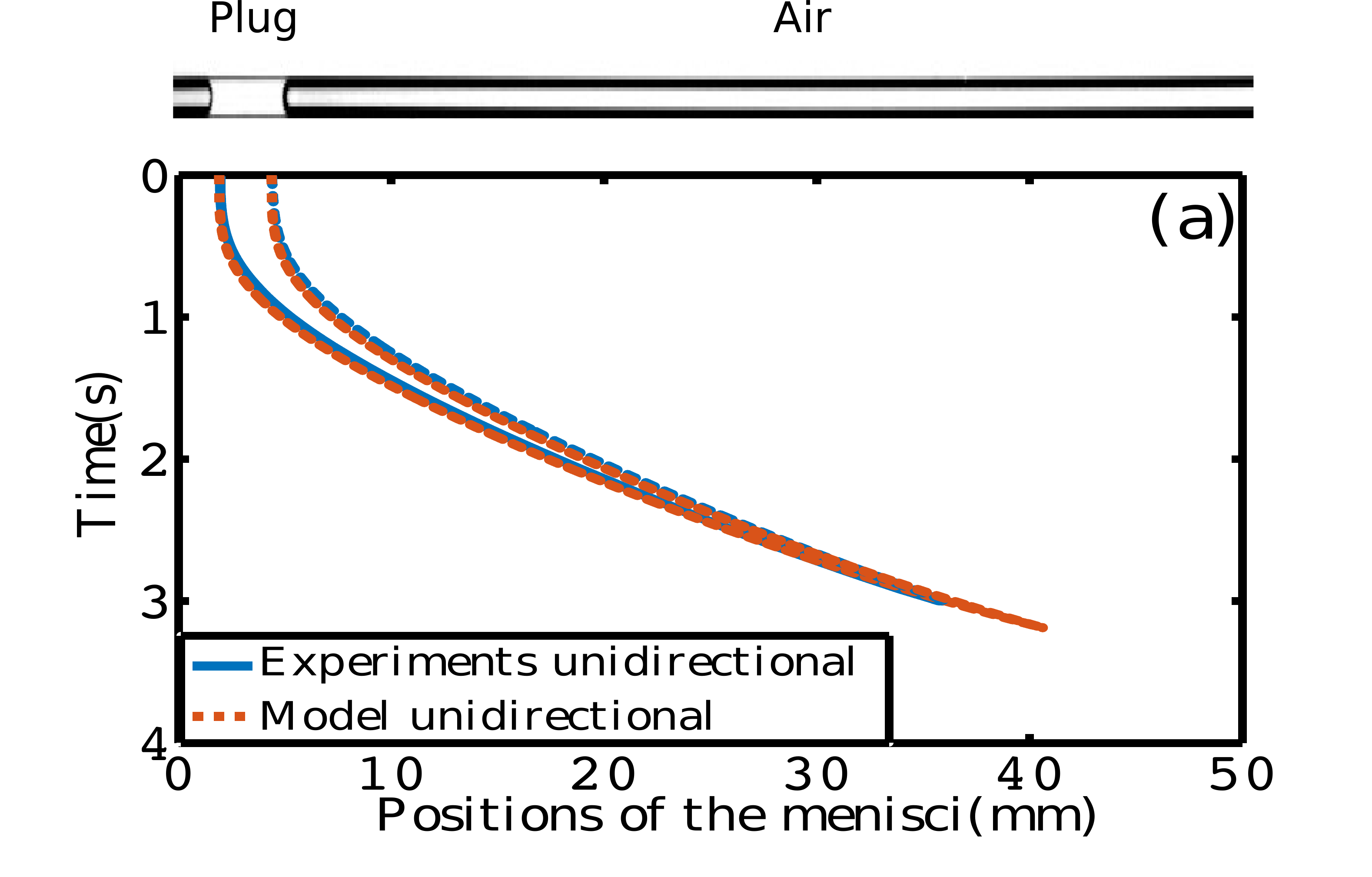}
		%\caption{First}\label{subfig-1:dummy}
	\end{subfigure}
	\hfill
	\begin{minipage}[htbp]{0.45\textwidth}
		\begin{subfigure}[b]{\linewidth}
			\includegraphics[width=\linewidth,height=3cm]{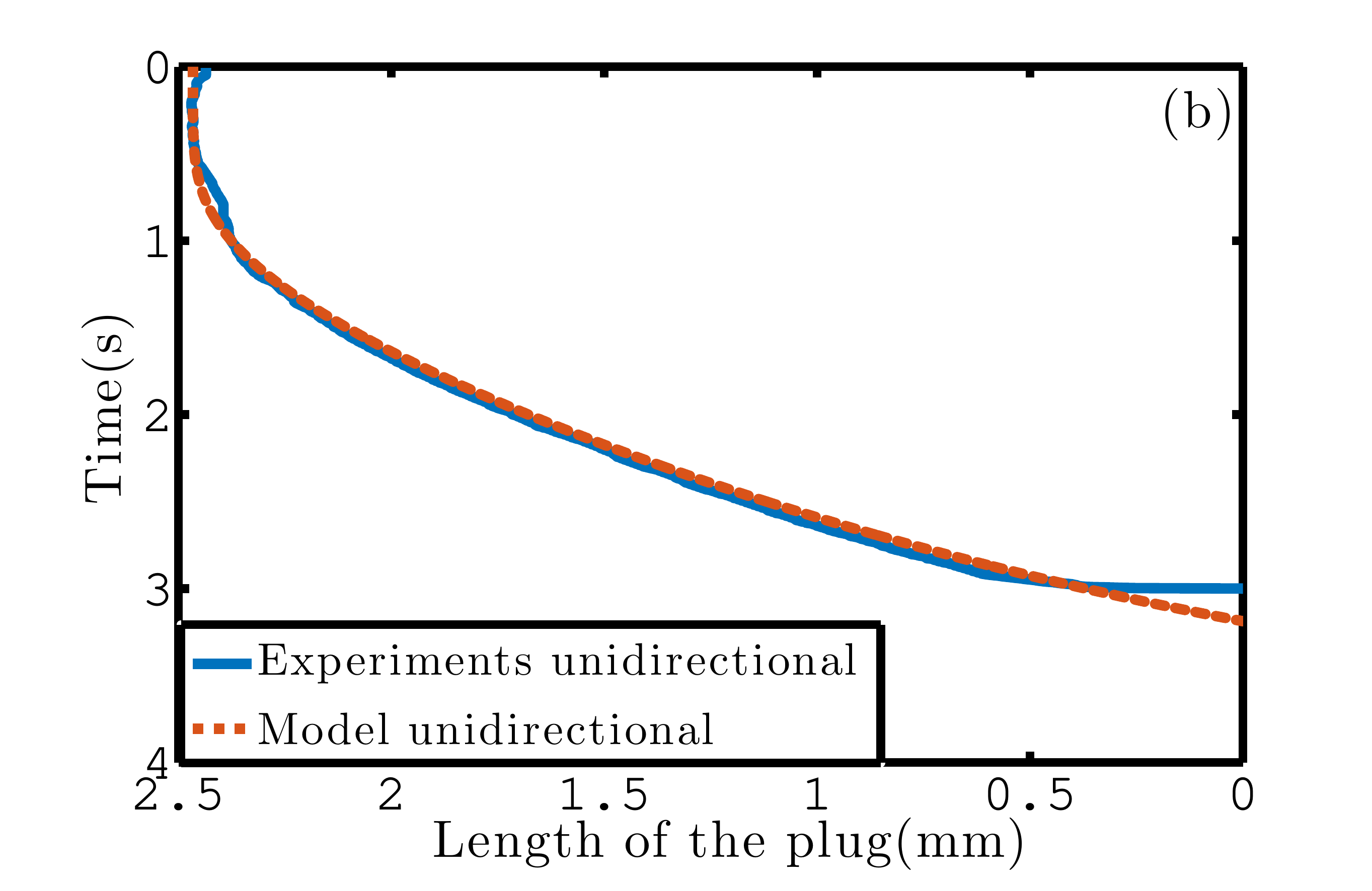}
			%\caption{Second}\label{subfig-2:dummy}
		\end{subfigure}\\[\baselineskip]
		\begin{subfigure}[b]{\linewidth}
			\includegraphics[width=\linewidth,height=3cm]{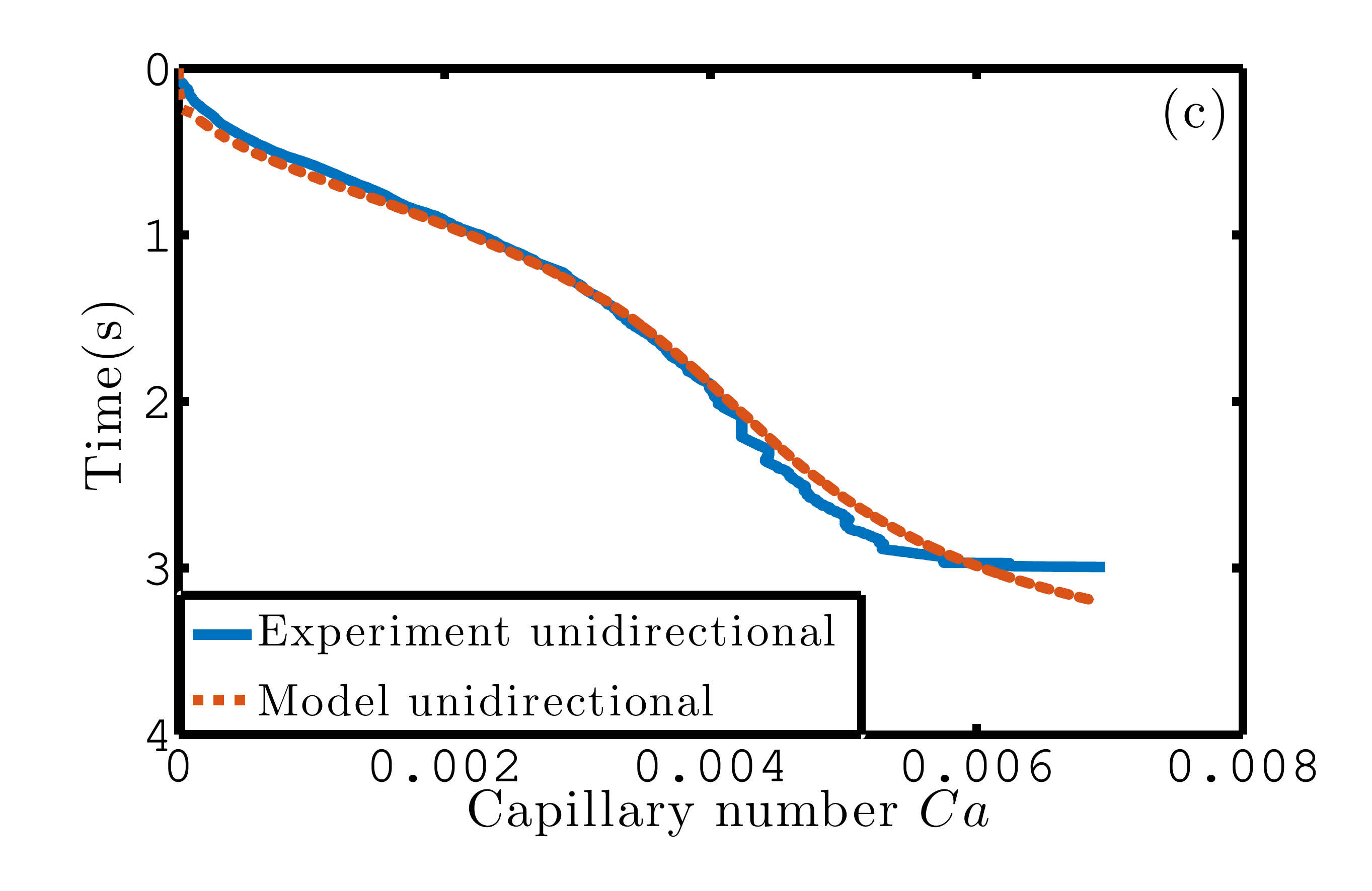}
			%\caption{Third}\label{subfig-3:dummy}
		\end{subfigure}
	\end{minipage}
	\caption{Temporal evolution of a liquid plug of initial length $L_{0} = 2.5 {mm}$ pushed with a unidirectional driving pressure (represented on \cref{pressure_driving}) in a dry capillary tube. (a) Position of the rear and front meniscus. (b) Evolution of the plug length. (c) Evolution of the plug dimensionless speed, i.e. the capillary number $Ca$. Blue curves correspond to experiments and the red dashed curves are obtained from simulations of equations \ref{eq:ptdry} to \ref{eq:mem}.  }  
	\label{Normal_cascade} 
\end{figure}

In such configuration (\cref{Normal_cascade}), the deposition of a trailing film behind the plug leads to the reduction of the plug length (\cref{Normal_cascade}(b)) and eventually its rupture when the front and rear interface meet (see \cref{Normal_cascade}(a) at time $t = 3$s). This process is unsteady and highly accelerative  as seen on \cref{Normal_cascade}(c). From $t = 0$ to $t = 2 s$ this acceleration is mostly related to the increase in the pressure head (see \cref{pressure_driving}).  After $t = 2$ s, the acceleration goes on and is even exacerbated close to the plug rupture, while the pressure head reaches a plateau. This behaviour can be understood by rewriting equation \ref{eq:ptdry} under a form reminiscent of Ohm's law: $ \Delta \tilde{P_t}  =  \tilde{R_t} Ca$, with $\tilde{R_t} = (\tilde{R_v} +\tilde{R}_i^r+\tilde{R}_i^f)$ the dimensionless global resistance to the flow, $\tilde{R_v} = 4 \tilde{L_p}$,  $\tilde{R}_i^f = E^2 / 2 \; Ca^{-1/3}$ and $\tilde{R}_i^r= 3.72 \; Ca^{-1/3}$ the viscous, front and rear interface resistances respectively. From this form of the pressure balance, we see that the reduction of the plug length $\tilde{L_p}$ leads to a reduction of the viscous resistance $\tilde{R_v}$ and thus, at constant pressure driving $\Delta \tilde{P_t}$, to an increase of the capillary number. This increase in the capillary number is strengthened by a decrease of the interfacial resistance $\tilde{R}_i = \tilde{R}_i^f + \tilde{R}_i^r$, since $\tilde{R}_i$ is proportional to $Ca^{-1/3}$. Finally, the increase of the trailing film thickness with the capillary number (equation \ref{eq:hr}) implies that the whole process (fluid deposition and plug motion) accelerate progressively as can be seen on  \cref{Normal_cascade}.  \textcolor{black}{It is important to note that in these experiments the acceleration of the plug does not rely on inertial effects (which can be neglected according to the dimensional analysis provided in section \ref{ssection:da}) but rather on the intimate relation between the plug size and velocity.}

\begin{figure}
	%\vspace*{-0.3cm} 
	
	\begin{subfigure}{0.5\textwidth}
		\hspace*{0.4cm} 
		\includegraphics[width=\linewidth, height=5cm]{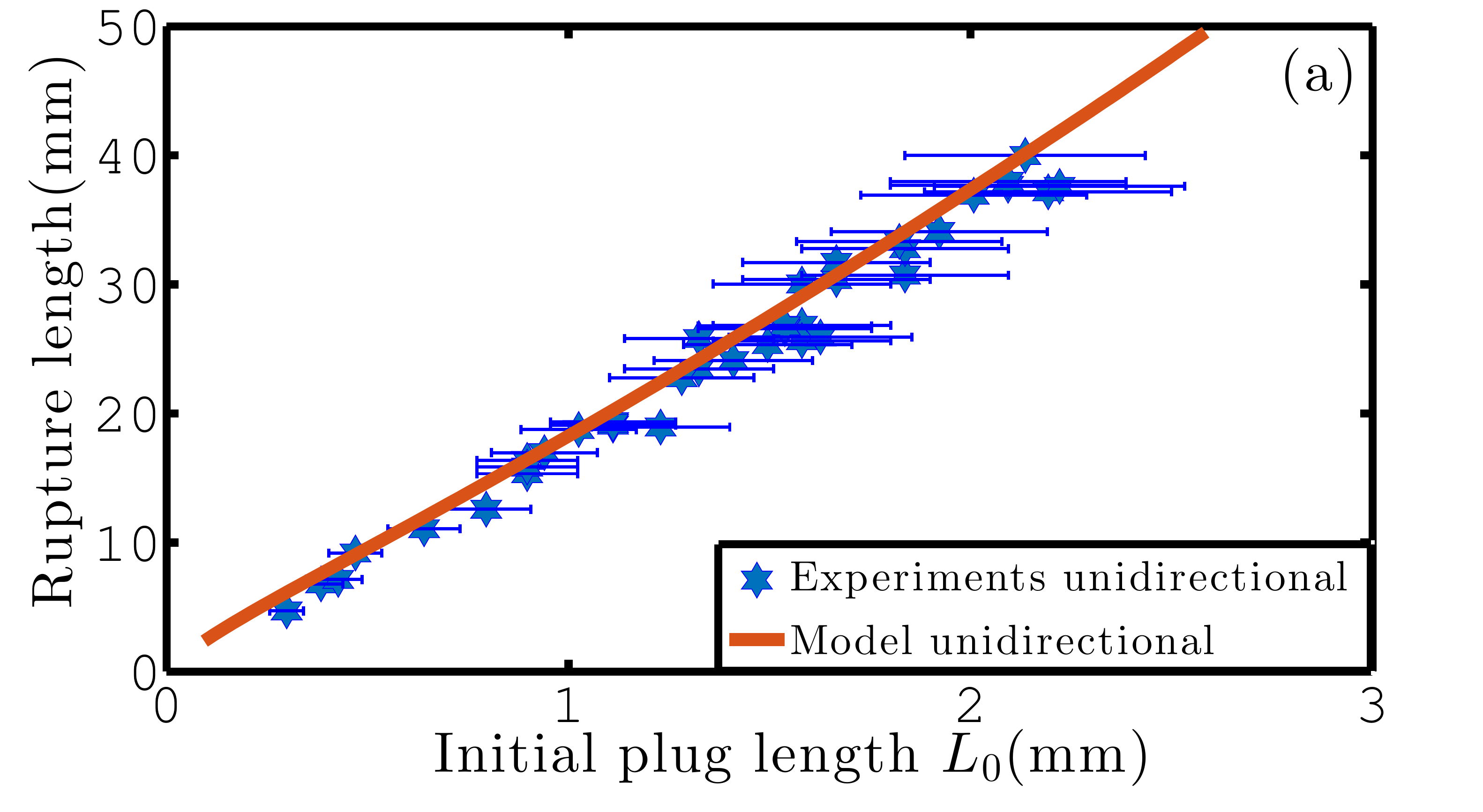} 
	\end{subfigure}
	\begin{subfigure}{0.5\textwidth}
		\hspace*{0.2cm} 
		\includegraphics[width=\linewidth, height=5cm]{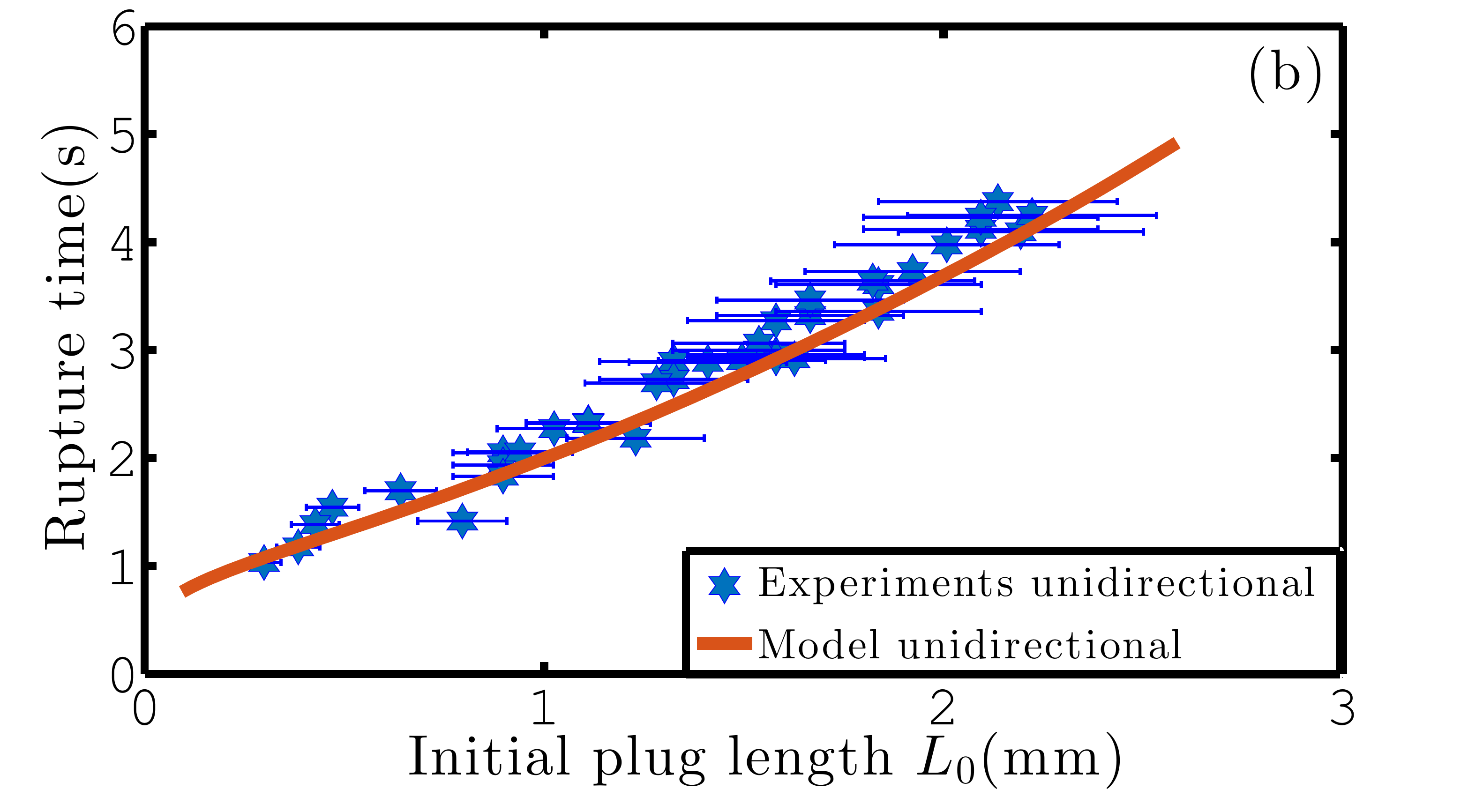}
	\end{subfigure}
	
	\caption{
		Evolution of (a) the rupture length and (b) the rupture time of a liquid plug pushed with a unidirectional pressure head of the form \textcolor{black}{$\Delta P_t = 60 \exp(-6 \exp(-3.5 t))$} in a dry capillary tube as a function of the initial plug length $L_0$. The blue stars represent experiments. \textcolor{black}{The red curve is obtained numerically by simulating the evolution for 107 initial plug lengths.}}
	\label{Normal_cascade_cloud}
\end{figure}

Many experiments have been performed for different initial plug lengths and compared with the numerical solutions of equations  (\ref{eq:ptdry})  to (\ref{eq:mem}) (dry version with $\tilde{h_f} = 0$). In the simulations, a Gompertz function \textcolor{black}{$\Delta P_t = 78 \exp(-6 \exp(-3t))$} was used as the driving pressure head due to its excellent match with the pressure head measured experimentally at the exit of the the pressure controller with an integrated pressure sensor (see \cref{pressure_driving}, blue line corresponds to experimental signal and red line to best fit with Gompertz function). The complex shape of the pressure head is the result of the pressure controller response time (the command is a constant pressure $P_o = 80 \; Pa$ starting at $t=0$). The only adjustable parameter in the model is the Hoffman-Tanner constant $E$ appearing in equation (\ref{eq:ptdry}). The best fit between experiments and theory was achieved for $E=4.4$, a value close to the coefficient $4.3$ obtained by \cite{bico2001falling} in their experiments on falling of liquid slugs in vertical dry capillary tubes. With this value, an excellent prediction of the plug dynamics is achieved for all experiments (see e.g. \cref{Normal_cascade} where blue solid lines correspond to experiments and red dashed lines to simulations). In particular, this model enables a quantitative prediction  of the rupture length, defined as the portion of the tube $L_d = max(X_f)-min(X_r)$ visited by the liquid plug before its rupture (\cref{Normal_cascade_cloud}(a)), and the rupture time, which is the total time elapsed between the beginning of the experiment and the plug rupture  (\cref{Normal_cascade_cloud}(b)). 

\section{Cyclic forcing of liquid plugs}\label{Periodic_forcings} 

This section is dedicated to the dynamics of liquid plugs under cyclic forcing. In the first subsection, the influence of the forcing configuration (pressure of flow rate) is examined. The second subsection enlightens the fundamental role played by hysteretic effects resulting from fluid deposition on the walls.

\subsection{Influence of the driving condition: pressure head versus flow rate}

	\begin{figure}
	%\vspace*{-0.3cm} 
	
	\begin{subfigure}{0.5\textwidth}
		\hspace*{0.4cm} 
		\includegraphics[width=\linewidth, height=5cm]{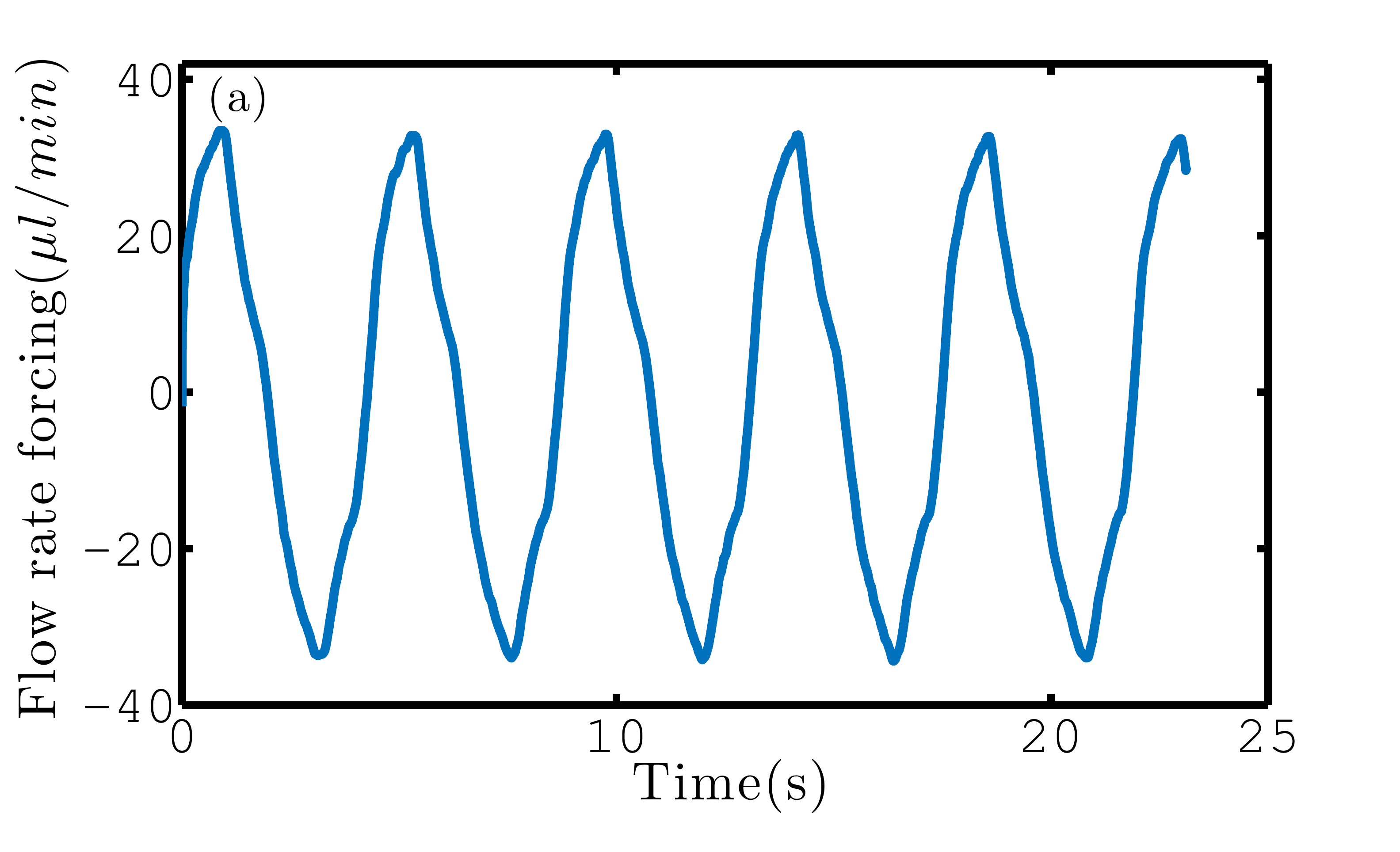} 
	\end{subfigure}
	\begin{subfigure}{0.5\textwidth}
		\hspace*{0.2cm} 
		\includegraphics[width=\linewidth, height=5cm]{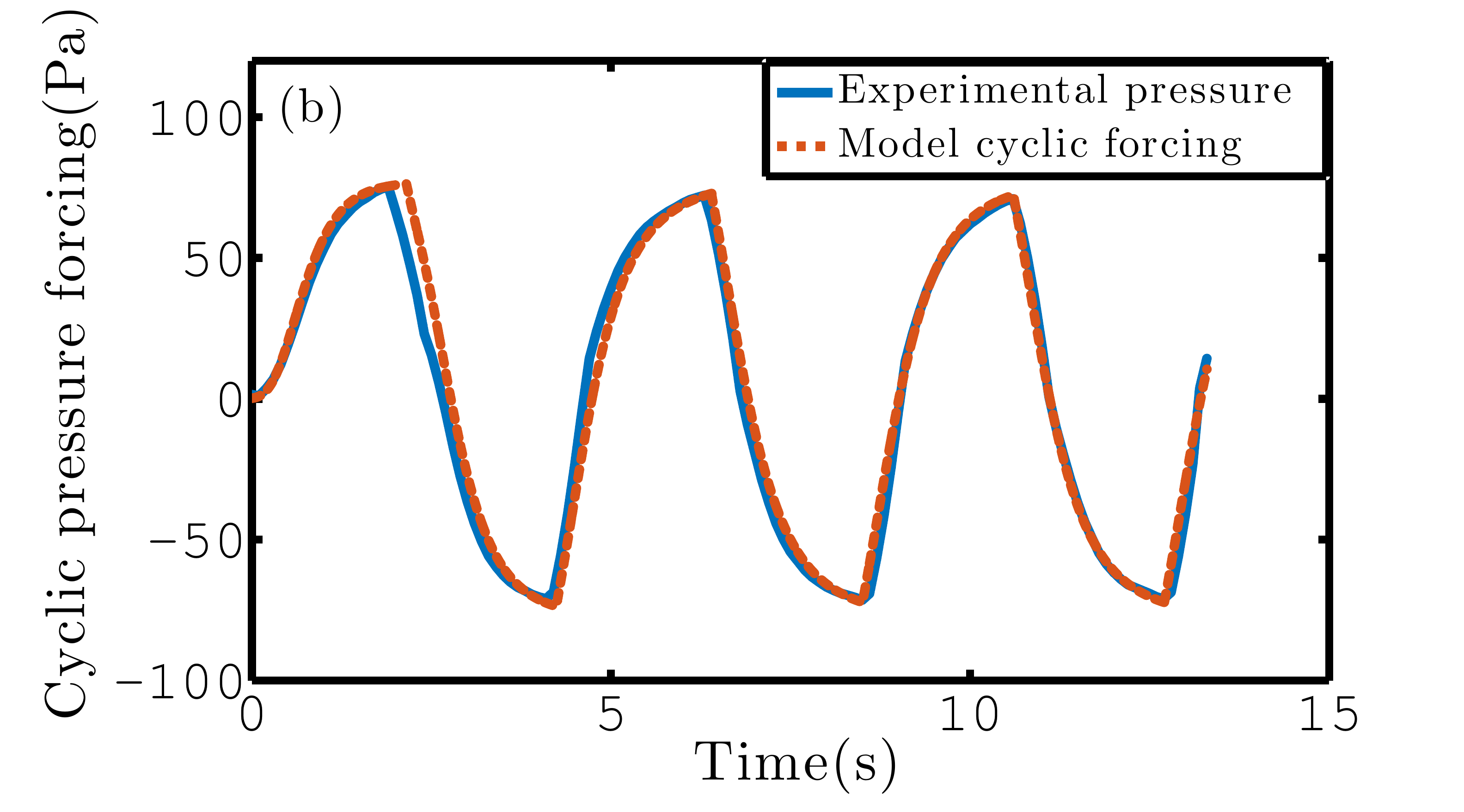}
	\end{subfigure}
		
		\caption{
			(a) Flow rate and (b) pressure cyclic forcing imposed experimentally with respectively a syringe pump and a pressure controller. Blue solid line: experimental values measured with sensors. Red dashed line: fit of the pressure cyclic forcing measured experimentally with the analytical expression: \textcolor{black}{$\Delta P_t = 78 \exp(-6 \exp(-3t))$ for $t \in [0,T]$, $\Delta P_t = (-1)^n (P_c - P_d)$ for $t \in [nT,(n+1)T]$ with $P_c= 78 \exp(-3 \exp(-3 (t-nT)))$ and $P_d = 78 \exp(-1.4 (t-nT)) \exp(-0.02  \exp(-1.4*(t-nT)))$}, $T = 2.15$ s the half period and $n \in \mathbb{N^*}$.}
		\label{Periodic_functions}
	\end{figure}

In this first subsection, the responses of liquid plugs to two types of forcings are compared: (i) a cyclic flow rate imposed by a syringe pump (represented on \cref{Periodic_functions}(a), blue line) and (ii) a pressure cycle imposed by a pressure controller (represented on \cref{Periodic_functions}(b), blue line). These forcing have a complex temporal shape owing to the response time of the syringe pump and pressure controller. \textcolor{black}{In appendix \ref{method}, we describe precisely how this forcing conditions are obtained.} The pressure forcing is well approximated  by the following \textcolor{black}{analytical expression, which is a} combination of Gompertz functions (see \cref{Periodic_functions}(b), red dashed line):
\textcolor{black}{
\begin{equation}
\left\{ \begin{array}{l}
\Delta P_t = 78 \exp(-6 \exp(-3t)) \mbox{ for } t \in [0,T] \\
\Delta P_t = (-1)^n (P_c - P_d) \mbox{ for } t \in [n,(n+1)T] \\
P_c= 78 \exp(-3 \exp(-3 (t-nT))) \\
P_d = 78 \exp(-1.4 (t-nT)) \exp(-0.02 * \exp(-1.4*(t-nT)))
\end{array}
\right.
\label{eq:deltapt_c}
\end{equation}}
with $T = 2.15$ s the half period and $n \in \mathbb{N^*}$ for cyclic forcing. 

\begin{figure}
		\vspace*{+0.3cm} 
		\centering
		\begin{subfigure}[b]{0.33\textwidth}
			\includegraphics[width=\linewidth, height=4cm]{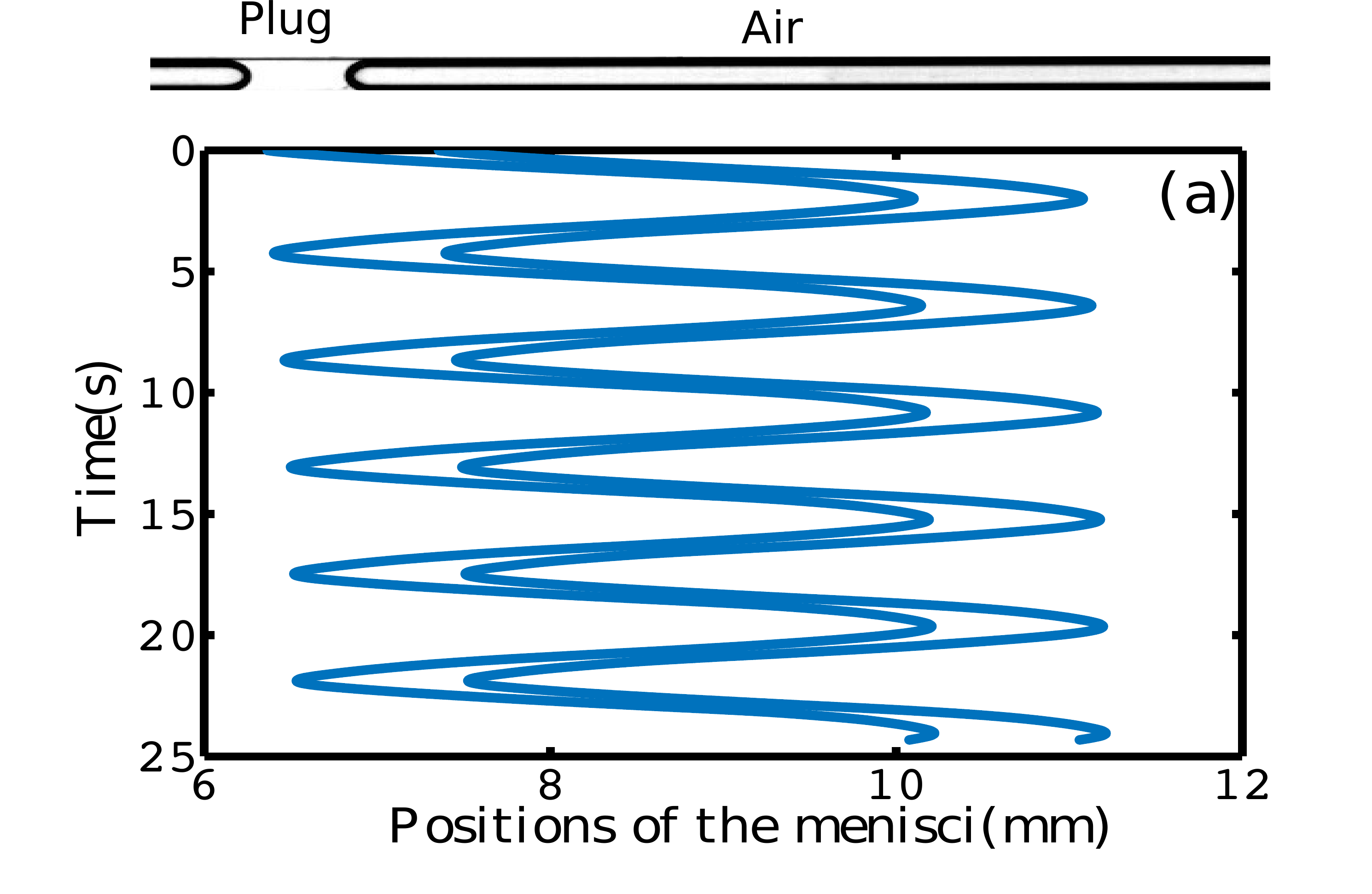}
			%  \caption{A gull}
			%  \label{fig:gull}
		\end{subfigure}%
		\begin{subfigure}[b]{0.33\textwidth}
			\includegraphics[width=\linewidth, height=3.75cm]{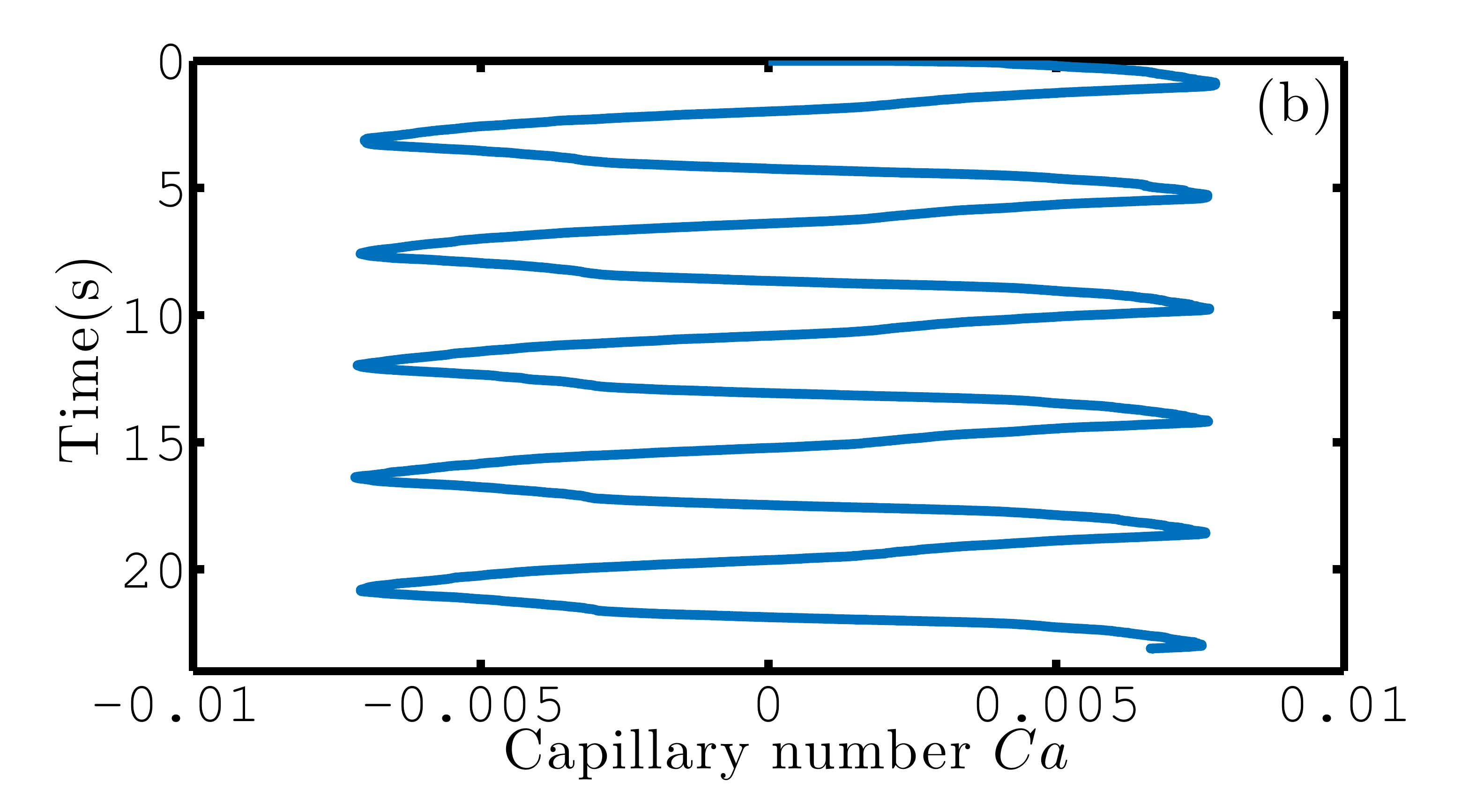}
			% \caption{A gull2}
			% \label{fig:gull2}
		\end{subfigure}%
		\begin{subfigure}[b]{0.33\textwidth}
			\includegraphics[width=\linewidth, height=3.75cm]{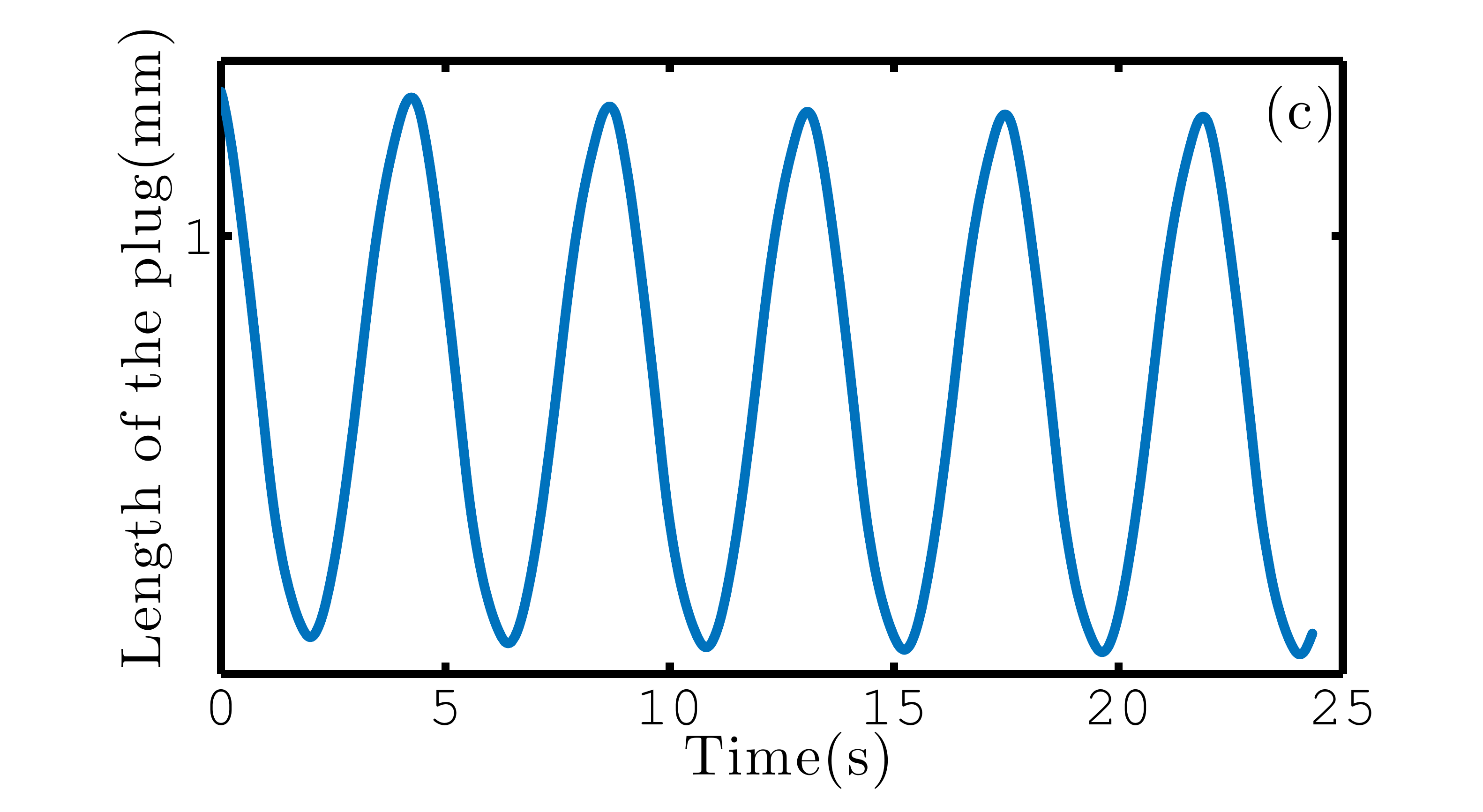}
			%\caption{A tiger}
			%\label{fig:tiger}
		\end{subfigure}% 
		\\
		\centering
		\begin{subfigure}[b]{0.33\textwidth}
			\includegraphics[width=\linewidth, height=4cm]{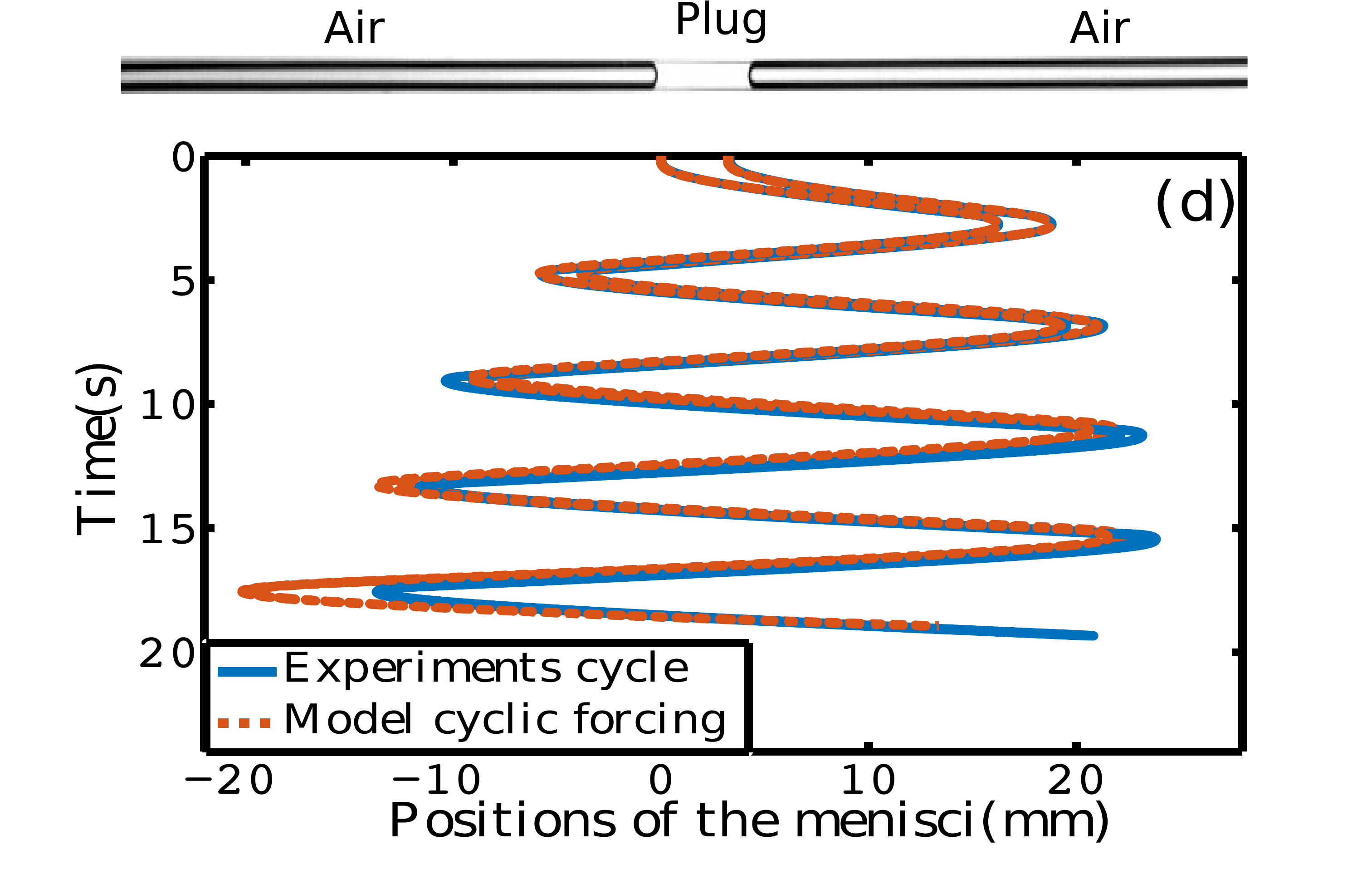}
			%  \caption{A gull}
			%  \label{fig:gull}
		\end{subfigure}%
		\begin{subfigure}[b]{0.33\textwidth}
			\includegraphics[width=\linewidth, height=3.8cm]{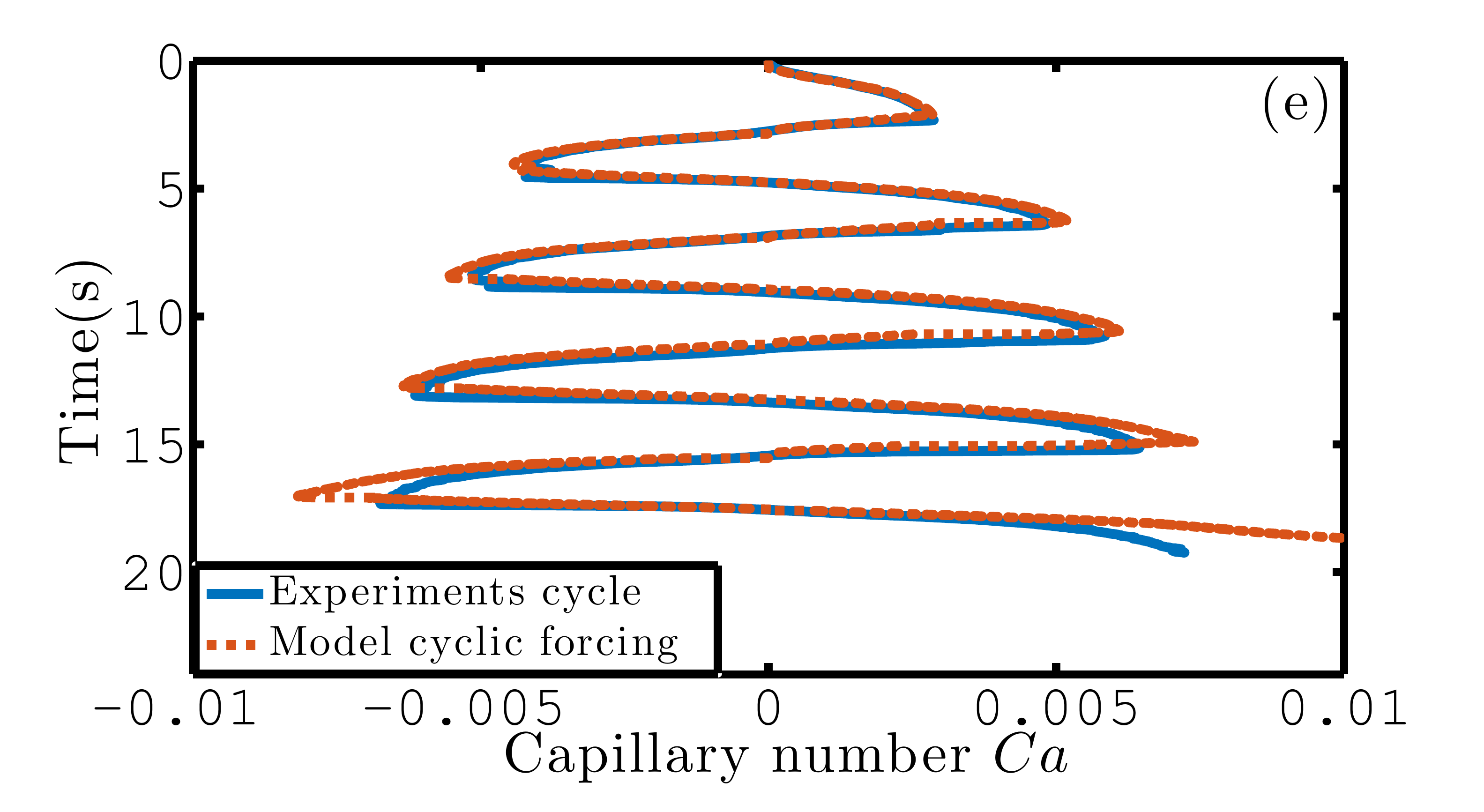}
			% \caption{A gull2}
			% \label{fig:gull2}
		\end{subfigure}%
		\begin{subfigure}[b]{0.33\textwidth}
			\includegraphics[width=\linewidth, height=3.8cm]{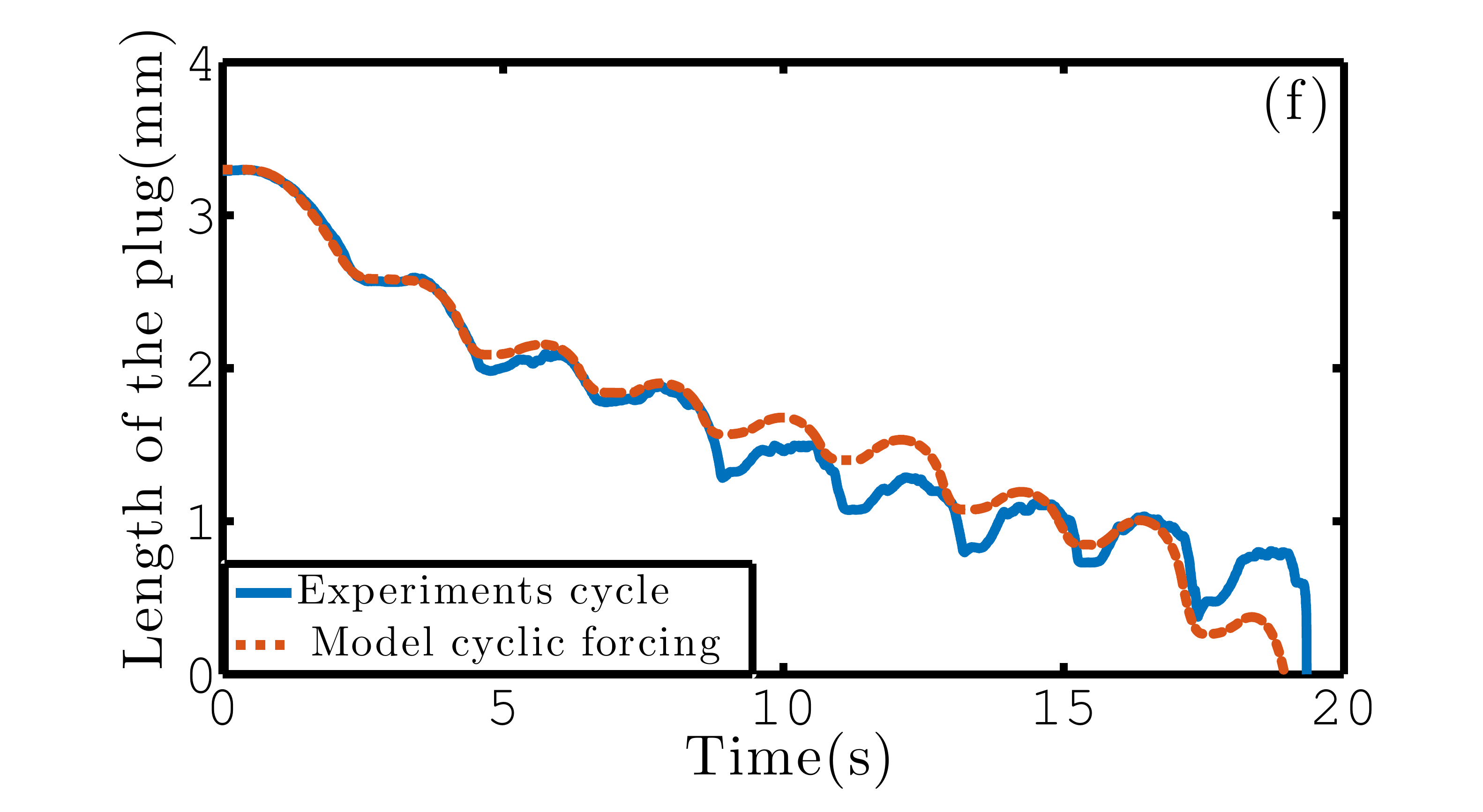}
			%\caption{A tiger}
			%\label{fig:tiger}
		\end{subfigure}% 
	
	\caption{ \label{Spatiotemporal_cyclic_motion}  (a), (b), (c) Temporal evolution of a single liquid plug of initial length $L_0=1.05$ mm pushed with the cyclic flow rate forcing represented on \cref{Periodic_functions}(a), see also movie S1. (a) Positions of the menisci. (b) Evolution of the dimensionless velocity of the left meniscus (capillary number). (c) Evolution of the plug length. (d), (e), (f) Temporal evolution of a single liquid plug of initial length $L_0 = 3.3$ mm pushed with the pressure cyclic forcing represented on \cref{Periodic_functions}(b), see also movie S2. (d) Position of the menisci. (e) Evolution of the dimensionless velocity of the left meniscus (capillary number). (f) Evolution of the plug length. In all these figures blue lines represent experiments and red dashed lines simulations obtained from equations (\ref{eq:ptdry}) to (\ref{eq:mem}).}  
\end{figure}

\begin{figure}
	\centering
	\begin{subfigure}[htbp]{0.5\textwidth}
		\includegraphics[width=\linewidth, height=5cm]{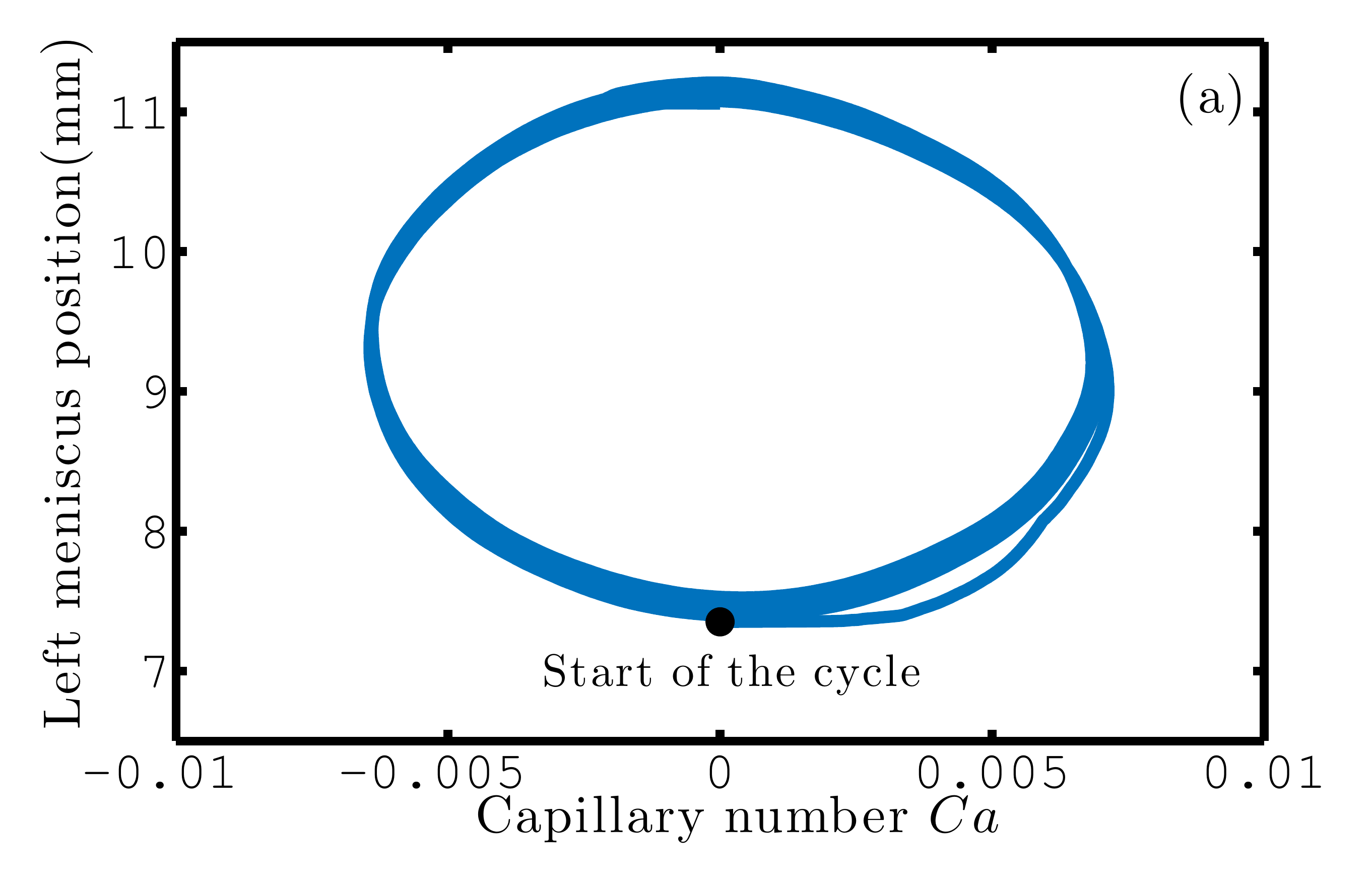}
		%  \caption{A gull}
		%  \label{fig:gull}
	\end{subfigure}%
	\begin{subfigure}[htbp]{0.5\textwidth}
		\includegraphics[width=\linewidth, height=5cm]{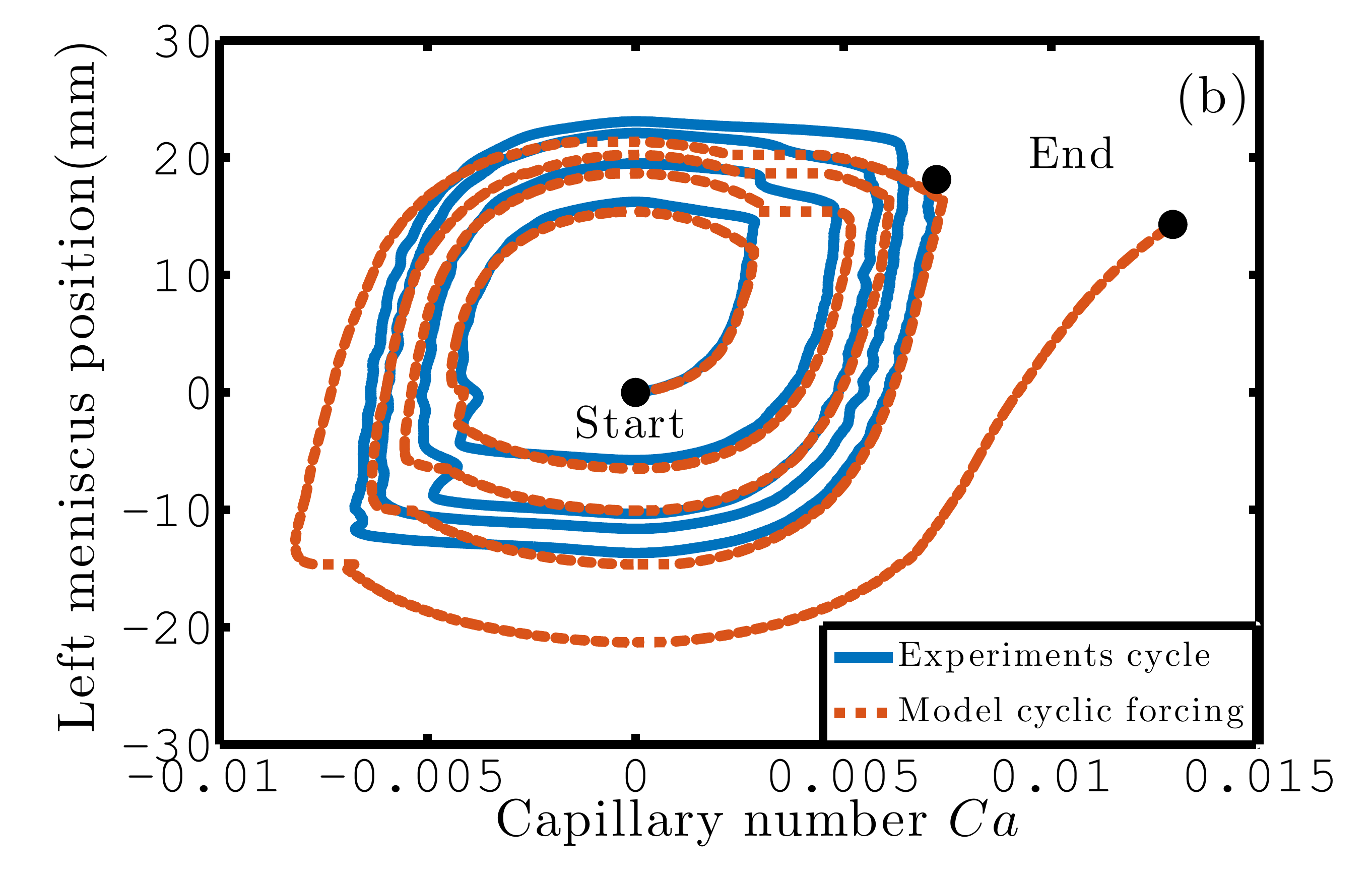}
		% \caption{A gull2}
		% \label{fig:gull2}
	\end{subfigure}%
	\caption{ \label{phase_portrait}  Phase portrait showing the evolution of the position of the left meniscus as a function of the capillary number $Ca$. (a) Cyclic flow rate forcing and (b) cyclic pressure forcing. The blue curves correspond to experiments and the red dashed line to simulations.}  
\end{figure}

Two extremely different behaviors are evidenced in these two cases: For a cyclic \textit{flow rate} forcing (\cref{Spatiotemporal_cyclic_motion}(a),(b),(c) and movie S1), the liquid plug dynamics is periodic and stable (see phase portrait on \cref{phase_portrait}(a)). Indeed, the plug velocity and positions are directly imposed by the motion of the syringe pump, thus: $U(t+2T) = U(t)$ (\cref{Spatiotemporal_cyclic_motion}(b)) and $X_r(t+2T) = X_r(t)$ (\cref{Spatiotemporal_cyclic_motion}(a)). Moreover, since (i) the film deposition process solely depends on the plug velocity and (ii) the fluid recovery at half cycle $N$ depends on the fluid deposition at half cycle $N-1$, the mass balance is null over each cycle and the evolution of the plug size is also periodic: $L_p(t+2T) = L_p(t)$ (\cref{Spatiotemporal_cyclic_motion}(c)). It is interesting to note that the initial wall wetting condition plays little role in this process; it only affects the mass balance during the first half cycle and thus determines the plug size $L_p((2n+1) T)$ with $n \in \mathbb{N}$. This wetting condition is indeed erased by the backward motion during the second half cycle and the plug evolution is then only dictated by the temporal shape of the flow rate cycle. 

The liquid plug undergoes a very different evolution for a periodic \textit{pressure} forcing (\cref{Spatiotemporal_cyclic_motion}(d),(e),(f) and movie S2). In this case, the plug velocity and position are no more enforced by the driving condition and depend only on the evolution of the resistance of the plug to motion. For the forcing condition represented on \cref{Periodic_functions}(b), it is observed that (i) the plug travels on a longer portion of the tube at each cycle (\cref{Spatiotemporal_cyclic_motion}(d)), (ii) the dimensionless velocity of the plug (the capillary number) is no more cyclic but increases progressively at each cycle, $U(t+2T) > U(t)$ (\cref{Spatiotemporal_cyclic_motion}(e)) and (iii) the size of the plug diminishes ($L_p(t+2T) < L_p(t)$), eventually leading to its rupture (\cref{Spatiotemporal_cyclic_motion}(f)). These phenomena are of course related since a larger plug velocity  leads to more liquid deposition and thus a diminution of the plug size. Conversely, the cyclic diminution of the plug size leads to a decrease in the viscous resistance (the same process as described in subsection \ref{ss:uni}). Nevertheless this \textcolor{black}{sole} mechanism \textcolor{black}{is not sufficient to} explain \textcolor{black}{the cyclic acceleration of the plug} evidenced in the phase portrait (\cref{phase_portrait}(b)) \textcolor{black}{as demonstrated in the next sections}.

\subsection{Memory effects and hysteretic behaviour}

	\begin{figure}
		\centerline{\includegraphics[width=8cm, height=5.5cm]{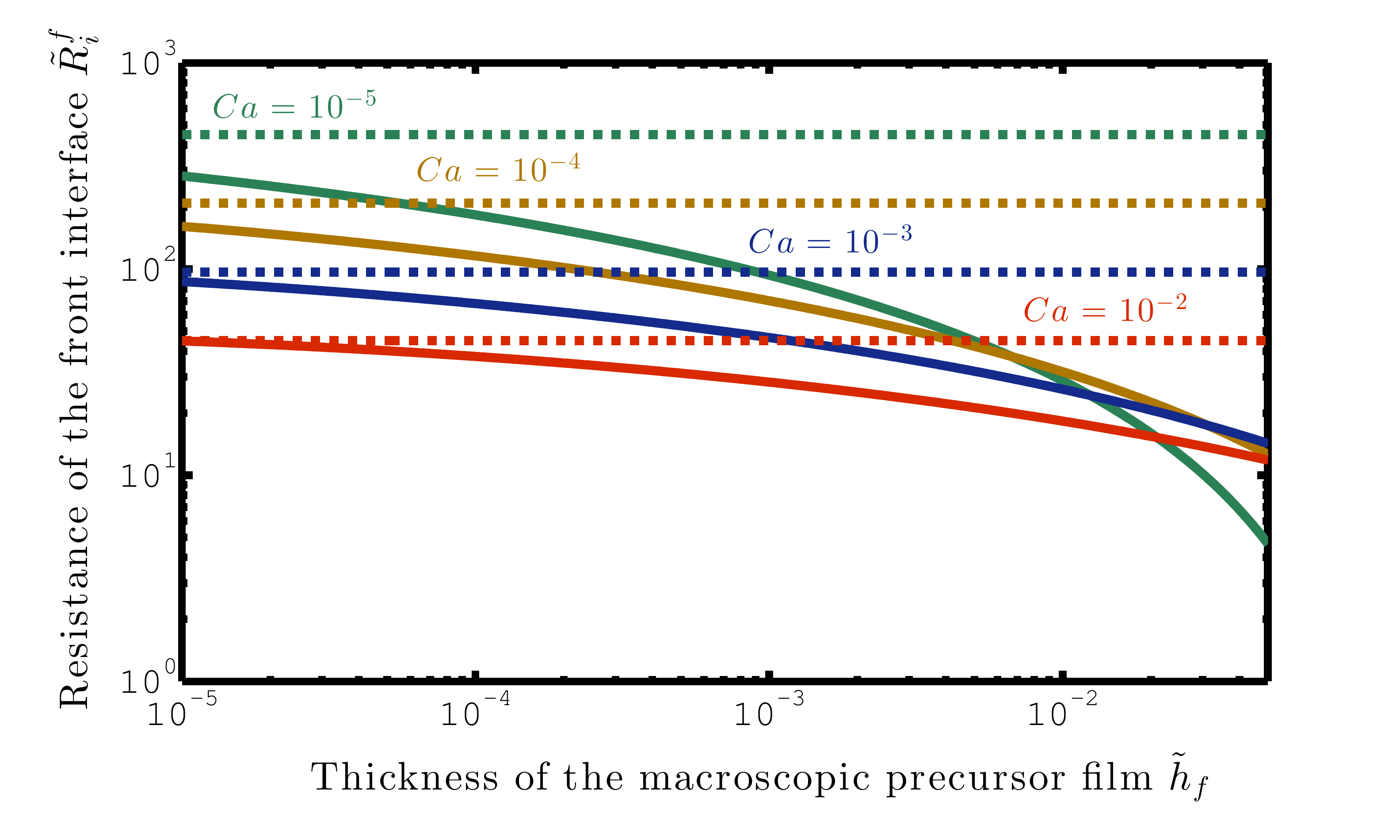}}
		\caption{Solid lines: dimensionless front interface resistance $\tilde{R}^{f}_i$ at different capillary numbers $Ca$ as a function of the prewetting film thickness $\tilde{h}_f$. Dashed lines: dimensionless front interface resistance for dry capillary tubes. These curves show (i) that the interfacial resistance is systematically lower for prewetted capillary tubes than for  dry capillary tubes and (ii) that for prewetted capillary tubes, the interfacial resistance decreases with the thickness of the prewetting film $\tilde{h}_f$}
		\label{Front_meniscus}
	\end{figure}

\textcolor{black}{In this section, the model introduced in section \ref{ssection:model} (equations (\ref{eq:ptdry})to (\ref{eq:mem})) is used to analyse the origin of the departure from a periodic evolution for a pressure  forcing and in particular analyse the contributions of the different terms. Indeed, this model quantitatively reproduces the liquid plug dynamics (see \cref{Spatiotemporal_cyclic_motion}(d), (e), (f), red dashed lines and \cref{cycle_cloud} and \ref{figapp} for comparisons of the simulations and experiments for a large set of initial conditions.)} \\

\textcolor{black}{The response of a liquid plug to a cyclic forcing is periodic since (i) the plug velocity is enforced and does not rely on the evolution of the plug resistance to motion and (ii) the liquid deposition on the walls, and thus the size of the plug solely depends on the plug velocity. This leads to a zero mass balance at each cycle.} 

	\begin{figure}
	\begin{center}
			\includegraphics[width=0.7 \textwidth]{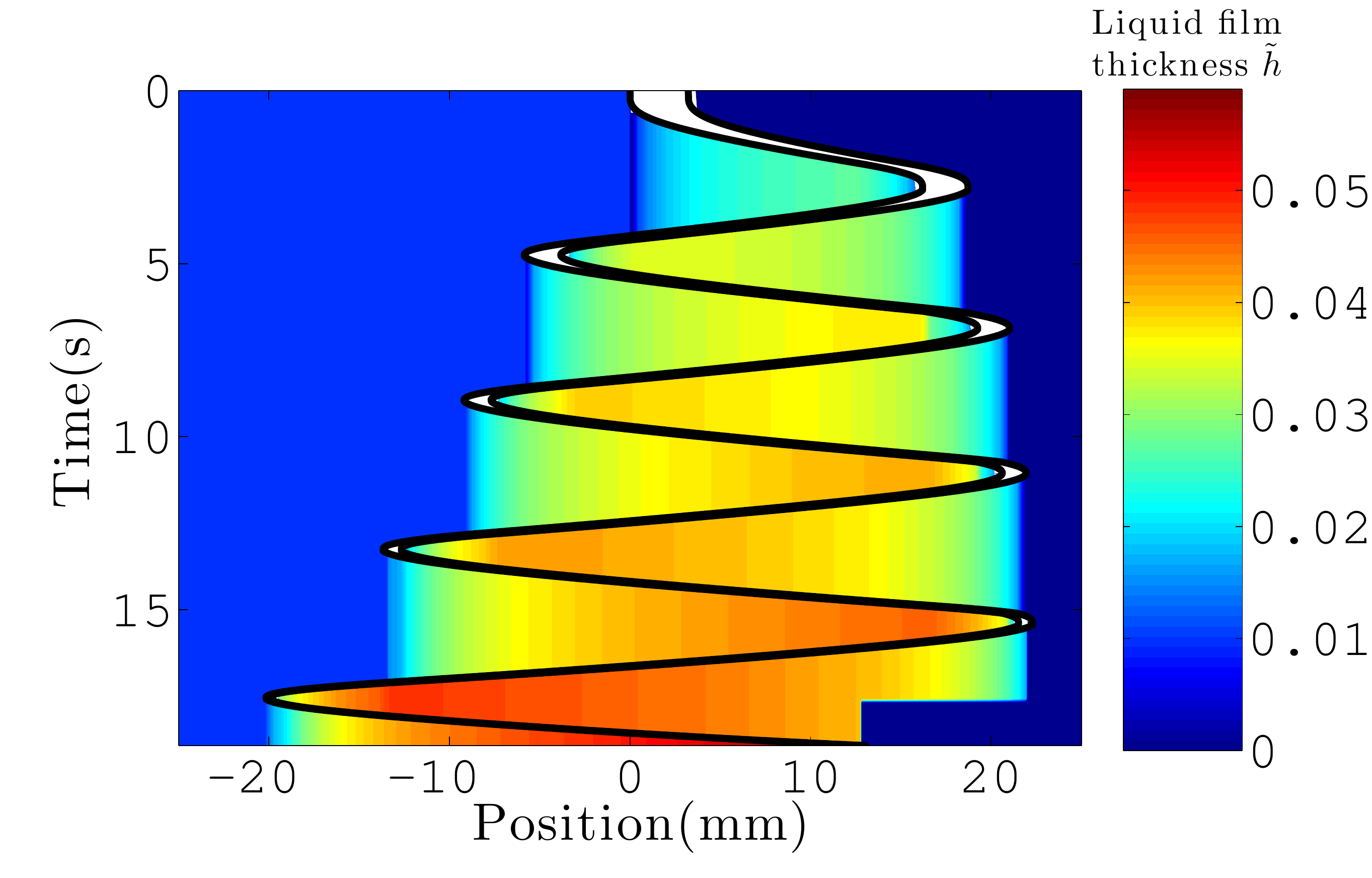} 
		\caption{\textcolor{black}{Spatiotemporal diagram showing the evolution of the thickness of the liquid film deposited on the walls obtained from simulations of equations (\ref{eq:ptdry})to (\ref{eq:mem}) with the same initial and forcing conditions as \cref{Spatiotemporal_cyclic_motion}. The dimensionless film thickness $\tilde{h} = h/R$  is represented in color at each time and position in the tube.}}
		\label{liquidfilm}
		\end{center}
	\end{figure}	

\textcolor{black}{For pressure driven cyclic forcing, the departure from this periodic behaviour thus relies on the evolution of the plug resistance at each cycle due to the existence of flow memories since the flow is quasi-static. } \textcolor{black}{\textit{A first memory}, which is already at hand in unidirectional pressure forcing is simply the evolution of the plug length. Indeed, the plug size at a given time $t$ depends on the history of the plug velocity at times $t^* < t$. In turn, the plug length modifies the viscous resistance $\tilde{R_v} = 4 \tilde{L_p}$ and thus the plug velocity. The mass balance which is relatively simple for a unidirectional driving in a dry tube (the liquid left at time $t$ on the walls only depends on the velocity of the plug at time $t$) becomes much more complex for a  cyclic forcing. Indeed, at each back and forth motion, the liquid plug leaves on the walls a film layer whose thickness keeps a memory of the plug velocity during the corresponding half-cycle (since $\tilde{h_r}$ depends on $Ca$). Thus, the mass balance both depends on the velocity of the plug at time $t$ and its velocity at the same position in the previous cycle. The progressive transfer of mass from the liquid plug to the liquid film is clearly evidenced on \cref{liquidfilm}. This graph shows that both the portion of the tube covered by the liquid film and the film thickness increase at each cycle. This graph also exhibits the complexity of the mass transfer observed in the movie S2: While the size of the plug gradually decreases at each cycle, its evolution is not monotonous during a half-cycle. Indeed, when the flow direction is changed the plug moves at first slowly (due to the response time of the pressure controller) and thus the thickness of the liquid film deposited on the walls behind the plug is smaller than the one in front of the plug, leading to a growth of the liquid plug. Then the plug accelerates and progressively the tendency is inverted leading to a reduction of the plug size. The transition between the growing and decreasing phases correspond on the graph to the times when the thickness (color) on each side of the plug is the same.}

\textcolor{black}{\textit{A second memory} originates from the lubrication of the plug motion by the liquid film, i.e. the reduction of the front interface resistance $\tilde{R}_i^f = (F^2/2) Ca^{-1/3}$ as the thickness of the prewetting film $\tilde{h_f}$ is increased (see equation (\ref{eq:lubri}) and (\cref{Front_meniscus})). Indeed, during the first half cycle the liquid plug moves on a dry capillary tube and leaves a liquid film behind it on the walls whose thickness increases with the speed of the liquid plug (see \cref{liquidfilm}). This liquid film lubricates the passage of the plug during the back motion, leading to a drastic reduction of the front interface resistance (\cref{Front_meniscus}) and thus, a higher plug speed. Then the same mechanism is reproduced during the following cycles: Since the speed is increased at each cycle, the plug leaves more liquid on the walls, leading again to a reduction of the interfacial resistance through a lubrication effect.}
	\begin{figure}
	%\vspace*{-0.3cm} 
	
	\begin{subfigure}{0.32\textwidth}
		\hspace*{0.1cm} 
		\includegraphics[width=0.9\linewidth, height=4.5cm]{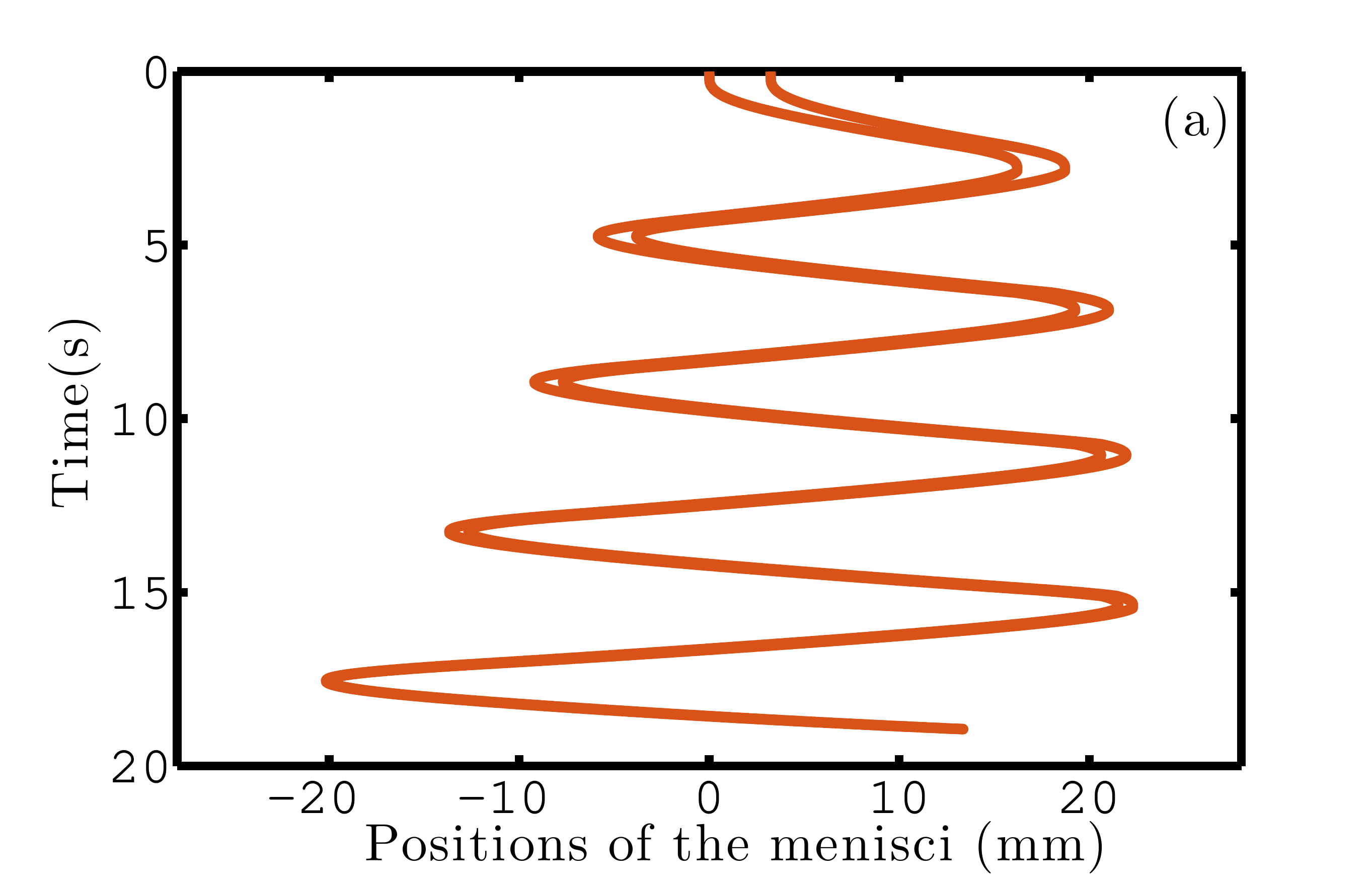} 
	\end{subfigure}
	\begin{subfigure}{0.32\textwidth}
		\hspace*{0.1cm} 
		\includegraphics[width=0.9\linewidth, height=4.5cm]{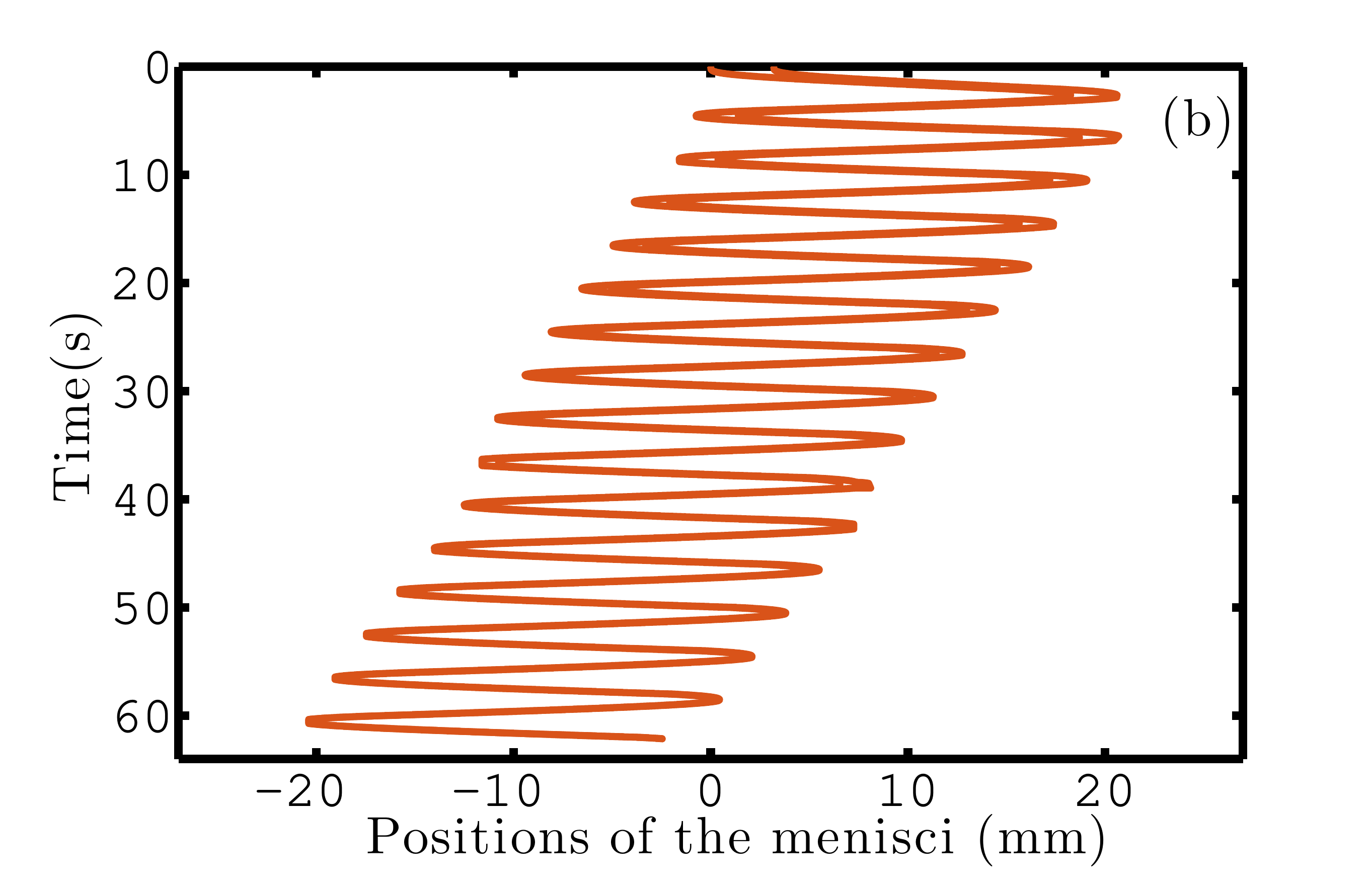}
	\end{subfigure}
	\begin{subfigure}{0.32\textwidth}
		\hspace*{0.1cm} 
		\includegraphics[width=0.9\linewidth, height=4.5cm]{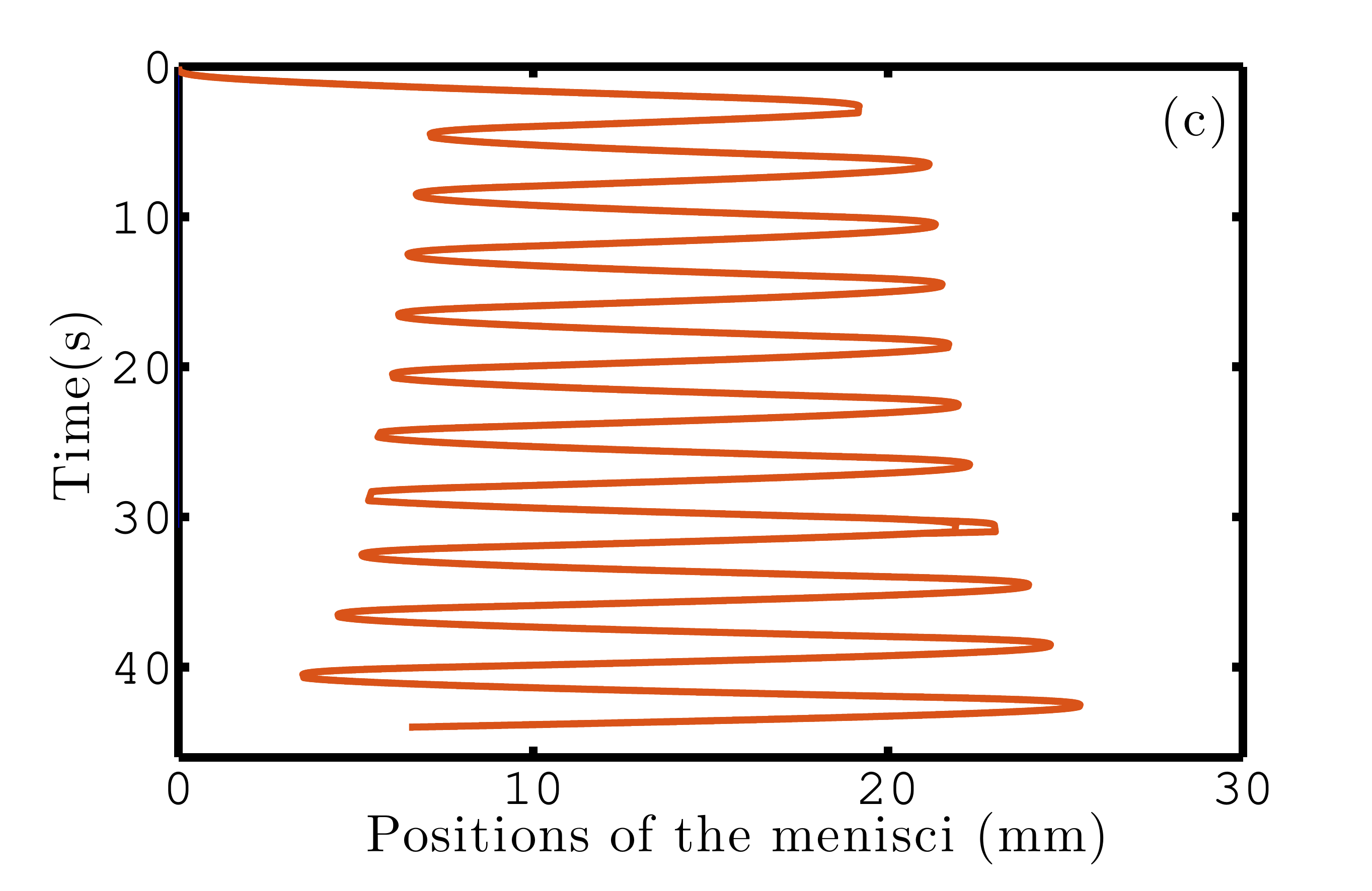}
	\end{subfigure}
		
		\caption{Simulations with equations (\ref{eq:ptdry}) to (\ref{eq:mem}) of the positions of the menisci of a liquid plug of initial of size $L_0 = 3.3$ mm pushed with the cyclic pressure driving represented on \cref{Periodic_functions}(b). (a) Complete models (equations (\ref{eq:lubri2}) and (\ref{eq:lubri})). (b) The viscous  resistance interface is kept constant $\tilde{R}_v = cste$. (c) The front interface resistance is kept constant $(F=E)$.}
		\label{Comparision_Lub_Ecst}
	\end{figure}
	
\begin{figure}
	\begin{center}
		\includegraphics[width=0.7 \textwidth]{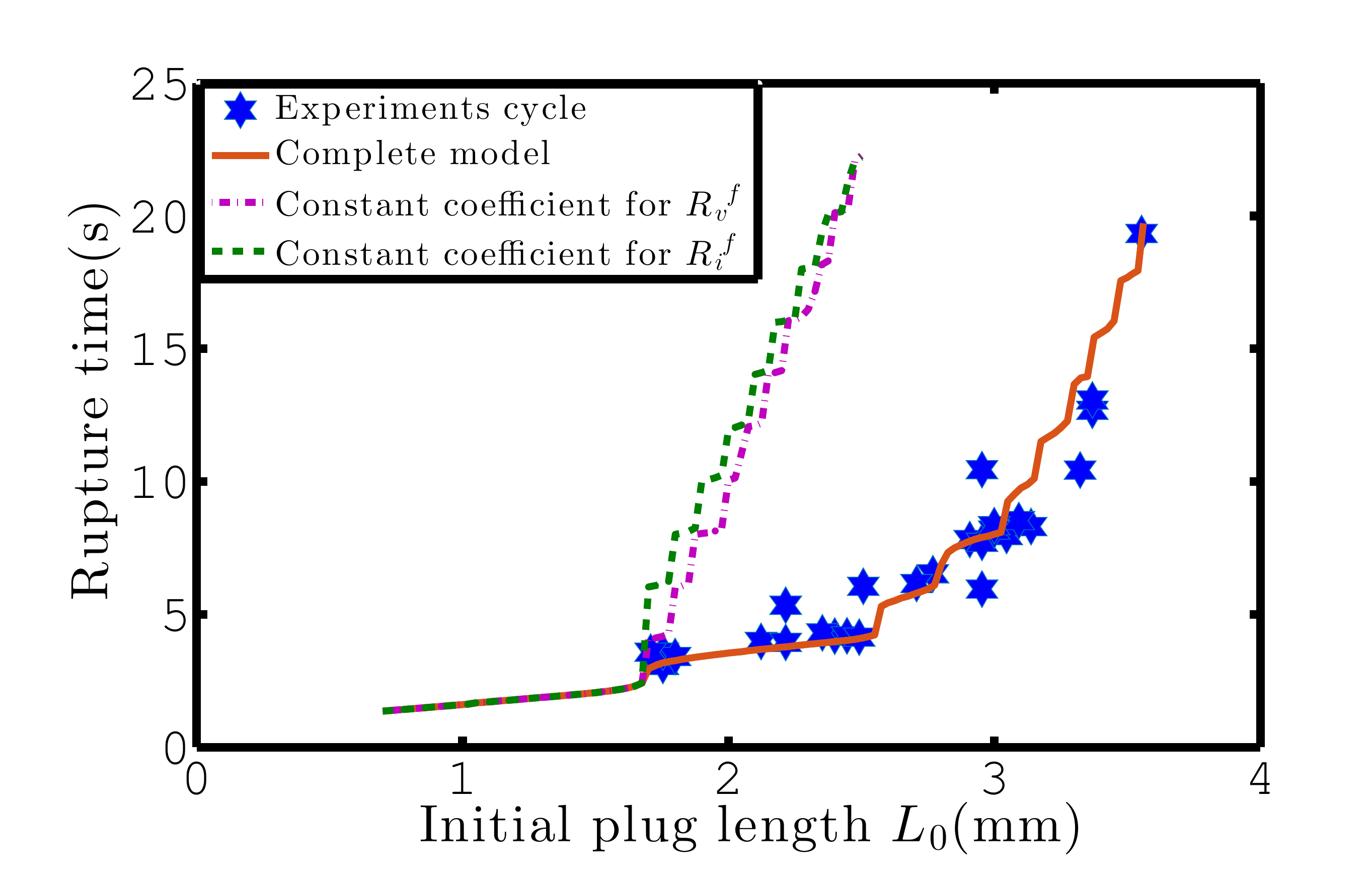} 
		\caption{\textcolor{black}{Rupture time of liquid plugs pushed with a cyclic pressure driving given by equation (\ref{eq:deltapt_c}) as a function of their initial lengths $L_0$. Blue stars correspond to experiments, the red solid curve to simulations with the complete model, the purple dashed-dotted line to simulations when the viscous resistance $\tilde{R}_v$ is kept constant and the green dashed-dotted line to simulations when the front interface resistance $\tilde{R}_i^f$ is kept constant and thus lubrication effects are discarded. }}
		\label{lubrication_cloud}
		\end{center}
	\end{figure}

\textcolor{black}{Of course, these two memory effects are coupled. To quantify the relative contribution of  these two effects, we simulated the plug behaviour when the viscous resistance $\tilde{R}_v$ is kept constant (\cref{Comparision_Lub_Ecst}(b)) and when the front interface resistance $\tilde{R}_i^f$ is kept constant (\cref{Comparision_Lub_Ecst}(c)). The simulations show that in these two cases the plug acceleration and rupture still occurs but that the rupture time is substantially increased. This tendency is confirmed on figure \ref{lubrication_cloud} where we compared the time necessary for the plug to rupture (called the "rupture time") to simulations for a large number of initial plug sizes when the whole model is considered (red solid line), when the first memory effect (length effect) is discarded (purple dashed line) and when the second memory effect (lubrication effect) is discarded (green dashed-dotted line). While the complete model quantitatively reproduces the tendencies, the other two simulations largely overestimate the rupture time.} 

This analysis also shows the central role played by the initial wetting condition: The successive accelerations at each half-cycle all originate from the transition between a dry and a prewetted capillary tube during the first cycle, which led to a massive acceleration of the plug in the back motion. In theory, the opposite behavior (plug cyclic slow down and growth) might be observed in a prewetted capillary tube depending on the thickness of the prewetting film and the amplitude of the pressure driving as was already observed by \cite{magniez2016dynamics} for unidirectional constant pressure forcing. \\

\section{Cyclic motion vs direct rupture of the plug under pressure forcing} \label{cyclic_motion}

\begin{figure}
	\centering
	\begin{subfigure}[b]{0.33\textwidth}
		\includegraphics[width=\linewidth, height=4cm]{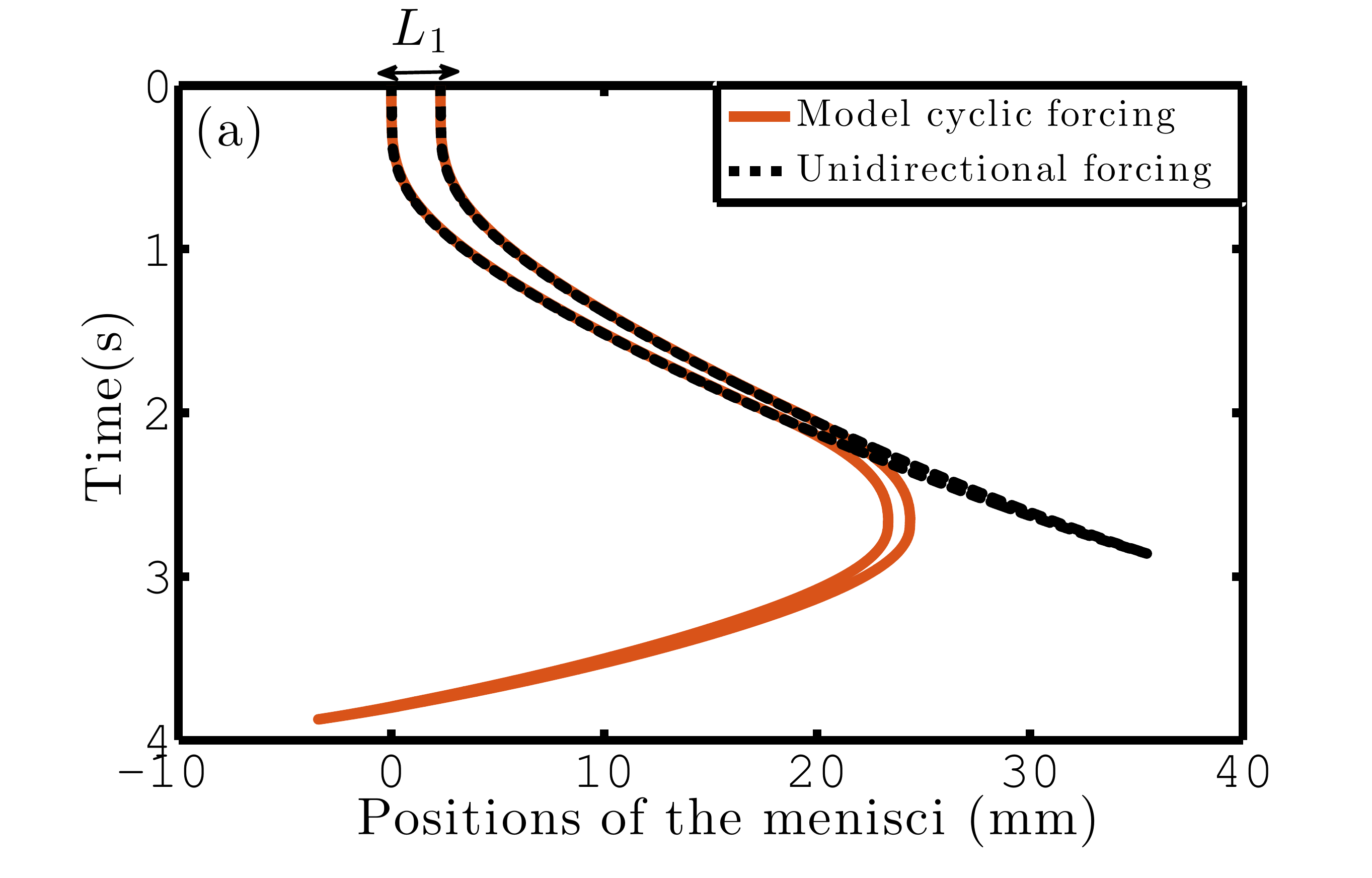}
		%  \caption{A gull}
		%  \label{fig:gull}
	\end{subfigure}%
	\begin{subfigure}[b]{0.33\textwidth}
		\includegraphics[width=\linewidth, height=4cm]{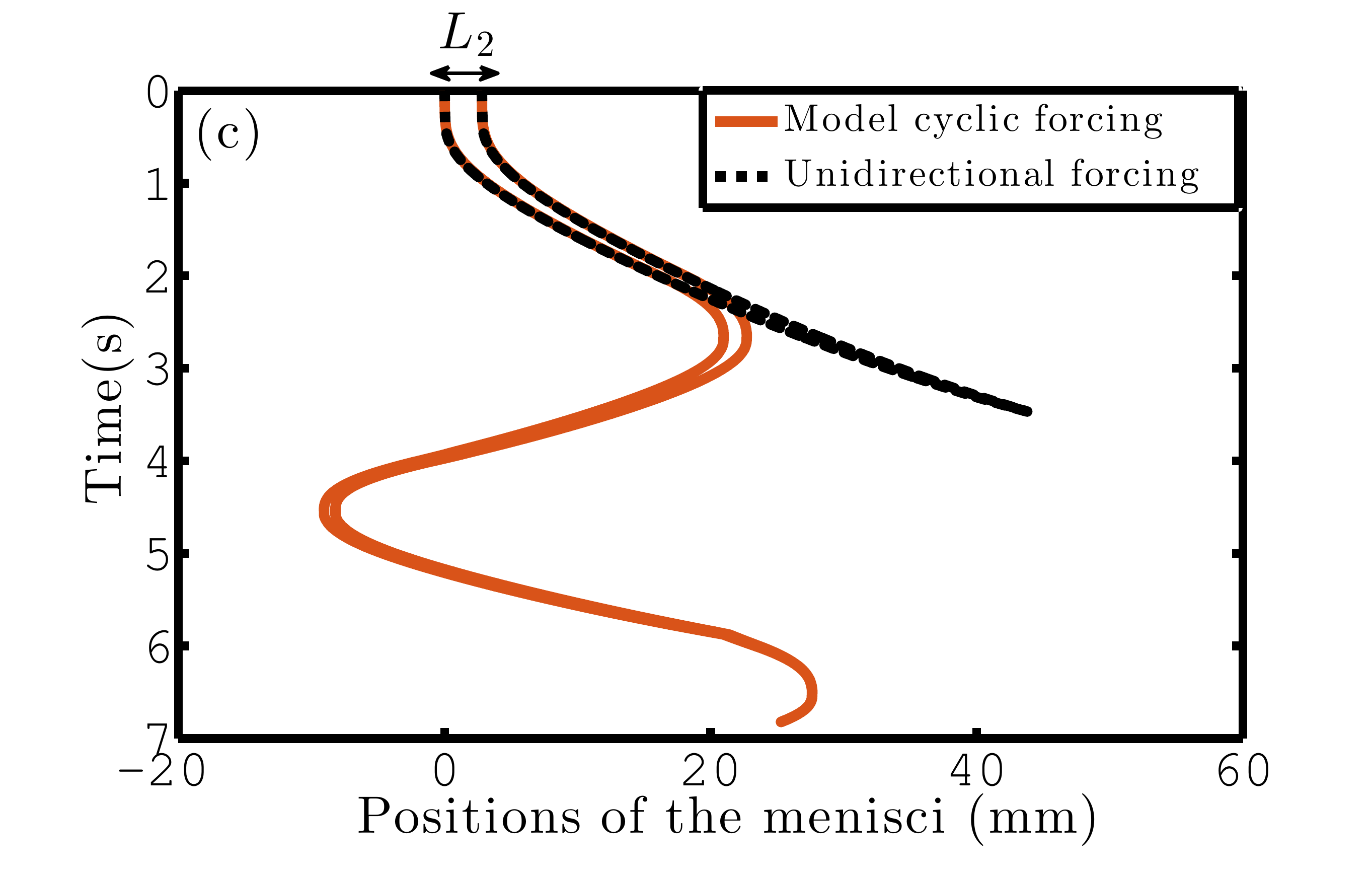}
		% \caption{A gull2}
		% \label{fig:gull2}
	\end{subfigure}%
	\begin{subfigure}[b]{0.33\textwidth}
		\includegraphics[width=\linewidth, height=4cm]{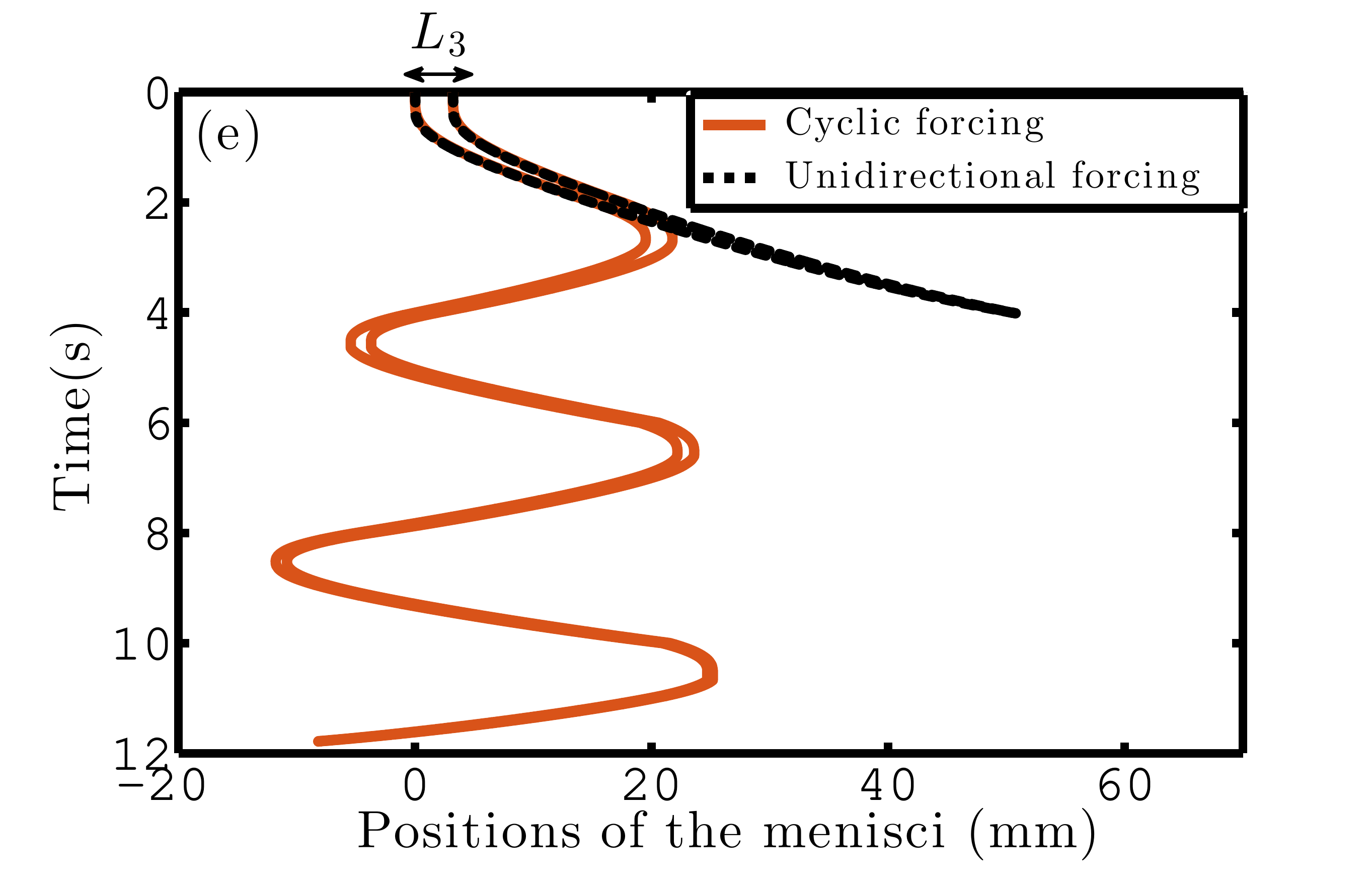}
		%\caption{A tiger}
		%\label{fig:tiger}
	\end{subfigure}% 
	\\
	\centering
	\begin{subfigure}[b]{0.33\textwidth}
		\includegraphics[width=\linewidth, height=4cm]{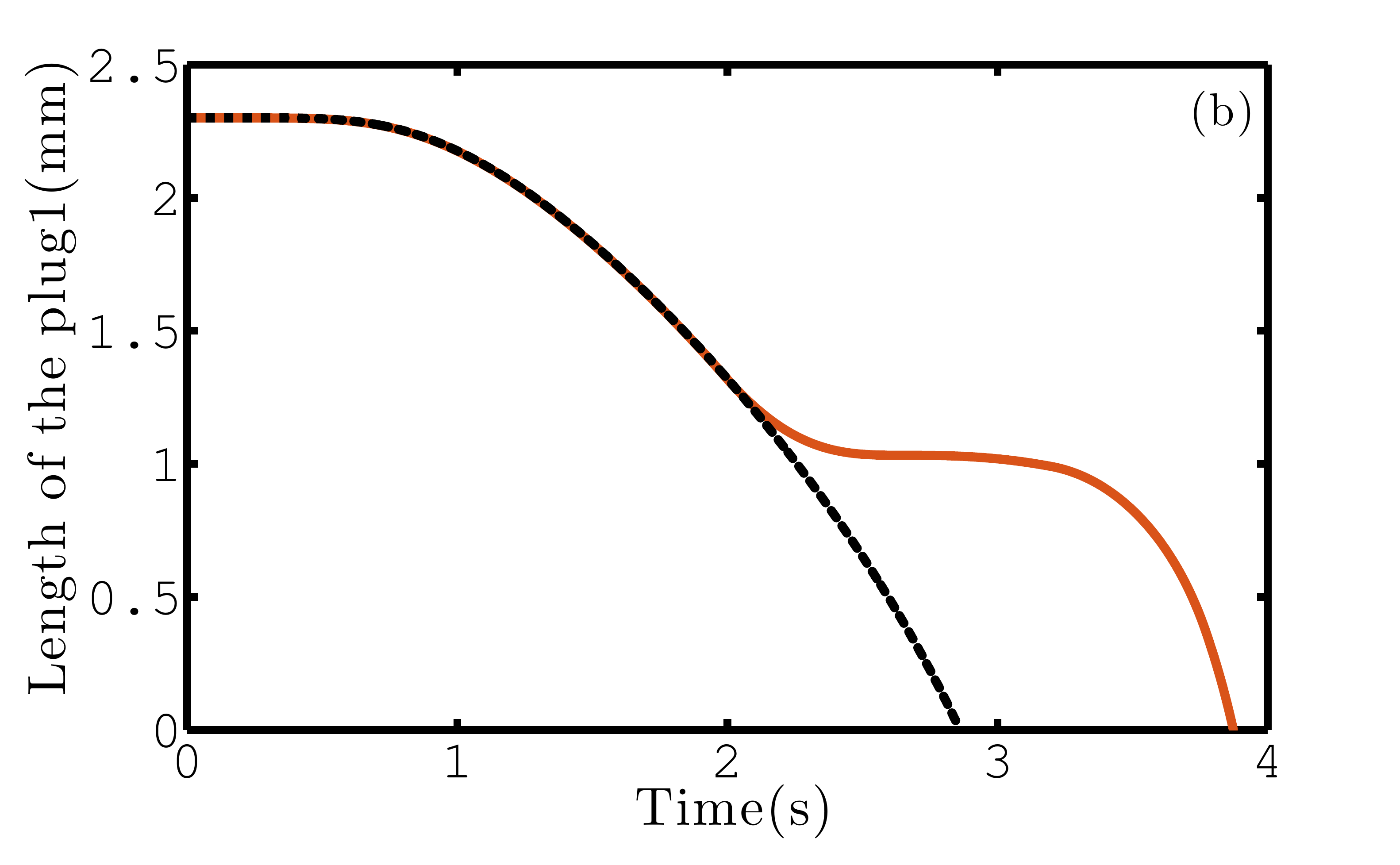}
		%  \caption{A gull}
		%  \label{fig:gull}
	\end{subfigure}%
	\begin{subfigure}[b]{0.33\textwidth}
		\includegraphics[width=\linewidth, height=4cm]{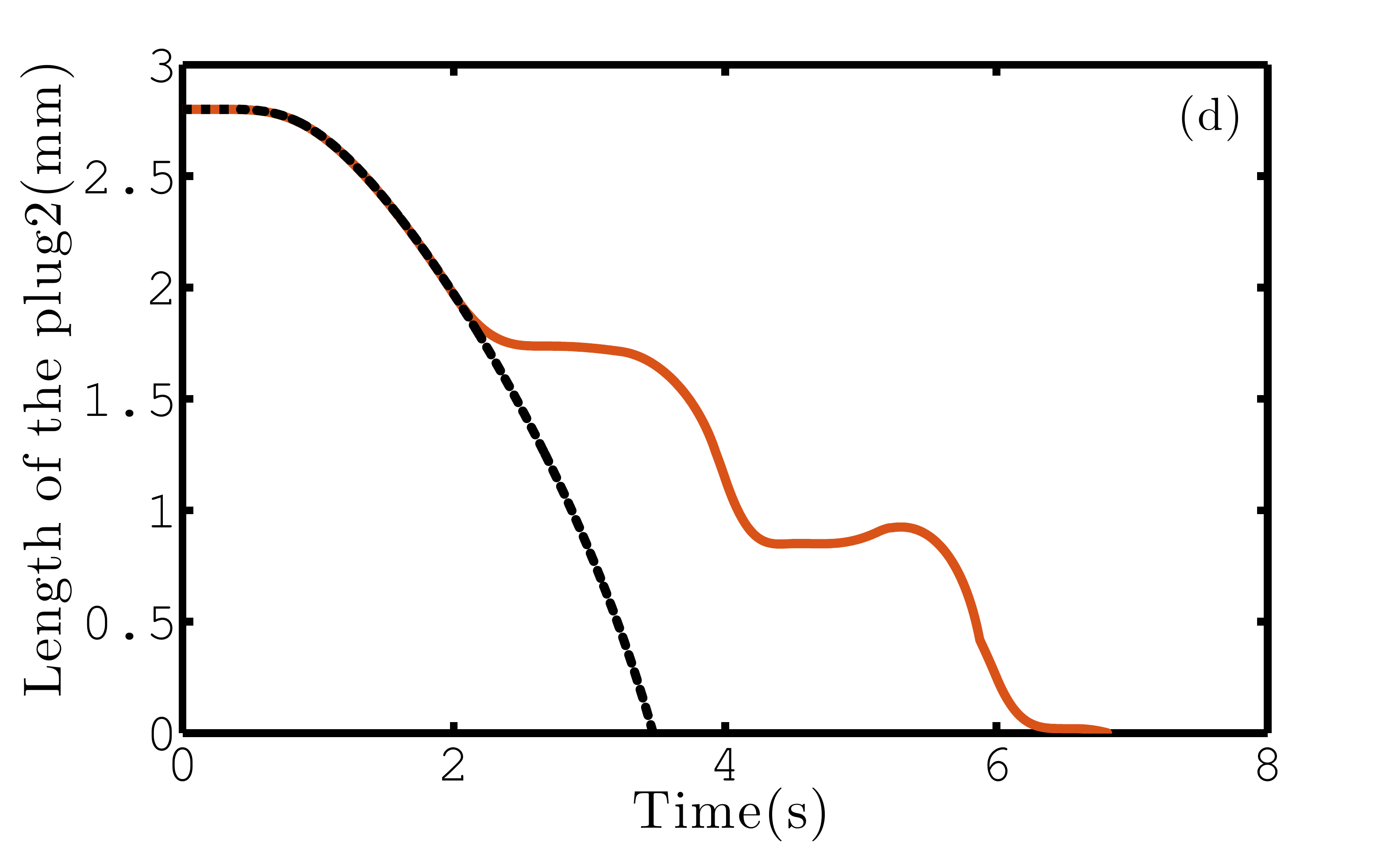}
		% \caption{A gull2}
		% \label{fig:gull2}
	\end{subfigure}%
	\begin{subfigure}[b]{0.33\textwidth}
		\includegraphics[width=\linewidth, height=4cm]{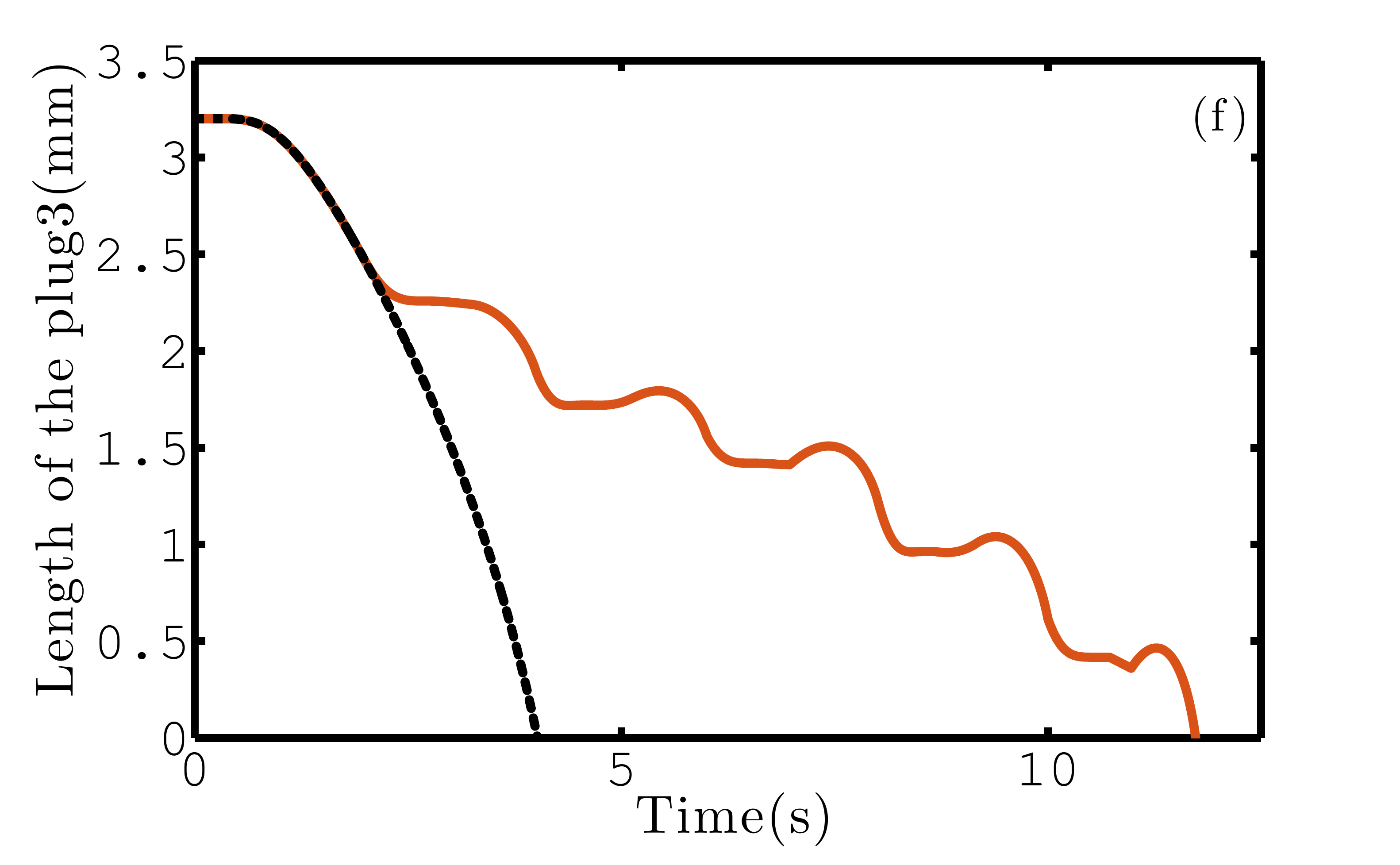}
		%\caption{A tiger}
		%\label{fig:tiger}
	\end{subfigure}% 
	\caption{ \label{Normcas_cycle}  Spatiotemporal evolution of plugs of initial lengths  \textcolor{black}{$L_1=2.2$ mm (a,b), $L_2=2.8$ mm (c,d)  and  $L_3=3.2$ mm} (e,f) pushed either with a unidirectional pressure driving (black dotted line) or with a cyclic pressure driving (\textcolor{black}{red solid line}). (a), (c), (e): Position of the rear and front menisci. (b), (d), (f): Evolution of the plug size.}
\end{figure}

In this last section, we compare experimentally and theoretically the time and space required to break a liquid plug with either a unidirectional or a cyclic pressure forcing with the same magnitude. The two driving conditions used for this comparison are represented respectively on \cref{pressure_driving} and \cref{Periodic_functions}(b). As  previously mentioned, their temporal evolution can be approximated respectively by the Gompertz function \textcolor{black}{$\Delta P_t = 78 \exp(-6 \exp(-3t))$} and the equations (\ref{eq:deltapt_c}).

\begin{figure}
	%\vspace*{-0.3cm} 
	
	\begin{subfigure}{0.51\textwidth}
		\hspace*{-0.5cm} 
		\includegraphics[width=\linewidth, height=5cm]{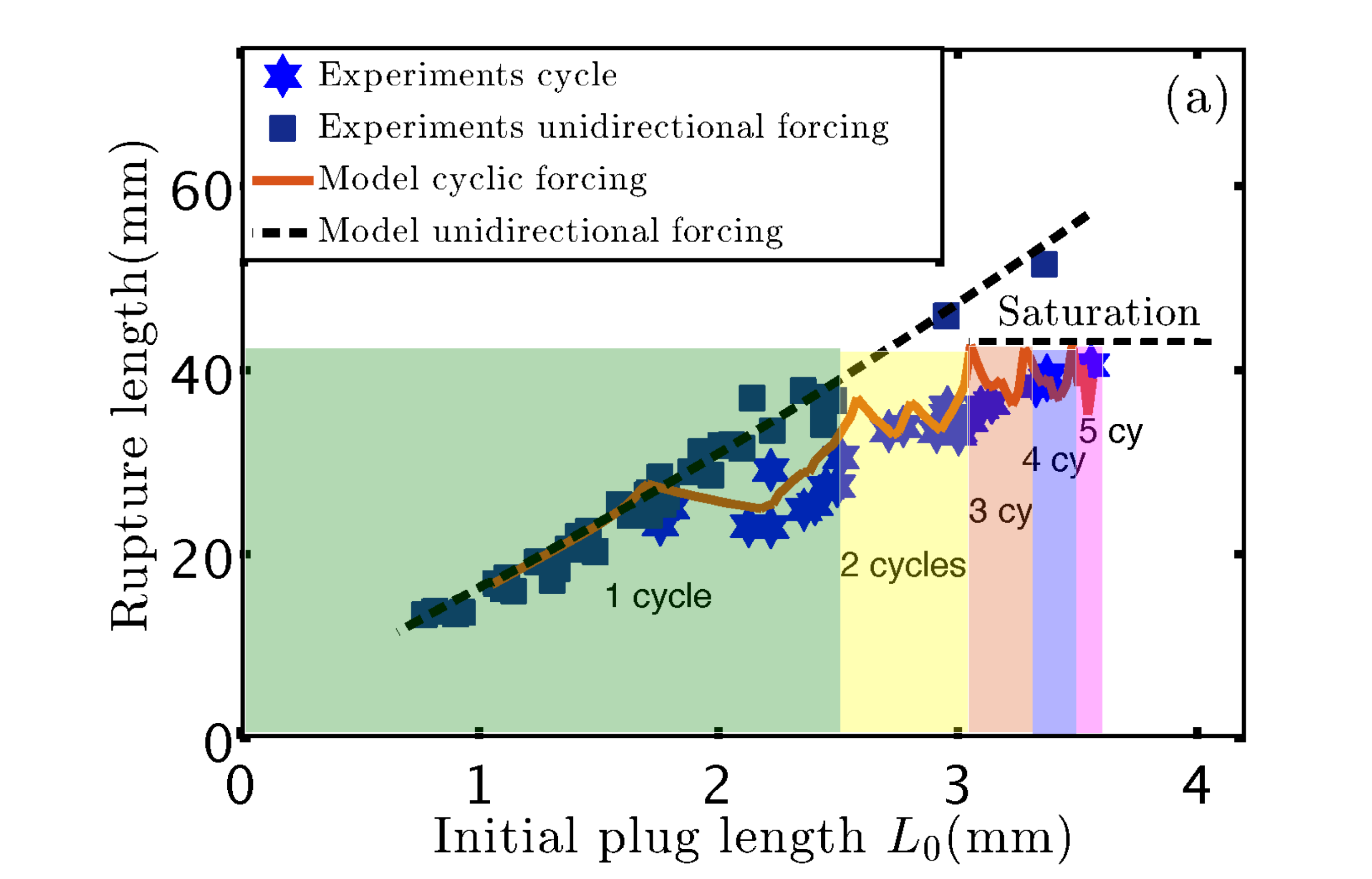} 
	\end{subfigure}
	\begin{subfigure}{0.51\textwidth}
		\hspace*{-0.5cm} 
		\includegraphics[width=\linewidth, height=5cm]{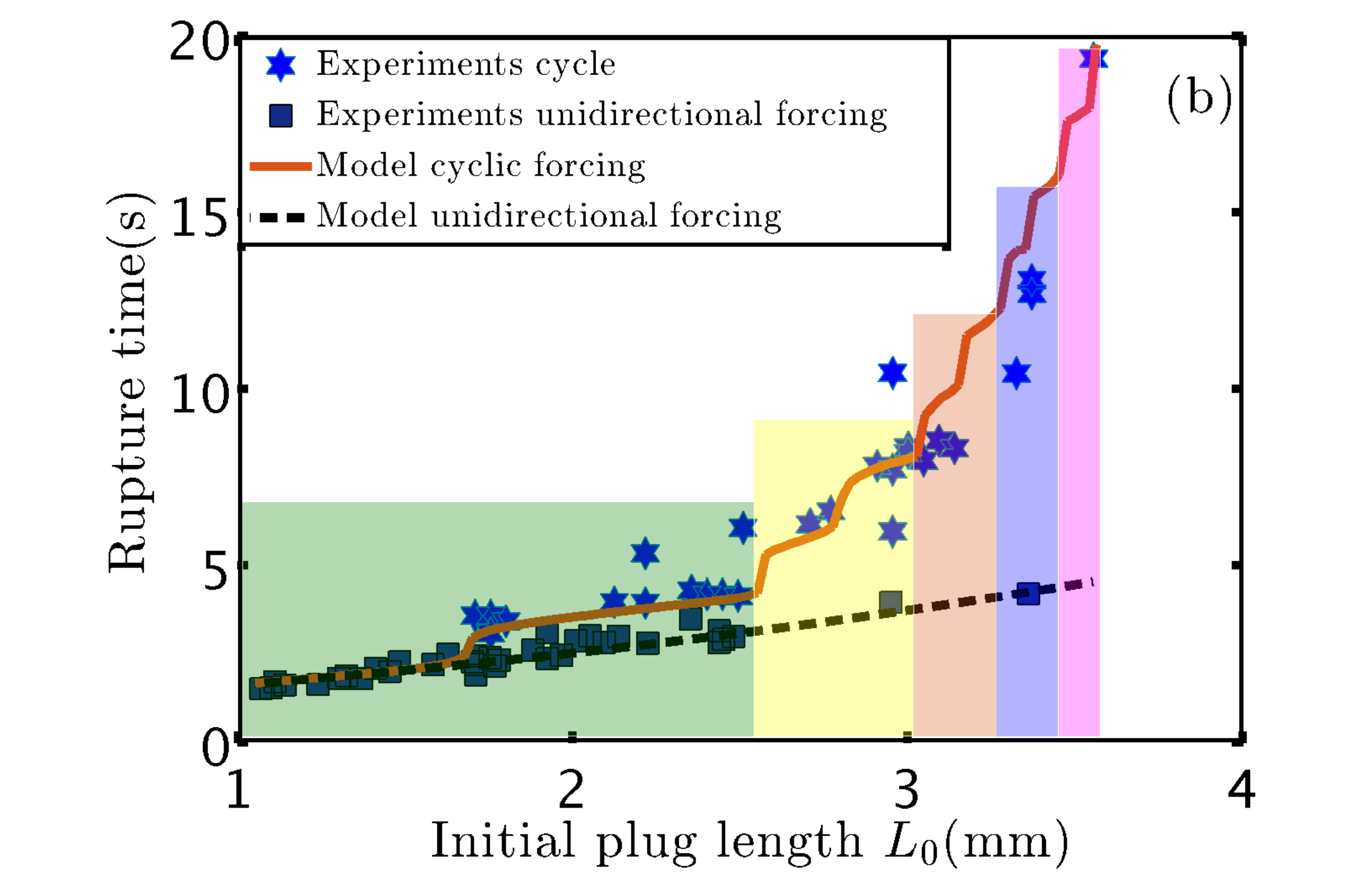}
	\end{subfigure}
	
	\caption{ \label{cycle_cloud}
		(a) Rupture length and (b) rupture time of liquid plugs pushed with a cyclic pressure driving given by equation (\ref{eq:deltapt_c}) as a function of their initial lengths $L_0$. Blue stars correspond to cyclic experiments and the red solid curve to simulations. The black dashed line and the blue square dots correspond respectively to simulations and experiments for a unidirectional pressure driving.}
	
\end{figure}

Figure \ref{Normcas_cycle} compares theoretically the dynamics of liquid plugs of increasing sizes for unidirectional and cyclic pressure driving. This figure shows (i) that plug rupture and thus airways reopening is obtained in a longer time but in a more confined space with a cyclic forcing compared to a unidirectional pressure forcing and  (ii) that the difference between these two driving conditions increases with the number of cycles and hence with the initial size of the liquid plug. This tendency has been verified experimentally and theoretically on a large number of initial plug lengths. The results are summarized in \cref{cycle_cloud}. Figures \ref{cycle_cloud}(a) and (b) show respectively the rupture length (the portion of the tube visited by the liquid plug before its rupture) and the rupture time (the time required for the plug to rupture) as a function of the plug initial length, $L_0$. In these two figures, the blue stars and the solid red line correspond respectively to experiments and simulations for a \textit{cyclic} pressure driving, while the black dots corresponds to simulations with a \textit{unidirectional} pressure driving. The successive cycles are highlighted with different colors. This figure shows again excellent agreement between experimental data and numerical predictions for up to 5 cycles (\cref{cycle_cloud}), underlining that the model summarized in equations (\ref{eq:ptdry}) to (\ref{eq:mem}) captures the main physics.

As long as the liquid plug breaks during the first half cycle, the cyclic forcing (red solid line) and the unidirectional forcing (black dotted line) are of course equivalent. When the plug starts going back (for initial length $L_0 \approx 1.7$ mm) brutal changes in the tendencies are observed: the rupture length starts decreasing (\cref{cycle_cloud}(a)), while the increase in the rupture time is on the contrary exacerbated (\cref{cycle_cloud}(b)). For larger plug lengths, the number of cycles required to achieve plug rupture increases rapidly. Since each change in the plug flow direction is associated with some sharp fluctuations of the rupture length, this increase in the number of cycles leads to a saturation of the rupture length (\cref{cycle_cloud}(a)). This is very different from the relatively linear trend predicted by our simulations (black dots) and observed experimentally on \cref{Normal_cascade_cloud} for a unidirectional forcing. This saturation means that there is a maximal distance that a liquid plug can travel regardless of its size for a prescribed pressure cycle. An interesting point is that, despite this confinement,  the plug rupture remains possible due to the hysteretic effects that enable a progressive acceleration of the liquid plug at each cycle, even if the liquid plug moves on the same portion of the tube. 

\begin{figure}
	\centering
	\begin{subfigure}[b]{0.5\textwidth}
		\includegraphics[width=\linewidth, height=5cm]{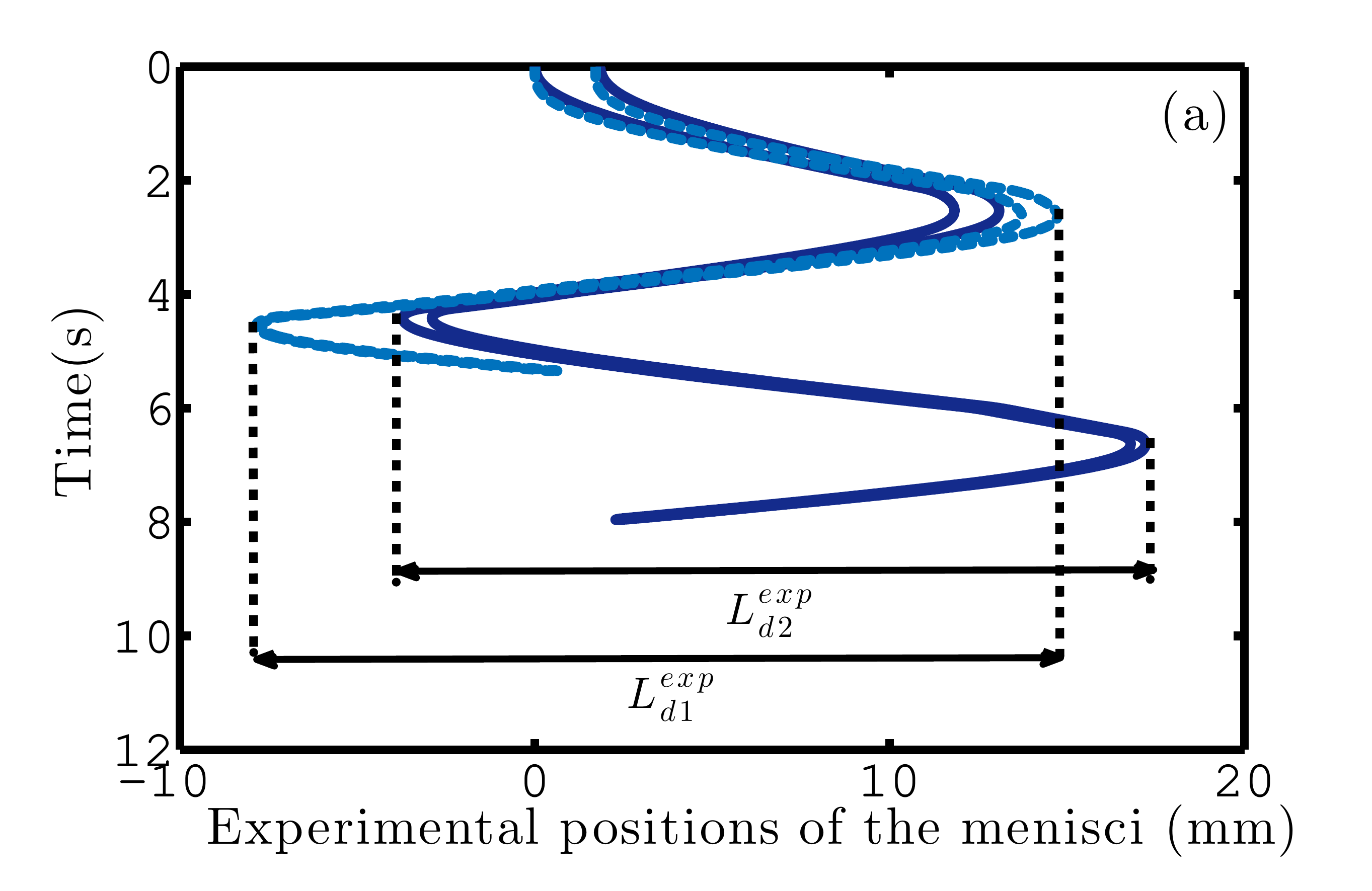}
	\end{subfigure}%
	\begin{subfigure}[b]{0.5\textwidth}
		\includegraphics[width=\linewidth, height=5cm]{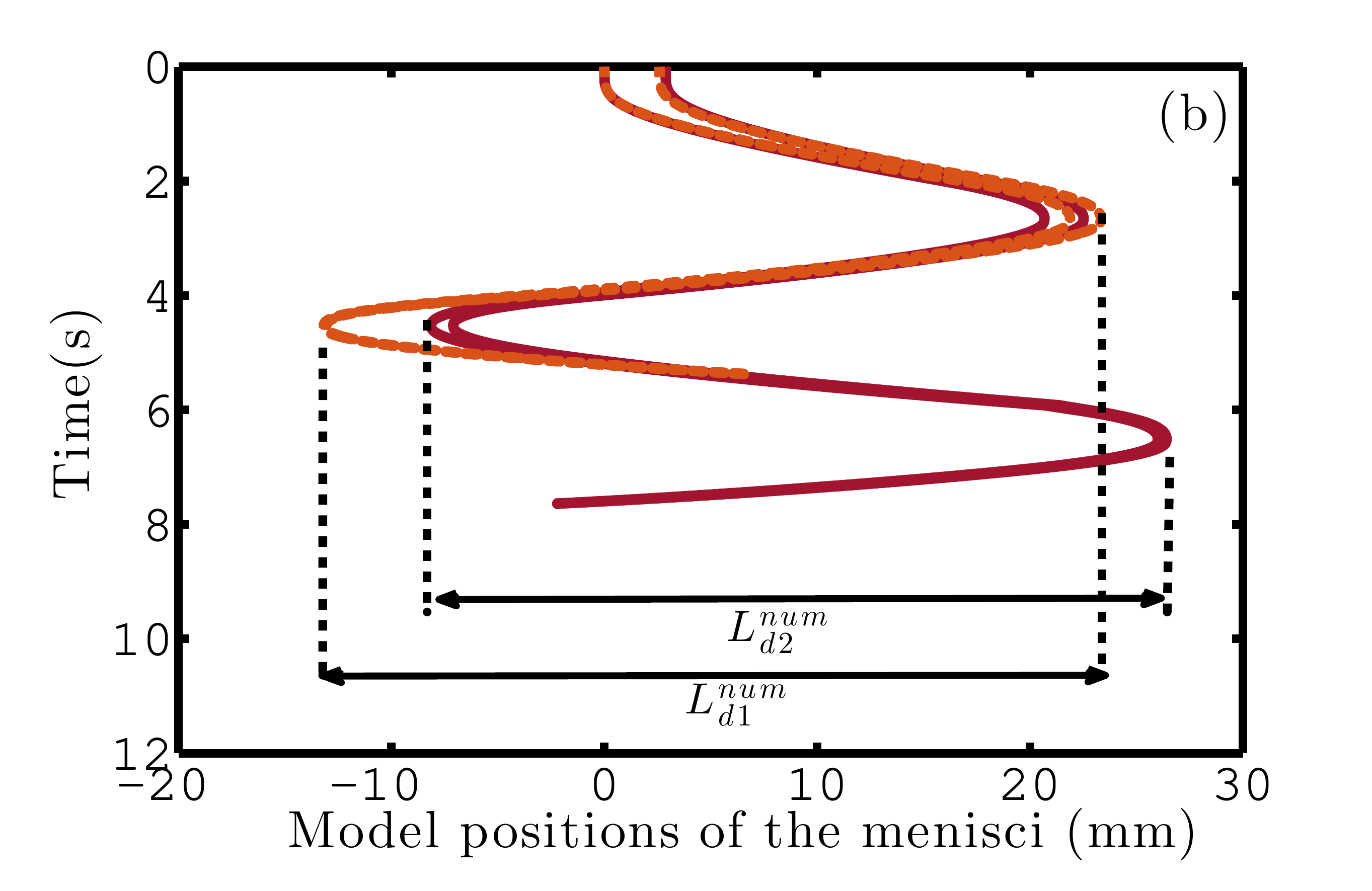}
	\end{subfigure}%
	\caption{ \label{destruc_length}  (a) Experimental investigation and (b) numerical investigation of the rupture length of two liquid plugs of close initial lengths. The initial length of the plug in dashed lines  $L_1=2.5$ mm is smaller than the initial length of the plug in solid line $L_2=2.85$ mm but nevertheless travels on a larger distance $L_{d1} > L_{d2}$  before its rupture. The evolution of these two plugs can also be seen on supplementary movies S3 and S4 respectively.}	
\end{figure}

To understand the decrease in the rupture length observed when the flow direction is changed, we plotted the experimentally observed  (\cref{destruc_length}(a)) and numerically predicted (\cref{destruc_length}(b)) spatiotemporal diagrams of the evolution of two plugs with initial lengths $L1=2.5$ mm (movie S3) and $L2=2.85$ mm (movie S4) in a region where the rupture length is decreasing when the initial plug length is increased. The experiments were performed with a different pressure driving magnitude $P_o = 60$ Pa than for \cref{cycle_cloud}. So the positions of these two points in the rupture length graph are represented on \cref{figapp} in Appendix \ref{append} (encircled points). In \cref{destruc_length} the dashed lines corresponds to $L1=2.5mm$ and the solid lines to $L2=2.85$ mm.  The experimental (\cref{destruc_length}(a), blue line)  and numerical trends (\cref{destruc_length}(b), red line)  are similar. These figures show that the largest plug requires less space to break than the smallest plug $L_{d1} > L_{d2}$. The origin of this rather counterintuitive behaviour again lies in memory effects. Since these two plugs are pushed with the same pressure head, the smallest plug with the lowest bulk resistance moves faster, leaves more liquid on the walls than the bigger one and thus goes further during the first half-cycle. When the sign of the pressure head is inverted, the smallest liquid plug will move on a more prewetted channel and thus (i) it will travel faster (since lubrication effects reduce its resistance to motion) and (ii) it will recover more liquid, thus slowing down the plug size decrease through the mass balance. The combination of these two effects enables the plug to reach a deeper location in the tube.

On the other hand, a comparison between unidirectional and cyclic forcing indicates that more time is required to break liquid plugs for cyclic motion than straight motion (\cref{cycle_cloud}(b)). This is simply the result of the mass balance. As the liquid plug moves back and forth on prewetted capillary tubes, it recovers some liquid while it doesn't when it moves only on a dry capillary tube. This slows down the plug size evolution. 

\begin{figure}
	\centering
	\begin{subfigure}[b]{0.5\textwidth}
		\includegraphics[width=\linewidth, height=5cm]{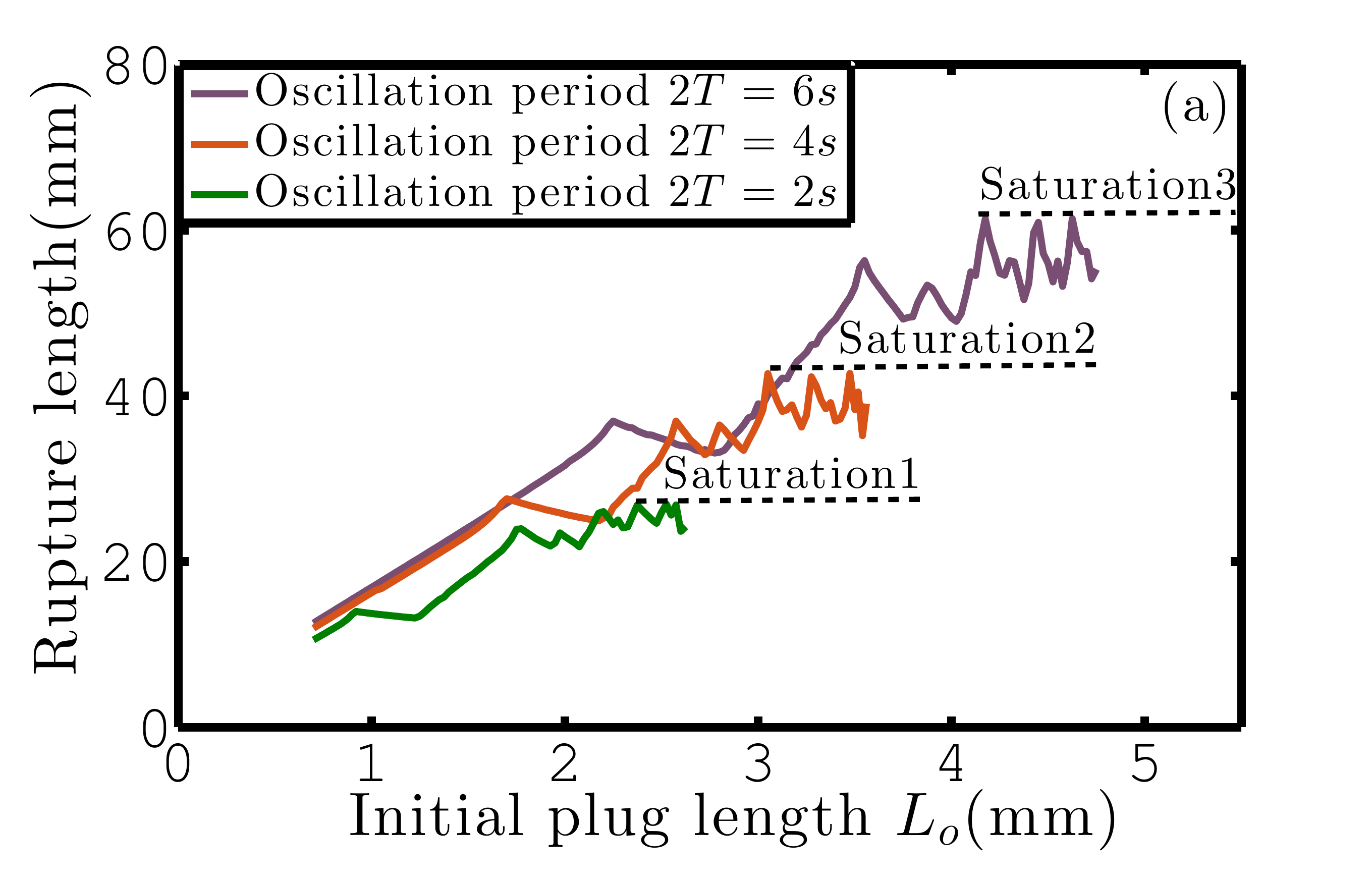}
	\end{subfigure}%
	\begin{subfigure}[b]{0.5\textwidth}
		\includegraphics[width=\linewidth, height=5cm]{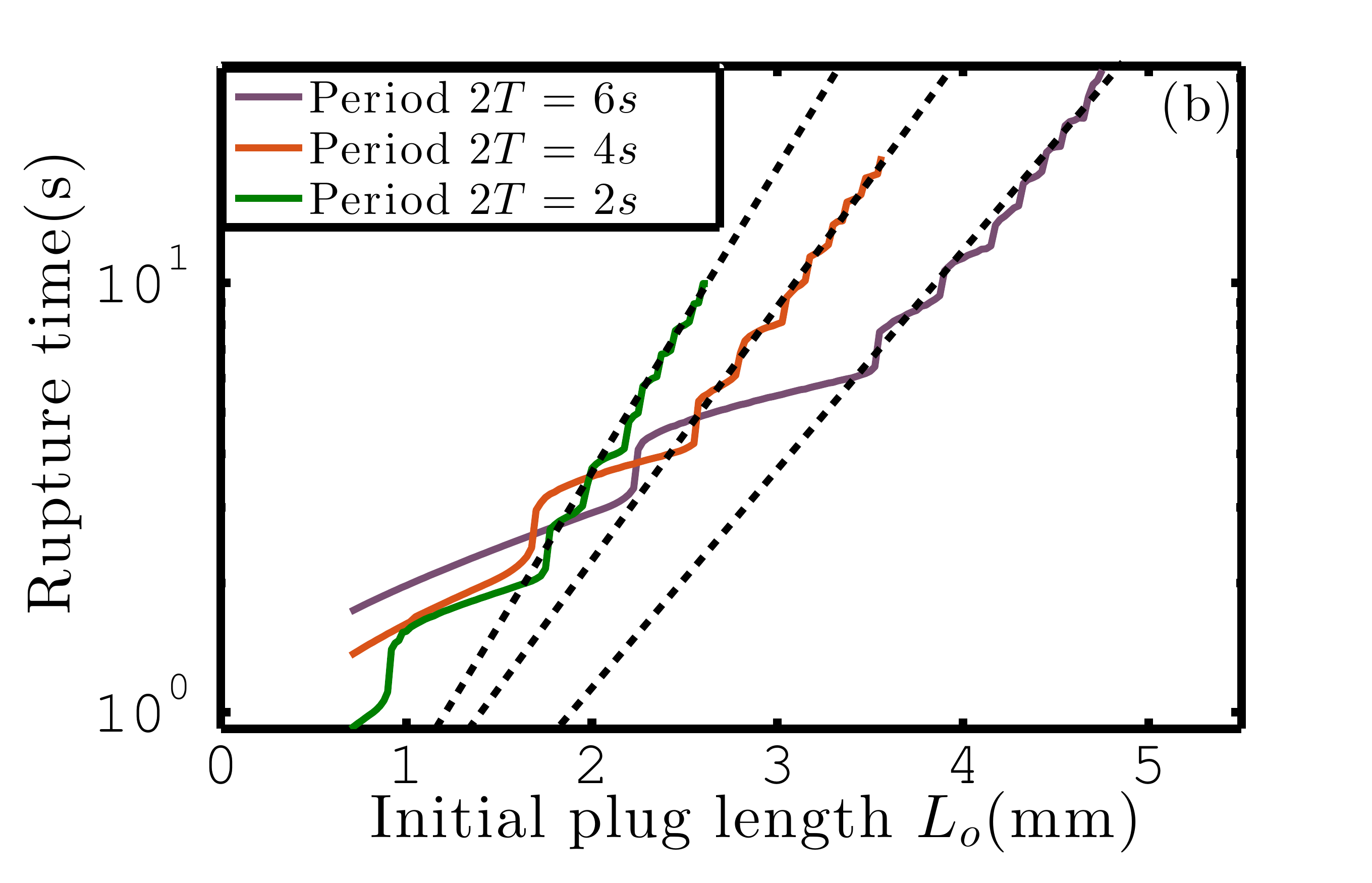}
	\end{subfigure}%
	\caption{ \label{period} \textcolor{black}{(a) Rupture length and (b) rupture time of liquid plugs pushed with a cyclic pressure driving given by equation (\ref{eq:deltapt_c}) as a function of their initial lengths $L_0$ for 3 different time periods $T =$ $2$s, $4$s and $6$s.}}	
\end{figure}

\textcolor{black}{Finally, a comparison of the evolutions of the rupture time and rupture length on \cref{cycle_cloud} shows that the rupture time undergoes an exponential-like growth when the rupture length approaches the saturation zone (for length $L_o > L_s$) with $L_s \approx 2.4$ mm. To confirm this trend, we performed some numerical simulations of the rupture time and rupture length for three different time periods (see \cref{period}). For each period, the evolution is relatively similar and the semilog graph (\cref{period} (b)) indeed underlines that the rupture time follows an exponential growth (rupture time $\propto e^{x/L_c}$) for initial plug lengths larger than a critical length $L_s$. We calculated both the critical saturation length $L_s$ and the characteristic length $L_c$ for the three time periods $T =$ $2$s, $4$s and $6$s and found the values: }
\begin{eqnarray}
 \mbox{Critical saturation lengths: } &  & L_{s2} \approx 1.6 \; mm \quad L_{s4} \approx  2.4 \; mm \quad L_{s6} \approx  3.4 \; mm \\
   \mbox{Characteristic lengths: } & & L_{c2} \approx  0.6 \; mm, \quad L_{c4} \approx  0.7 \; mm \quad L_{c6} \approx 0.8  \; mm
\end{eqnarray}
\textcolor{black}{These two factors depend on the time period $T$. This means that for each time period and pressure forcing, the rupture time becomes exponentially long when $L_o - L_s \gg L_c$ and the pressure driven plug dynamics asymptotes a stable periodic propagation.}

As a conclusion of this section, large liquid plug breaking is achieved in a more confined space but in longer time with a cyclic forcing than with a unidirectional pressure forcing. \textcolor{black}{Moreover, the rupture times grows exponentially when the plug initial length exceeds a critical length $L_s$, whose value depends on the cycle period. In this regime,  the liquid plug dynamics becomes quasi-periodic.}

\section{Conclusion} \label{conclusion}
Despite its occurence in practical situations such as pulmonary flows in pathological conditions, the specificity of the response of liquid plugs to cyclic driving has not been studied so far experimentally and theoretically. The present results show that the dynamics and rupture of a liquid plug strongly depends on the type of forcing. A flow rate cyclic forcing results in periodic oscillations of the plug and no rupture. On the contrary, a pressure cyclic forcing enables airway reopening through a progressive acceleration of the liquid plug dynamics and reduction of its size. This departure from a periodic response originates from \textcolor{black}{two memory effects which decrease the resistance of the plug to motion at each cycle: (i) the cyclic reduction of the plug size which reduces the viscous resistance and (ii) a lubrication effect which reduces the front interface resistance. These two coupled effects are strongly connected to the thickness of the liquid film lying on the walls, which keeps a memory of previous plug displacements.} In addition, this study shows that the rupture of a liquid plug with a prescribed pressure cycle is a spatially bounded phenomenon regardless of the initial plug length. In other words, large plug can be ruptured in a limited space with a cyclic forcing, while more and more space is required to break plugs of increasing size with a unidirectional forcing. The trade-off is that more time is nevertheless required and that this time grows exponentially above a critical length, which depends on the cycle period and the applied pressure.

The analysis of the underlying physics was achieved through a comparison of extensive experimental data to a reduced dimension model. This model quantitatively  predicts the plug behaviour for the numerous pressure cycles studied in this paper. \textcolor{black}{Moreover it is in principle valid for any pressure cycle in the visco-capillary regime (low capillary, Reynolds and Bond numbers)}. Combined with constitutive laws for the plug divisions at bifurcation, it might serve as a basis to simulate cyclic plug dynamics in more complex geometries, or even the dynamics of mucus plugs in distal pulmonary airways. In this last case however, complementary elements such as the influence of walls elasticity, the non-Newtonian fluid properties of mucus, \textcolor{black}{or the presence of an initial mucus layer on the walls} should be implemented to achieve realistic simulations. \textcolor{black}{In particular, it is envisioned that the presence of a prewetting film on the walls might lead to either plug ruptures or persistent occlusions as was demonstrated by \cite{magniez2016dynamics} for unidirectional driving.} Complete models of plug dynamics would open tremendous perspectives, such as the "virtual testing" of new strategies to improve airways clearance for patients suffering from chronic obstructive pulmonary disease or cystic fibrosis. \textcolor{black}{But it might also open perspectives to design robust pressure controllers that enable stable control of liquid plugs. Indeed, the instability to breaking is a major drawback to manipulate plugs with pressure controllers.}

\section*{Acknowledgements}
\textcolor{black}{The authors would like to express their profound gratitude to the four anonymous reviewers, for their extremely interesting suggestions that helped us substantially improve the quality of the manuscript. The author would also like to thank P. Favreau for performing the simulations of the viscous pressure drop in the liquid plug presented in appendix \ref{viscousdrop}. and S.V. Diwakar for his critical reading of the first version of the manuscript.} The author acknowledge the financial support from Universit\'{e} de Lille.

\appendix

\section{Method}

\label{method}

\textcolor{black}{In the two following sections, we describe how the flow rate and pressure cyclic drivings are enforced. }

\subsection{Flow rate cyclic driving}

\begin{figure}
		\begin{center}
			\begin{subfigure}{0.35\textwidth}
				\includegraphics[width=0.9\linewidth, height=3.5cm]{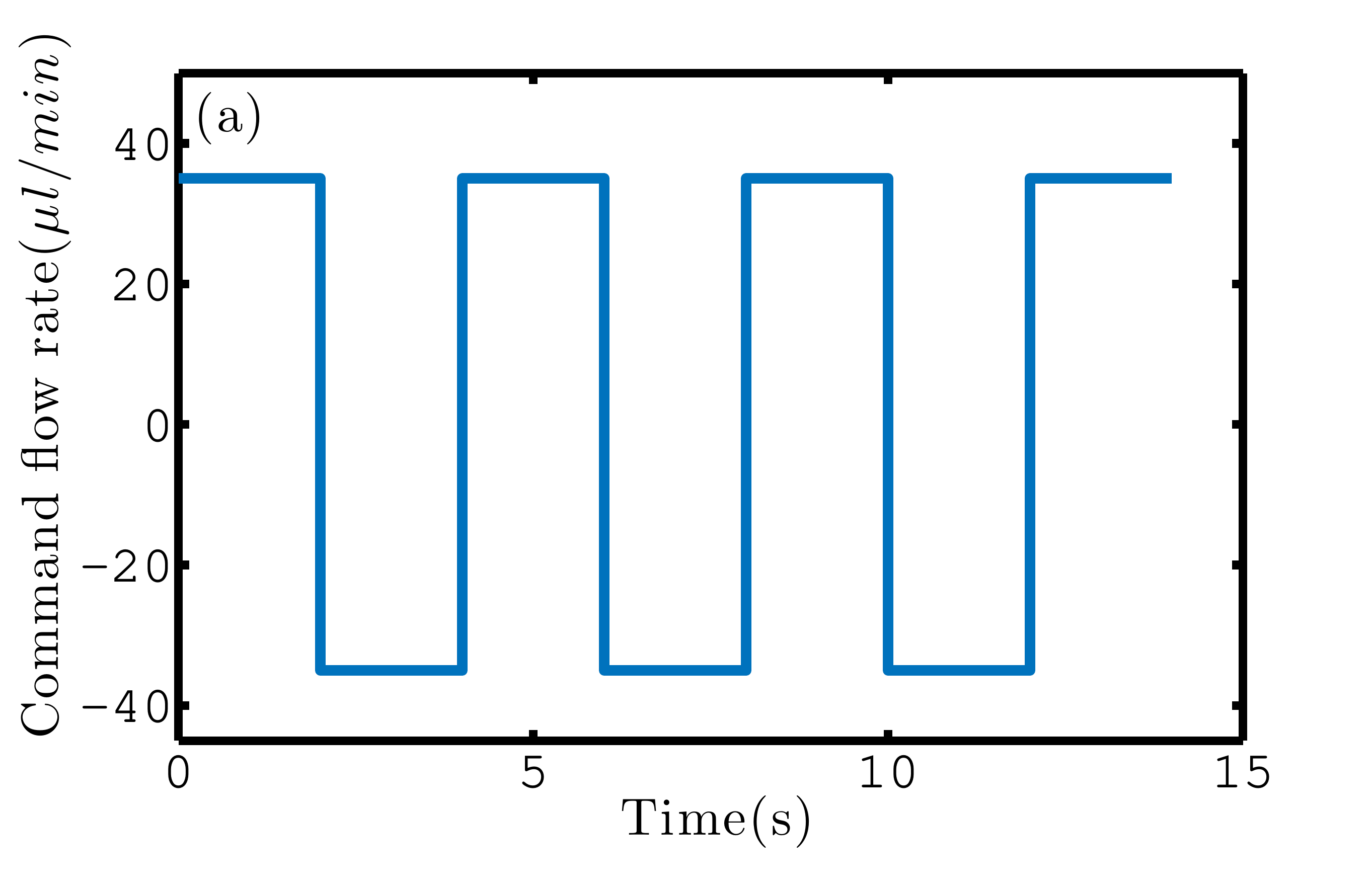} 
			\end{subfigure}
			\begin{subfigure}{0.35\textwidth}
				\includegraphics[width=0.9\linewidth, height=3.5cm]{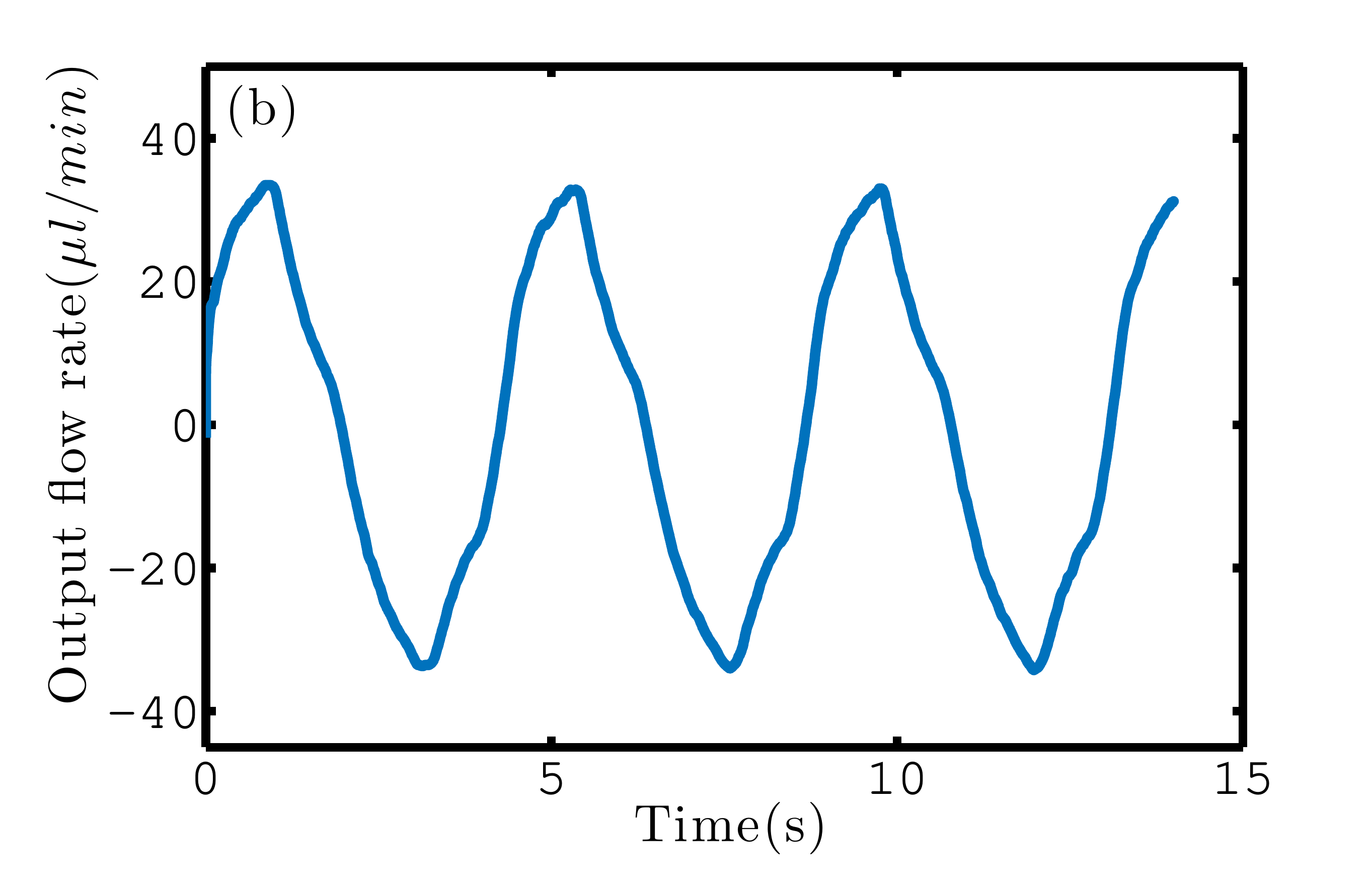}
			\end{subfigure}
		
		\caption{\textcolor{black}{(a) Command flow rate ordered to the programmable syringe pump. Positive values correspond to a motion from the left to the right. (b) Actual flow rate forcing measured by monitoring the motion of the left interface of the plug.}}
		\label{Flow_driving}
			\end{center}
	\end{figure}

\textcolor{black}{The flow rate forcing represented on \cref{Periodic_functions} (a) is obtained by connecting only one end (left side) of the capillary tube to a programmable syringe pump KdScientific 210. The command flow rate is a square signal with alternative motion in the right and left directions, see \cref{Flow_driving} (a). Owing to the response time of the syringe pump and compressibility effects, the actual flow rate imposed to the liquid plug may differ strongly. Thus, the imposed flow rate is monitored directly by measuring the motion of the left interface of the liquid plug. This signal is represented on \cref{Flow_driving} (b).}

\subsection{Pressure cyclic driving}

\begin{figure}
	\begin{center}
		\begin{subfigure}{0.32\textwidth}
			\includegraphics[width=\linewidth, height=3.5cm]{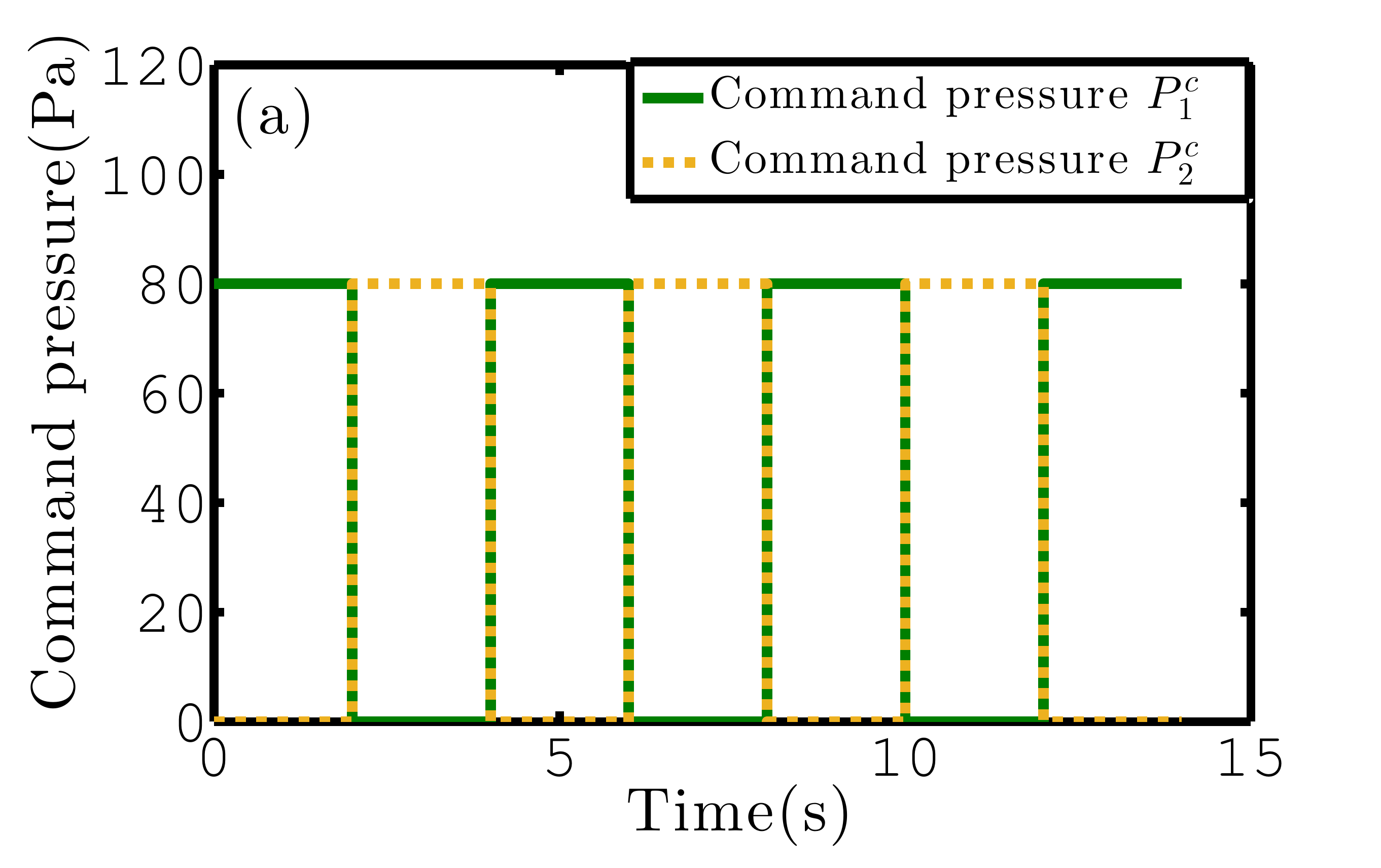} 
		\end{subfigure}
		\begin{subfigure}{0.32\textwidth}
			\includegraphics[width=\linewidth, height=3.5cm]{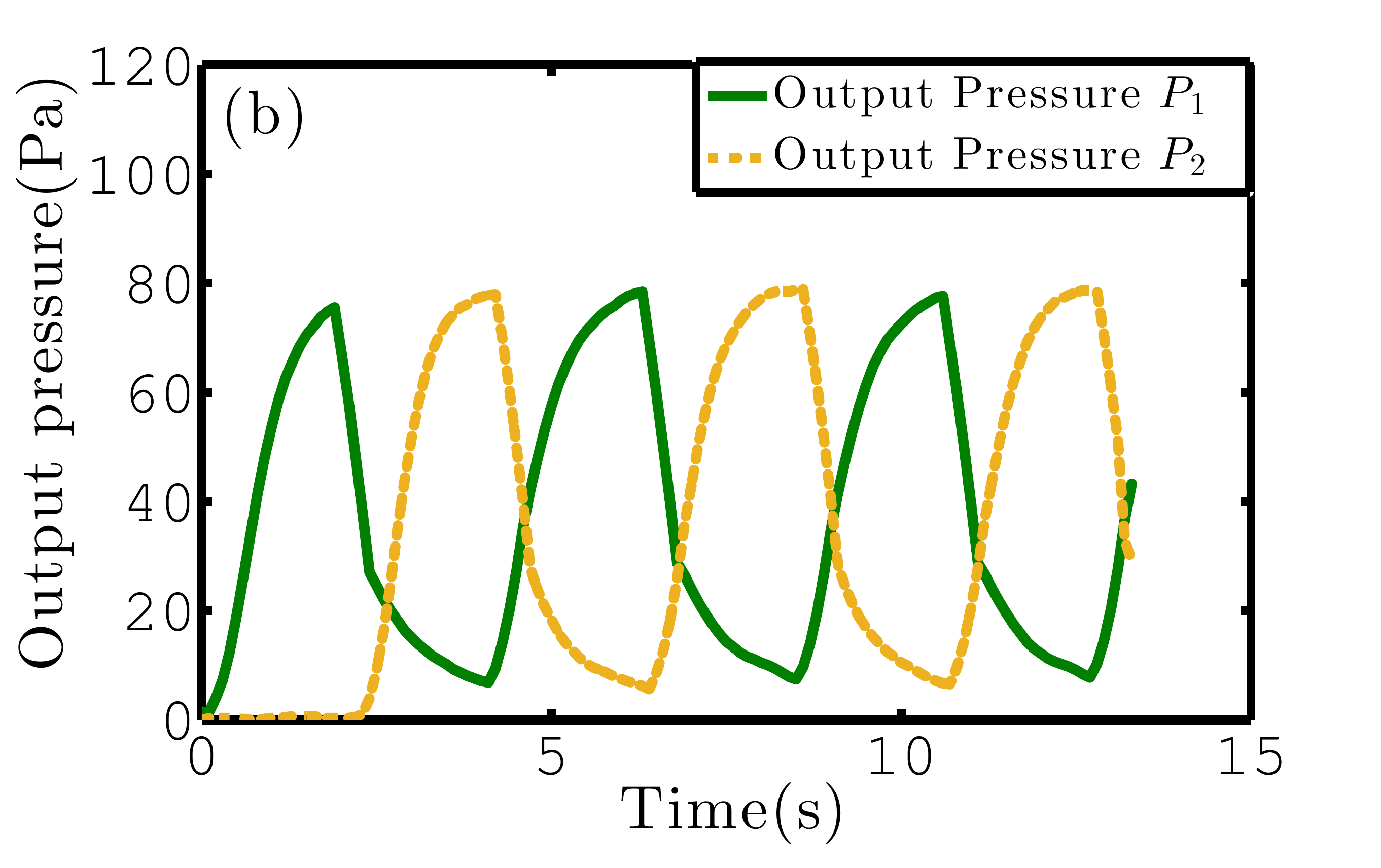}
		\end{subfigure}
		\begin{subfigure}{0.32\textwidth}
			\includegraphics[width=\linewidth, height=3.5cm]{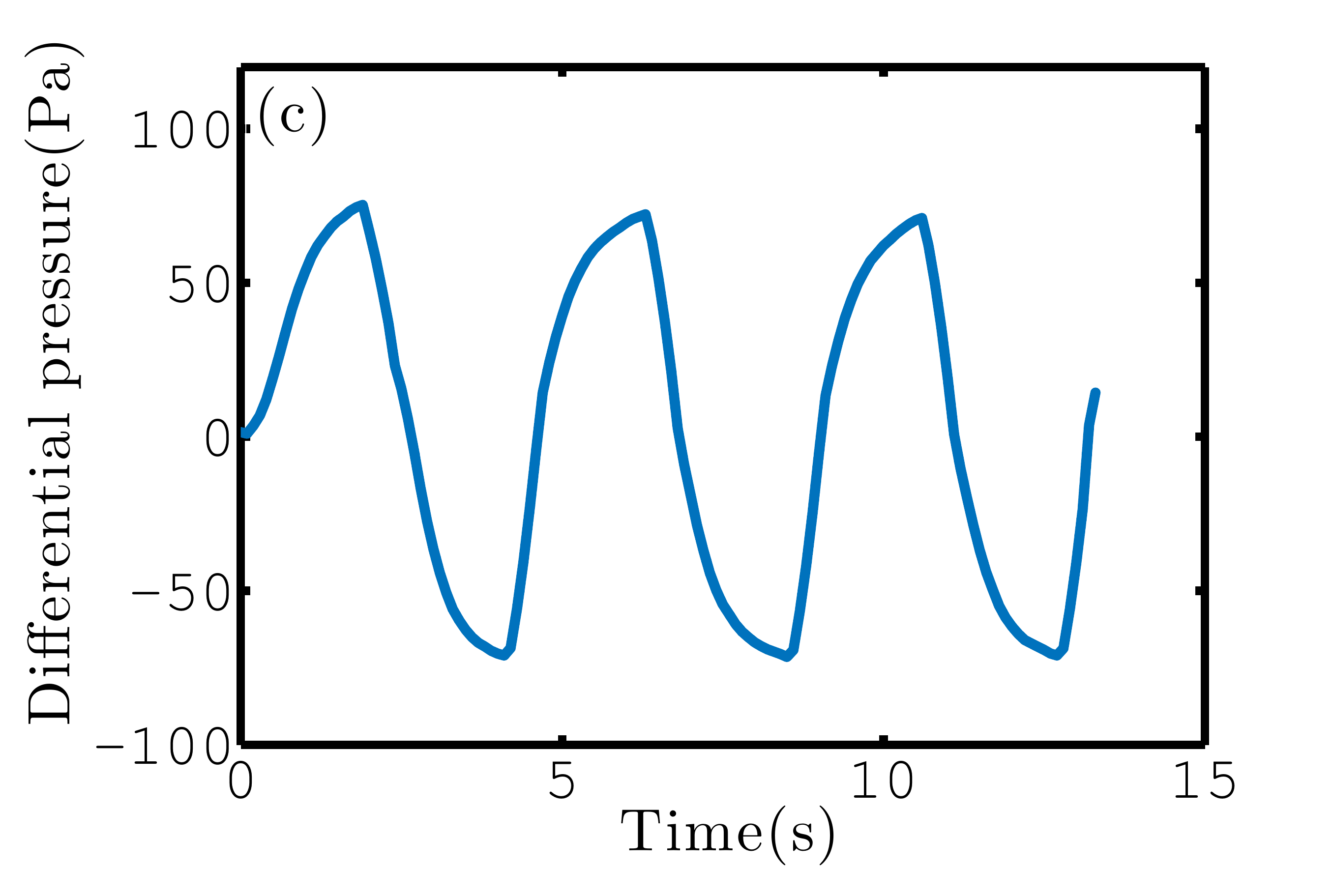}
		\end{subfigure}
		
		\caption{\textcolor{black}{(a) Command pressures $P_1^c$ and $P_2^c$ imposed via the software Maesflow to each channel of the MFCS Fluigent pressure controller, connected respectively to each extremity of the capillary tube. (b) Resulting output pressure signals effectively imposed to each end of the capillary tube, measured with an integrated sensor. The pressure represented on these figures correspond to overpressures compared to atmospheric pressure. (c) Pressure driving $\Delta P_t = P_1 - P_ 2$ imposed to the liquid plug.}}
		\label{Pressure_driving}
				\end{center}
	\end{figure}

\textcolor{black}{The pressure driving represented on \cref{Periodic_functions} (b) is obtained by connecting two channels of the MFCS programmable pressure controller to both ends of the capillary tube. This pressure controller based on valve and sensors enables automated control of the driving pressure. We impose alternatively a constant command overpressure (compared to atmospheric pressure) $P_1^c$ and $P_2^c$ to each channel of the pressure controller while the pressure of the other channel goes down to atmospheric pressure, as represented on \cref{Pressure_driving} (a). Due to the response time of the pressure controller (resulting from the response time of the valve, and the feedback loop, the actual overpressure imposed to each side of the pressure controller (measured by an integrated pressure sensor) is represented on \cref{Pressure_driving} (b). The final  pressure forcing  thus corresponds the difference of pressure  $\Delta P_t = P_1^c - P_ 2^c$ between the two ends of the channel (see \cref{Pressure_driving}) (c)}.

\subsection{Note on compressibility effects.}

\textcolor{black}{Compressibility effects are critical for flow rate driven experiments since they increase the response time of the syringe pump (difference between the piston motion and the actual motion of the fluid in the capillary). To reduce this response time, the syringes are filled with water. Moreover, since the imposed flow rate is measured directly by monitoring the displacement of the left interface, compressibility effects are accounted for in the forcing condition represented on \cref{Periodic_functions} (a). For pressure driven experiments however, the pressure is homogenised at the speed of sound (extremely rapidly) and the response time is mainly due to the valve and sensors response time. Thus, the pressure measured at the exit of the pressure controller with integrated pressure sensors is almost identical to the pressure imposed at both side of the capillary tube (if we neglect the pressure losses due to the air flow in the tubes compared to the pressure losses due to the presence of the liquid plug). }

\section{Direct numerical simulations of the viscous pressure drop inside the liquid plug}
\label{viscousdrop}

	\begin{figure}
	\begin{center}
			\includegraphics[width=0.9 \textwidth]{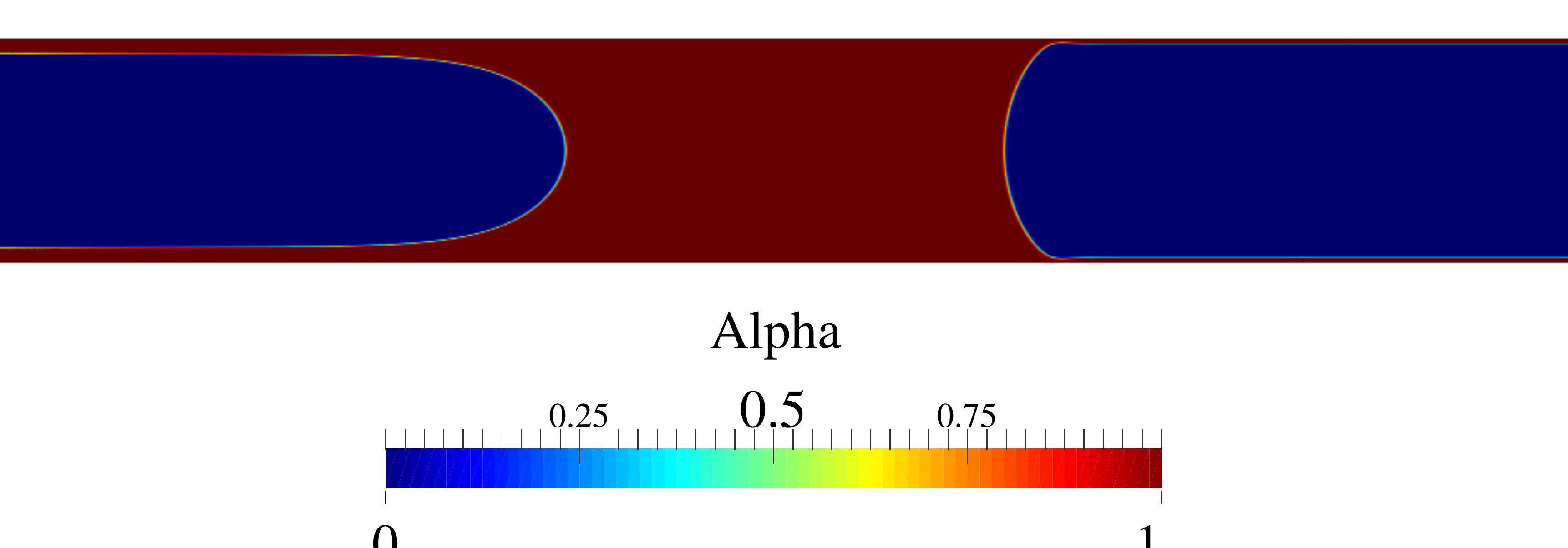}  
			\includegraphics[width=0.5 \textwidth]{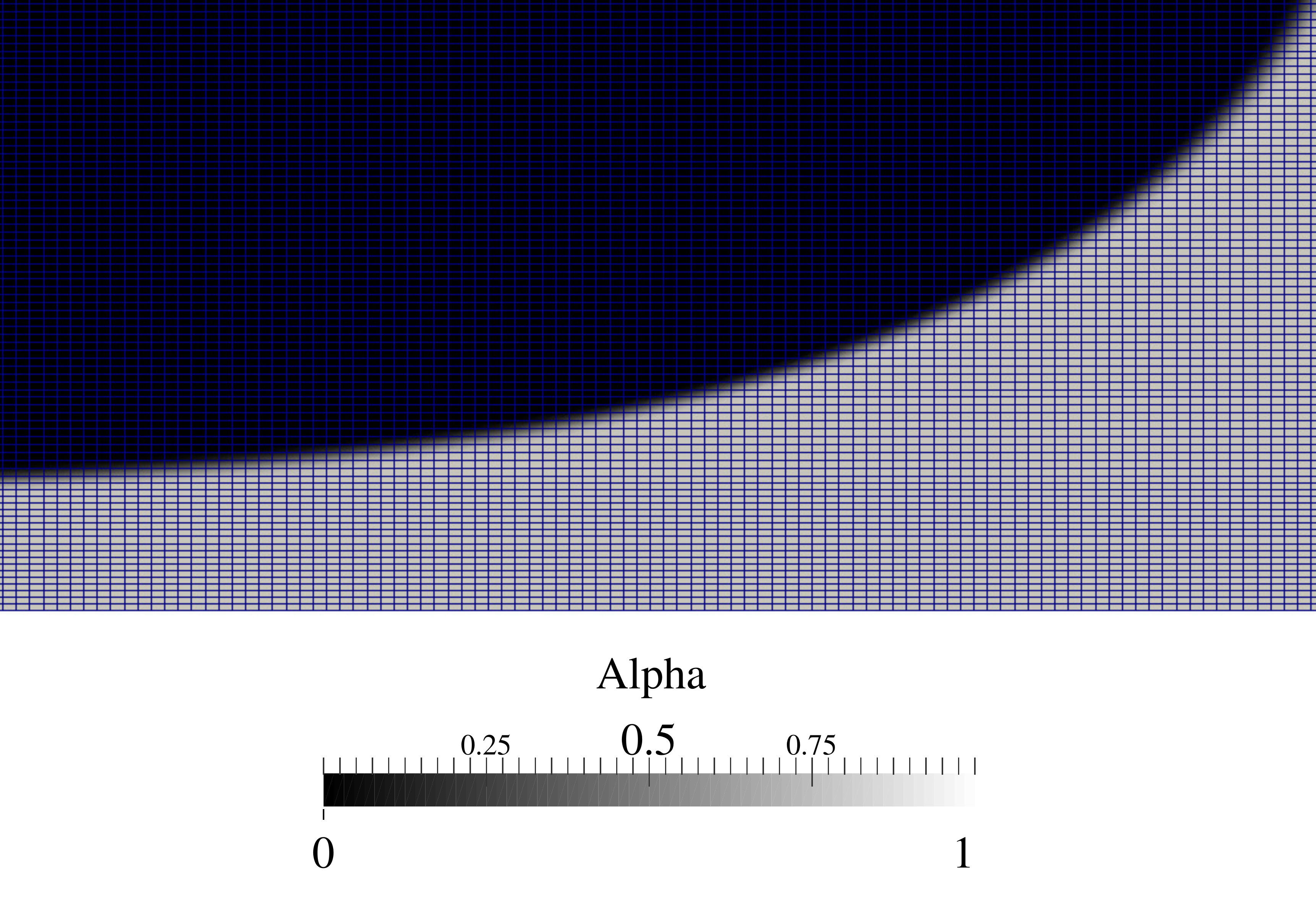}  
		\caption{\textcolor{black}{Configuration simulated with the VOF code Openfoam: a liquid plug is pushed at constant flow rate with an air finger inside a prewetted capillary tube. Up: Phase volume fraction $\alpha$. The air is represented in blue $(\alpha = 0)$ and the liquid in blue $(\alpha  = 1)$. Down: Zoom on the edge of the liquid plug to show the precision of the mesh. }}
		\label{phase}
		\end{center}
	\end{figure}
\begin{figure}
	\begin{center}
			\includegraphics[width=0.9 \textwidth]{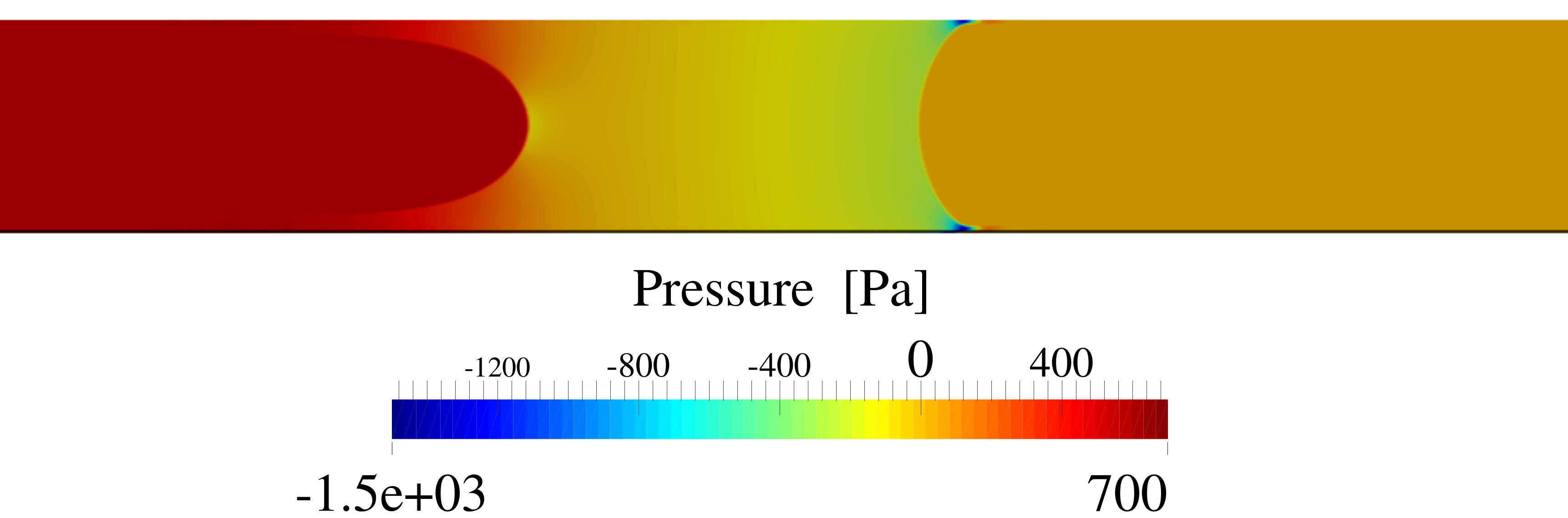} 
		\caption{\textcolor{black}{Pressure variation (in Pa) inside the plug and the air in the configuration represented on \cref{phase}.}}
		\label{pressure_simul}
		\end{center}
	\end{figure}
\begin{figure}
	\begin{center}
			\includegraphics[width=0.45 \textwidth]{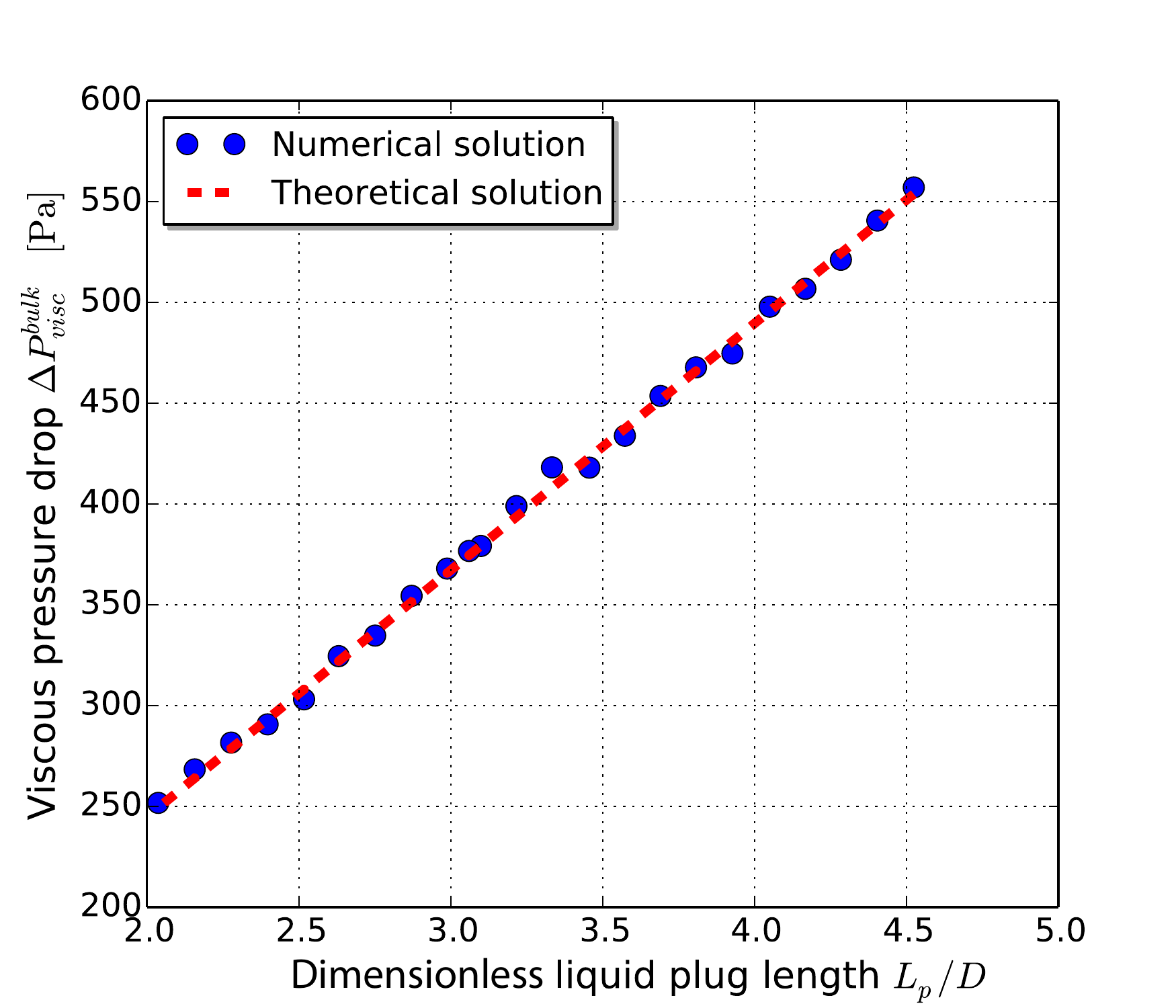}  
			\includegraphics[width=0.45 \textwidth]{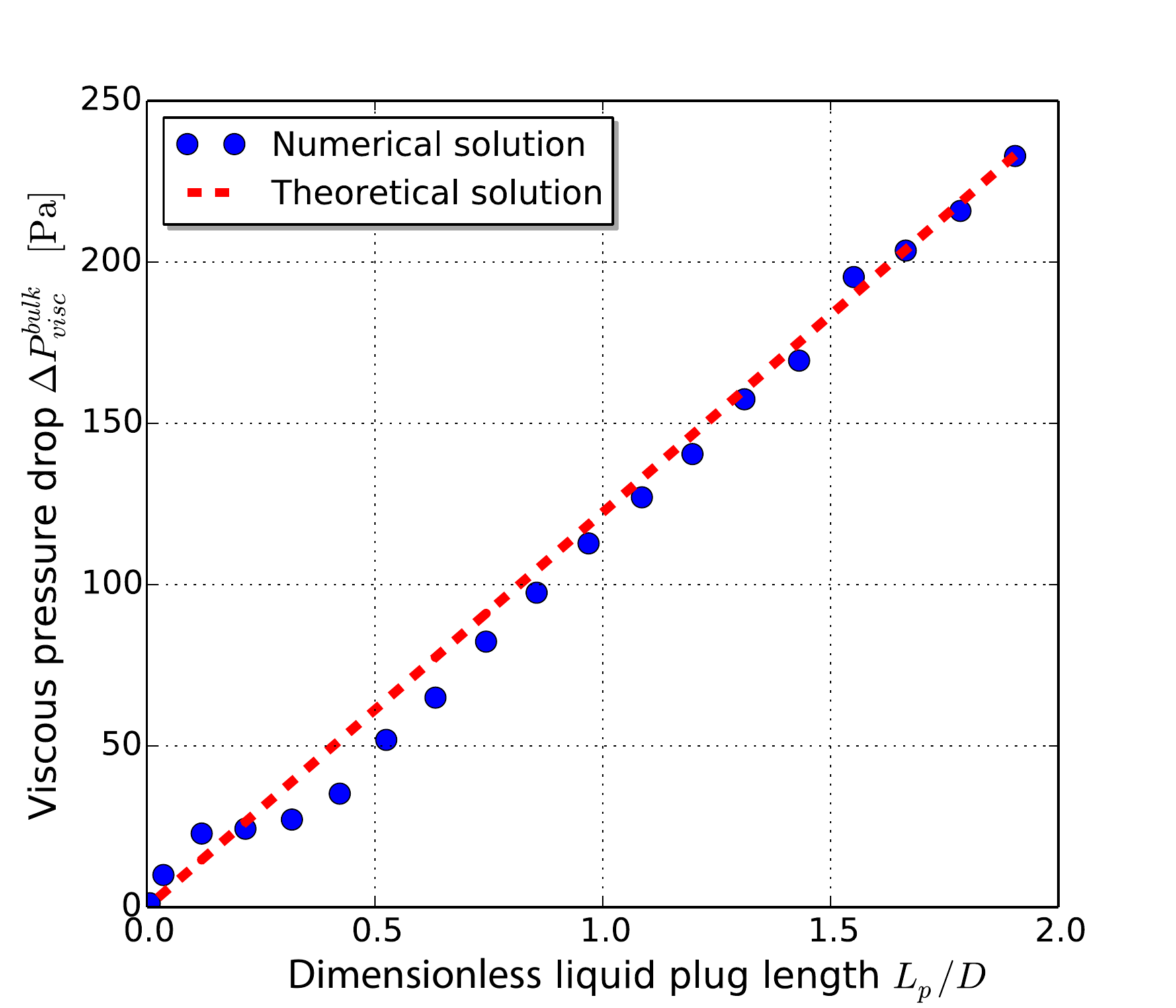}  
		\caption{\textcolor{black}{Comparison of the simulated pressure drop $\Delta P_{visc}^{bulk}$ (blue dots) and Poiseuille law (red dashed line) as a function of the plug dimensionless plug length $L_p / D$, with $D$ the tube diameter. Left: evolution for large plug. Right: zoom for small plugs.}}
		\label{poiseuille_simuls}
		\end{center}
	\end{figure}

\textcolor{black}{Equation (\ref{poiseuille}) relies on the assumption that the flow inside the liquid plug is a Poiseuille flow. In a Poiseuille flow, the liquid velocity is maximum at the center of the tube and decreases down to 0 at the walls. This flow structure is not compatible with the boundary conditions imposed by the two menisci: a constant velocity all over the liquid/air interface (in absence of interfacial deformation). Therefore, fluid recirculation will occur at the edges of the plug to match this boundary condition. For long plugs this recirculation is expected to play a minor role. Thus Equation (\ref{poiseuille}) should give a correct approximation of the pressure drop inside long plugs. However the accuracy of this approximation should decrease as the length of the plug drops. To test the validity of equation (\ref{poiseuille}), we performed 2D direct numerical simulations of a flow rate driven liquid plug in a capillary tube with the Openfoam Volume of Fluid (VOF) code (see \cref{phase}). This code was modified to include a regularisation technics \citep{cf_hoang_2013}, which reduces parasitic currents. The evolution of the viscous pressure drop as a function of the plug size was evaluated by pushing a liquid plug at a constant flow rate corresponding to $Re = 2$ and $Ca = 5 \times 10^{-2}$ in a prewetted tube (layer of thickness corresponding to $4 \%$ of the tube width $w$).  The properties of the liquid are the same as the one used in the experiments. A 825 000 points structured mesh was used with a refinement close to the walls (see  \cref{phase}). Figure \ref{pressure_simul} shows the computed pressure drop inside the liquid plug and the air. Since the input flow rate leads to a deposition of a trailing film thicker than the prewetting film, the length of the plug shrinks, which enables the evaluation of the pressure drop for various plug lengths with a single numerical simulation. The comparison between Poiseuille's law $\Delta P_{visc}^{bulk} = \frac{12 \mu L_p}{w^2} U$ and simulations is shown on figure \cref{poiseuille_simuls} (left) with a zoom on small values of the plug length on \cref{poiseuille_simuls} (right). This comparison shows that the discrepancy between the formula remains weak (below $4.5 \%$ for plugs with sizes $L_p > w$) but increases up to $25 \%$ for plugs whose size lies between $1/4 w$ and $w$. This larger discrepancy observed for small plugs is nevertheless not critical since, in this case, the viscous pressure drop is significantly smaller than the interfacial pressure drop. Thus this discrepancy will have a minor effect on the plug dynamics.}

\section{Supplementary figure.}

\begin{figure}
		\begin{center} 
			\includegraphics[height=6cm]{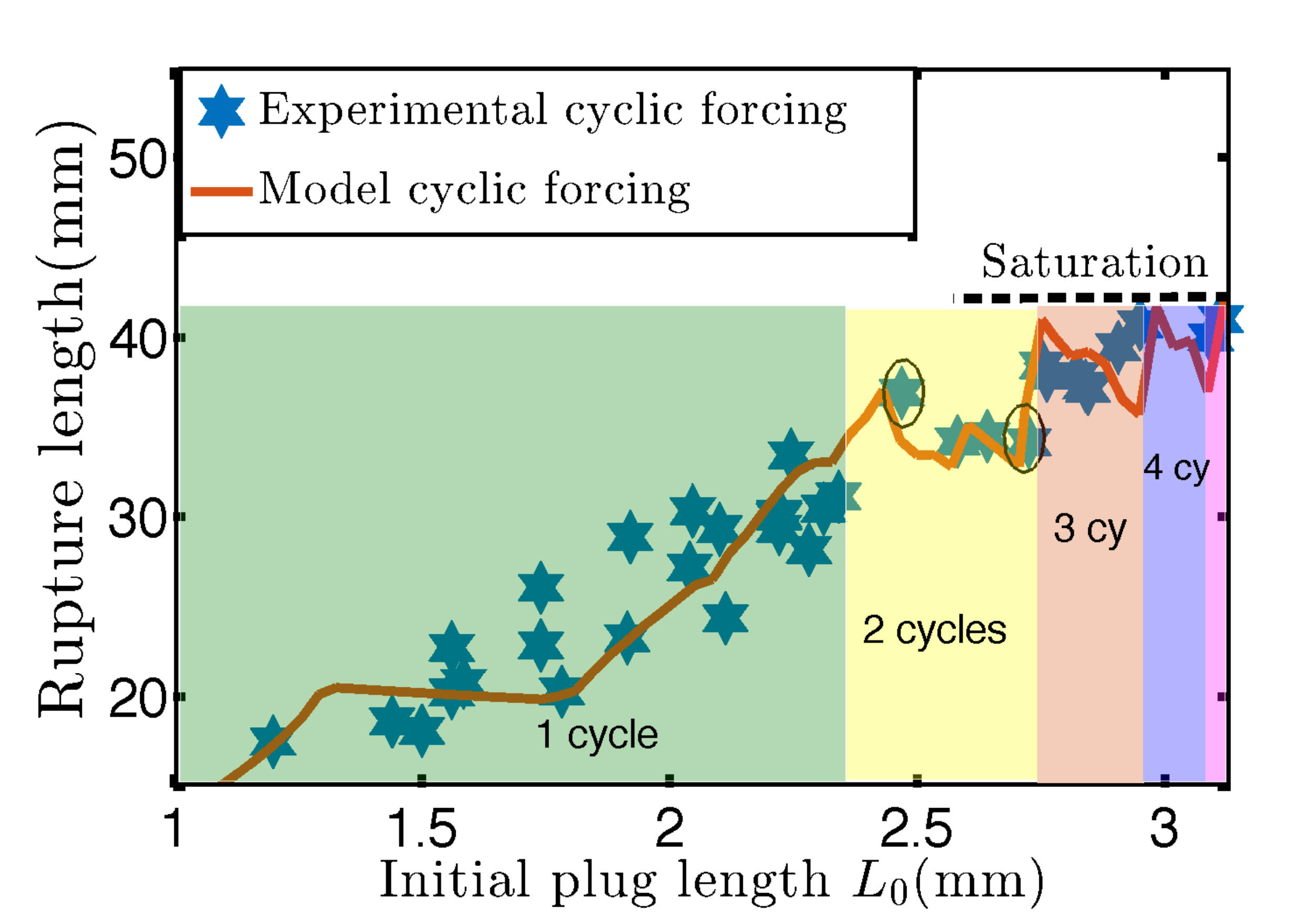}
		\end{center}
		\caption{\label{figapp} Rupture length of a liquid plug pushed with a cyclic pressure driving given by the analytical function \textcolor{black}{$\Delta P_t = 59 \exp(-6 \exp(-3.5t))$ for $t \in [0,T]$, $\Delta P_t = (-1)^n (P_c - P_d)$ for $t \in [nT,(n+1)T]$ with $P_c= 59 \exp(-2.2 \exp(-3.5 (t-nT)))$ and $P_d = 59 \exp(-1.1 (t-nT)) \exp(-0.06  \exp(-1.1*(t-nT)))$, $T = 2.1$} s the half period and $n \in \mathbb {N^*}$. Blue stars correspond to experiments, the red curve is the result from our simulations. The encircled experimental points correspond to plugs of initial lengths $L_1 = 2.5$ mm and $L_2 = 2.85$ mm, whose evolutions are compared in \cref{destruc_length}.}
	\end{figure}

The additional \cref{figapp} compares simulations and experiments of the rupture length for a different driving pressure from the one used in \cref{cycle_cloud}.

\label{append}

\newpage
 
\bibliographystyle{jfm}

\bibliography{Bibliography_jfm,rsc}

\end{document}